\documentclass[journal ]{new-aiaa}
\usepackage[utf8]{inputenc}
\usepackage{graphicx}
\usepackage{amsmath}
\usepackage[version=4]{mhchem}
\usepackage{siunitx}
\usepackage{longtable,tabularx}
\usepackage{subfloat}
\usepackage{setspace}
\usepackage{graphicx}
\usepackage{caption}
\usepackage{subcaption}
\usepackage{subcaption}
\usepackage{multirow}
\usepackage[dvipsnames]{xcolor}
\usepackage{lscape}
\usepackage{titlesec}
\usepackage[normalem]{ulem}

\title{Neural Network Flame Closure for a Turbulent Combustor with Unsteady Pressure}

\author{Zeinab Shadram\footnote{Graduate Research Assistant, Department of Mechanical and Aerospace Engineering, AIAA Student Member}}
\author{Tuan M. Nguyen\footnote{Assistant Specialist, Department of Mechanical and Aerospace Engineering, current position: Postdoctoral researcher at Sandia National Laboratories, AIAA Member}}
\author{ Athanasios Sideris \footnote{Professor, Department of Mechanical and Aerospace Engineering}}
\author{ William A. Sirignano \footnote{Professor, Department of Mechanical and Aerospace Engineering, AIAA Fellow} }
\affil{University of California Irvine, Irvine, CA, 92697}

\begin{document}
	\maketitle
		\begin{abstract}
			In this paper, neural network (NN)-based models are generated to replace flamelet tables for sub-grid modeling in large-eddy simulations of a single-injector liquid-propellant rocket engine. In the most accurate case, separate NNs for each of the flame variables are designed and tested by comparing the NN output values with the corresponding values in the table. The gas constant, internal flame energy, and flame heat capacity ratio are estimated with 0.0506\%, 0.0852\%, and 0.0778\% error, respectively. Flame temperature, thermal conductivity, and the coefficient of heat capacity ratio are estimated with 0.63\%, 0.68\%, and 0.86\% error, respectively. The progress variable reaction rate is also estimated with 3.59\% error. The errors are calculated based on mean square error over all points in the table. The developed NNs are successfully implemented within the CFD simulation, replacing the flamelet table entirely. The NN-based CFD is validated through comparison of its results with the table-based CFD.
	 \end{abstract}
	 \section*{Nomenclature} 
	 {\renewcommand\arraystretch{1.0}
	 	\noindent\begin{longtable*}{@{}l @{\quad=\quad} l@{}}
	 		$a_\gamma$ & ratio of specific heat capacity coefficient, \si{\per\kelvin}\\
	 		$C$ & progress variable\\
	 		$e$ & relative distance between two signals\\
	 		$e_m$ & relative distance between two signals mean value\\
	 		$e_f$& flame internal energy, \si{\meter \squared \per \second\squared}\\
	 		$P$ & pressure, \si{\kilo \pascal}\\
	 		$R$ & gas constant, \si{\joule \per \kilogram \per \kelvin}\\
	 	$\eta$& the ratio of one signal rms to the other one's\\
	 		$T_f$ & flame temperature, \si{\kelvin} \\
%	 		$\boldmath{x},\boldmath{y}$ & generic vectors ($x_i,y_i$; i-th the element of vectors)\\
	 		$Z$ & mixture fraction\\
	 	$Z''^2$ & mixture fraction variance from the mean\\
	 		$\gamma$ & ratio of specific heat capacity\\
	 		$\kappa$ & correlation between two signals\\
	 		$\lambda$ & thermal conductivity, \si{\watt \per \meter \kelvin}\\
	 		$\Omega$& vorticity, \si{\per \second}\\
	 		$\dot{\omega}_C$& progress variable reaction rate, \si{\kilogram \per
	 	 \meter\cubed\per \second}\\
	 		 			$\Phi$ & flamelet output set\\
	 		 			$\Psi$ & flamelet input set
	 		 				 	\end{longtable*}}	 
	 		 				 	\subsection*{Superscripts}
	 		 				 		 {\renewcommand\arraystretch{1.0}
	 		 				 		 	\noindent\begin{longtable*}{@{}l @{\quad=\quad} l@{}}
	 $\widetilde{\textcolor{white}{-}}$ & 		density-weighted Favre average\\
	 	$\overline{\textcolor{white}{-}}$& Reynolds average\\
	 	 $\widehat{\textcolor{white}{-}}$& empirical mean of a set
	 	\end{longtable*}}	 
\section{Introduction}
\lettrine{C}{omputationally} efficient and accurate flamelet models for a turbulent combustor are needed for useful large eddy simulations (LES). Promise is offered through the use of deep neural networks (NN). Here, a well-studied configuration through prior LES and experiment is used but now with NN providing the sub-grid model for the flamelets. It is of special interest to extend the use of NN modeling for unsteady behavior of mean pressure and velocity fields. Both transient and dynamic equilibrium oscillatory conditions are considered.

Combustion instability is an acoustical phenomenon caused by the high rate of energy release that increases pressure oscillation amplitude inside a combustion chamber; i.e., combustion excites and sustains an unstable high-amplitude pressure oscillation, which can be destructive. While most rockets and jet engines can oscillate, their unstable behaviors can differ. The configuration here is a single-injector liquid-propellant rocket engine (LPRE), see
\cite{CrocCheng2,culick1995overview}. 
Models for LPRE combustion instabilities has been a research subject for decades \cite{Poinsot2,SigPop2,SigPop3}.
 Computational fluid dynamics (CFD) models have been proposed for detecting combustion instabilities in LPREs. The Continuously Variable Resonance Combustor (CVRC) experiment of Purdue University provides a fundamental test case with available data \cite{Yu2,Yu3}.
 CVRC is a combustion chamber with a single-element injector/oxidizer post, in which the oxidizer post length can be varied, resulting in configurations with different stability characteristics. Although it uses gaseous injection, the CVRC is an experiment that is widely accepted as a valid benchmark for computational methods that address combustion instability in liquid-propellant rocket engines.
 
 Current computational capabilities for reacting flows in realistic combustion chambers do not allow for resolution of the smallest scales of importance. Therefore, LES rather than direct numerical simulations (DNS) are common \cite{Poinsot1,Harva3}. 
 This requires models for sub-grid phenomena, especially for the combustion process. For gaseous reactants, the flamelet model (\cite{peters_2000,pierce_moin_2004}) enjoys popularity. The sub-grid model has been employed through a look-up table approach (\cite{Tuan1, Tuan3,Tuan4}) that avoids the need for a computational time step determined by the chemical kinetics time scales, which typically are shorter
 than the ones from the numerically resolved physics.
 Nguyen et al. developed a hybrid LES/Reynolds-Averaged Navier-Stokes code, capturing combustion instabilities in CVRC at a much lower computational cost compared to prior works \cite{Tuan1}. The results from \cite{Tuan1} were compared favorably with experimental data and numerical simulations of CVRC experiment developed in Purdue University \cite{Harvazinskithesis,Yu2}. Implementing the flamelet model was a key step in reducing the computational cost.
 
 \citet{peters_2000} introduced the flamelet concept for turbulent combustion modeling. 
 The flamelet model for turbulent combustion, based on the non-premixed flame physics, has been developed further by \citet{pierce_moin_2004} for the integration of a sub-grid model with LES. The flamelet model is a sub-grid model, and its raison d'etre is that one cannot afford the grid resolution needed to incorporate combustion details in the LES. With the required resolution, the sub-grid model would not be needed. The flamelet model is based on the assumption that time scales for heat and mass diffusion, advection, and strain are larger than the chemical times; thereby, quasi-steadiness for chemistry is used. However, the chemistry is localized in narrow regions, and the stronger assumption of chemical equilibrium throughout the flow is not employed; i.e., there are narrow flame regions.
While using flamelet models results in computational efficiency, they are the product of simplifying assumptions and they have certain limitations. There are uncertainties associated with presumed sub-grid PDF distributions as well as the progress variable definition.
Flamelet theory assumes an axisymmetric strain field while three-dimensional behavior is commonly found in practice. A single diffusion flamelet occurs with fuel only on one side of the flame and oxygen only on the other side. Experience indicates that combustible mixtures can exist on one or both sides; premixed flames, diffusion flames, and multi-branched flames (e.g., triple flames) can occur. New flamelet models addressing these issues are in development (\cite{Sirignano_coufl2019, SirignanoWSSCI, ClaudiaWSSCI,Sirignano2019T&F}).
 
Neural networks have been used in many different applications for the purpose of classification, clustering, and regression analysis. Previous studies have also used NNs as a tool to develop closure models in fluid mechanics and even combustion problems. Recursive NNs were utilized in an early work to predict the unsteady boundary-layer development \cite{faller_unsteady_1997}. NN closure models have been incorporated for turbulence modeling in \cite{okstateSAN2018681, CMexample2, CMexample_2016}.
 Menon and co-workers (\cite{SEN2010566, SEN201062}) used NN for the coupling of the linear eddy mixing (LEM) sub-grid model with the LES. 
 \citet{IHME_SCHMITT_PITSCH_PCI32} had developed optimal artificial NNs to compute flame variables and compare their performances with flamelet tables with different resolutions for a stably burning flame. Among the findings in \cite{IHME_SCHMITT_PITSCH_PCI32} are that: (1) using NNs requires much less memory than a look-up flamelet table, (2) NNs can obtain a smoother flow field solution, and (3) NNs require higher computational cost than the flamelet table; yet the computational cost of using NNs is still significantly lower than solving the flame equations. Given the positive findings of \cite{SEN2010566, SEN201062,IHME_SCHMITT_PITSCH_PCI32}, we envision that future versions of the flamelet model might best be introduced through the use of NNs, allowing nonlinear interpolation that is not easily obtained with the table, which now uses linear interpolation. NN can also allow both experimental and theoretical data to be used.
Here, we use the data from the look-up table for training the NNs. In attempting to study the capability of the NNs not only to reproduce combustion behavior but also its integration into CFD calculation, flamelet tables provide a good stepping stone, as evidenced by the encouraging results shown in this work. In the future with new flamelet models, we propose moving directly to the flamelet calculations (possibly augmented by experimental data).

The objective of our work is to develop NN-based closure models suitable for studying combustion instability.
 The key factor in such a model is the coupled relation between pressure and flame variables, specifically the progress variable reaction rate (PVRR) and heat release rate (HRR). Combustion instability leads to a considerably huge change in pressure, which affects the flame behavior. Thus, it requires the flamelet table to cover a large range of pressure; in fact, \citet{Tuan1} needed to generate a flamelet table that takes pressure as an input. As an example, the maximum amount of HRR at 30 \si{atm} is 650.5\% of the maximum HRR in 8 \si{atm}; 8 \si{atm} and 30 \si{atm} determine the pressure range for the CFD simulation of CVRC model with a 14-cm oxidizer post after reaching dynamic equilibrium state. 
A goal is to develop NNs for calculating flame variables based on a pressure-dependent flamelet model. In the earlier stages of the work, the approach was to develop a model that calculates flame related variables from the CFD simulation, replacing the function described within the dashed line in the schematic in \figurename{~\ref{toplevel}}, \cite{MyPaper1_2019,ShadramWSSCI}. 
In this paper, however, the approach is to study the data inside the flamelet table directly and to develop NNs that focus only on the flamelet model outputs; therefore the NNs are trained on the flamelet table data, that was used before in \cite{Tuan1}. Several different NN structures are developed. In the most accurate case, separate NNs for each of the flame variables are designed and tested by comparing their output with the output of the similar variable in the table.
 
In Sec.~\ref{prelim}, our numerical simulation and combustion modeling method is discussed.
In Sec.~\ref{CMNN}, the design and evaluation of the proposed NNs are discussed. In Sec.~\ref{Res}, the NNs are tested on the flamelet table and are also implemented into different CFD simulations. The paper concludes with Sec.~\ref{conclusion}.

 \section{Background on Combustion Instability Analysis} \label{prelim} 
\subsection{More on the Numerical Simulation}
\begin{figure}[hbt!]
	\centering
	\begin{subfigure}{.6\textwidth}
		\centering
	\includegraphics[width=0.95\textwidth]{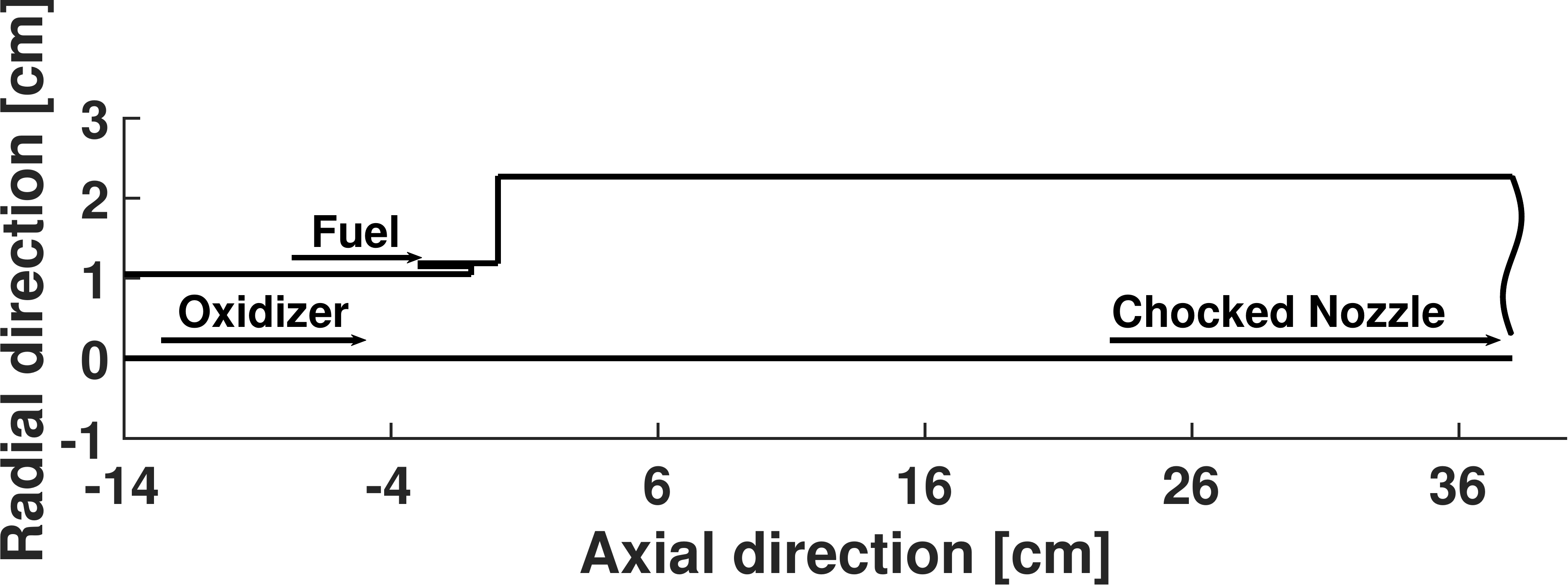}
\caption{ CVRC configuration \cite{Tuan1}}	
\label{CVRC}		
		\centering
		\includegraphics[width=0.95\textwidth]{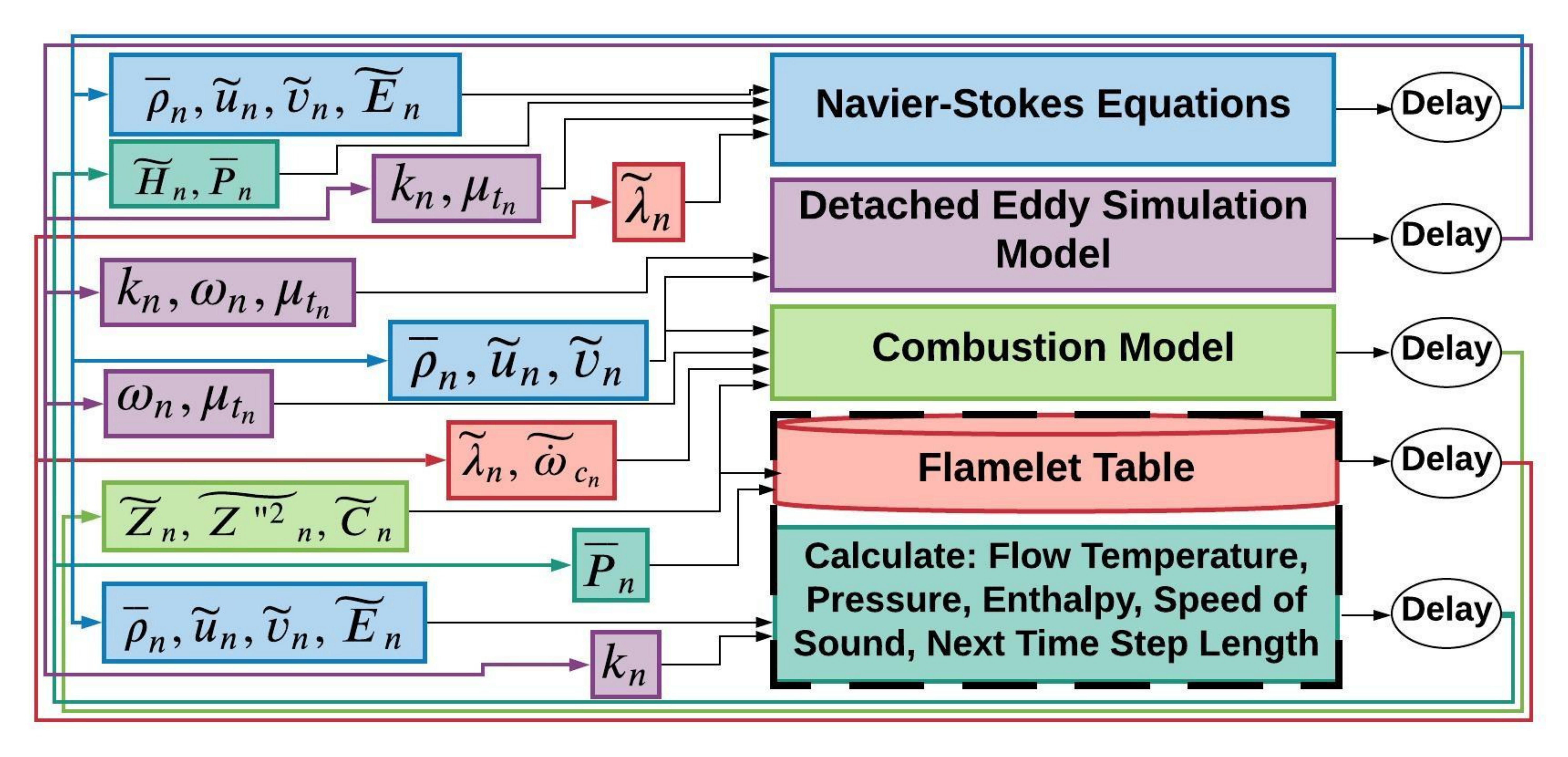}
		\caption{Top level architecture of CFD code}
		\label{toplevel}
	\end{subfigure}
		\caption{Overview of the computational domain for the CVRC experiment and its CFD code architecture}
\end{figure}

The CVRC experiment is a single-injector co-axial dump combustor \cite{Harva1}. Methane is injected through the outer concentric tube at 300~\si{\kelvin}. The oxidizer, which is injected through the inner tube, is composed of 58\% $H_2O$ and 42\% $O_2$ at 1030~\si{\kelvin}. Both reactants are injected at constant mass flow rates of $0.32$~\si{\kilogram\per\second} and $0.027$~\si{\kilogram\per\second} for the oxidizer and fuel, respectively. The 0.8 equivalence ratio makes the flow globally fuel-lean. Here, an unstable configuration with a 14-cm oxidizer post length and a 38-cm combustion chamber is the test case; see \figurename{~\ref{CVRC}}. Also, a stable configuration is discussed with a 9-cm oxidizer post length and a similar 38-cm combustion chamber.

A constant mass flow rate inlet boundary condition is implemented using the Navier-Stokes Characteristic Boundary Conditions \cite{Poinsot4} at both reactant inlets. To save computational resources, a short-choked-nozzle \cite{CrocSig} outlet boundary condition is used instead of an actual convergent-divergent nozzle computational domain; this promotes a high amplitude pressure oscillation. The mesh consists of 137,494 grid points. Its structure is based on the mesh used in the 3D calculations of Srinivasan et al. \cite{Menon1}. The smallest radial grid size is 0.05 \si{\milli\meter}, located around the mixing shear layer or any walls. The smallest axial grid size is 0.2 \si{\milli\meter}, located both upstream and downstream of the back step. The maximum grid stretching factor along any direction is 1.05, thus ensuring high-quality mesh. Our code is a multi-block, structured finite-difference solver with axisymmetric cylindrical coordinates. (See \figurename{~\ref{toplevel}} for a high-level view.) The overall accuracy is second-order in space and fourth-order in time. An important feature is the shock-capturing capability of the code longitudinal-mode.
 The entrance-to-throat area ratio is 5. On the wall, the no-slip boundary condition is applied.
 The walls are also assumed to be adiabatic and impermeable. Data is acquired from the CFD simulation at a 200 \si{\kilo\hertz} sampling rate from all points. 

\subsection{Combustion Modeling and Challenges}
The flamelet approach models the turbulent flame as a collection of laminar flamelets, where the chemical time scales are shorter than the turbulent time scales. Accordingly, the chemistry-related calculation occurs prior to the main flow simulation, through a quasi-steady flamelet table for a diffusion flame under normal compressive strain \cite{peters_2000, pierce_moin_2004,Tuan1}.  Decoupling flamelet solution and the resolved LES flow simulation allows the prediction of the complex mechanism at a much less computational cost, yet with much less computational cost, which is the advantage of flamelet modeling (\cite{Tuan3}). 
As shown in \figurename{~\ref{toplevel}}, 
the inputs to the flamelet model are the flame state variables ($\widetilde{C}$, $\widetilde{Z}$, and $\widetilde{Z"^2}$) and $\overline{P}$. 
%The combustion state variables are progress variable ($\widetilde{C}$), mixture fraction ($\widetilde{Z}$), and its variance ($\widetilde{Z"^2}$).
 $\widetilde{Z}$ indicates the mixture character at each point. $\widetilde{Z"^2}$ defines the deviation from the mean of $\widetilde{Z}$ and is important for turbulent combustion. Essentially, in turbulent non-premixed modeling, laminar flamelet solutions are convoluted by a subgrid beta probability density function (pdf) of Z to model turbulence-chemistry interaction. The beta pdf parameters are functions of $\widetilde{Z}$ and $\widetilde{Z"^2}$. $\widetilde{C}$ determines how much of the combustion process has been conducted at each time and point in space. A Dirac delta pdf is assumed to relate $C$ to $\widetilde{C}$ according to the discussions in \cite{pierce_moin_2004}. The inputs of the flamelet model are collected in $\Psi=[\widetilde{Z},B,\widetilde{C},\overline{P}]$, where $B=\frac{\widetilde{Z"^2}}{\widetilde{Z}-\widetilde{Z}^2}$ is a surrogate variable that replaces $\widetilde{Z"^2}$. In the input set, $\overline{P}$ varies between 1~\si{atm} and 30~\si{atm}, $B$ and $\widetilde{Z}$ range between 0 and 1, and $\widetilde{C}$ varies from 0 to 0.261. The table includes 30 grid points for $\overline{P}$, 85 grid points for $\widetilde{Z}$, 26 grid points for $\widetilde{Z"^2}$, and 156 grid points for $\widetilde{C}$.
The output set for this flamelet model is defined as $\Phi=[\widetilde{\dot{\omega}}_C,\widetilde{T}_f,\widetilde{e}_f, \widetilde{R},\widetilde{\lambda},\widetilde{\gamma},\widetilde{a}_\gamma]$. The proposed NN-based models in this work take exactly the same inputs and outputs as the inputs and outputs of the flamelet model stored in $\Psi$ and $\Phi$, respectively.
In the CFD structure, flow internal energy along with the above quantities are used to calculate flow temperature and enthalpy. Pressure is calculated based on the ideal gas law after extracting data from the flamelet table in the CFD. 
 
 The main difficulty with the flamelet model is that the HRR becomes a one-dimensional quantity, although embedded in a three-dimensional flow field. Furthermore, in the case of combustion instability, the HRR interactions with turbulence and acoustical phenomena result in extremely nonlinear behaviors. 
Implementing the flamelet model helps to preserve the multi-scale and highly nonlinear behavior of HRR in the CFD simulation. In the flamelet-based simulation, HRR is not playing a direct role in the governing equation; instead, the PVRR is the variable that affects the governing equations as the source term in the $\widetilde{C}$ transport equation.
 The flamelet solutions use a complex chemical mechanism (72 reactions with 27 species).
Heat release rate relates to the products of the reaction rates of all the species and their corresponding enthalpies; in contrast, the progress variable, defined as the summation of CO2 and H2, only represents the major global chemical reactions.
 HRR accounts for more detailed (including less significant) reactions than the PVRR, which indicates the progress of the global methane reaction,
using only simple global oxidation chemistry. However, it still has similar multi-dimensional, multi-scale, and highly nonlinear behavior, yet less severe and less costly to model than HRR. 

\subsection{Rayleigh Index as an Instability Criterion} 
The most important cause of the high-frequency instability is the coupling between the HRR 
 and the acoustic pressure wave. The Rayleigh Index (\textit{RI}) measures this coupling based on the HRR and pressure fluctuation correlation determining if the HRR drives or damps the pressure wave. The time-averaged local Rayleigh Index is defined over a time period ($\tau$), typically few cycles, starting from an initial time $t_o$ as:
 
\begin{equation}
\label{RIdef}
RI=\frac{1}{\tau}\int_{t_o}^{t_o+\tau}\frac{\widetilde{\gamma}+1}{\widetilde{\gamma}}\times\frac{p'}{\widehat{P}}\times\frac{HRR'}{\widehat{HRR}}dt 
\end{equation}

where $p'$, and $HRR'$ are the local fluctuations in $\overline{P}$ and HRR, respectively \cite{Tuan1}. Also, $\widehat{P}$ and $\widehat{HRR}$ are the global time averages of $\overline{P}$ and HRR.
A positive (negative) value of \textit{RI} conveys that the pressure oscillation is driven (damped) by HRR. The importance of \textit{RI} led us to discuss a similar measure for PVRR oscillations by replacing HRR with $\widetilde{\dot{\omega}}_C$ in Eq.~\eqref{RIdef} to get a modified Rayleigh Index (\textit{mRI}).

\section{Neural-Network-based Closure Model} \label{CMNN}
The objective here is to design 
a NN-based closure model to replace the flamelet table in \figurename{~\ref{toplevel}}.
Therefore, exactly the same sets of inputs and outputs of the flamelet model, i.e., $\Psi$ and $\Phi$, are selected as the NN-based model input and output sets.
A NN is a computational unit that replaces an input/output block in a system implementing its original task. After training, the NN learns to perform a task based on input-output examples without knowing the algorithm that led to those examples. A deep NN contains a series of layers, each of which has several nodes (neurons). 
Each neuron gets a linear combination of the outputs from neurons in the previous layer as its input, and provides an output, i.e., a specific nonlinear function (activation) of its input used for the next-layer calculations.
The objective of the training process is to find the coefficients of those linear combinations, i.e., the weights, through solving an optimization problem to minimize the square of absolute error between NN outputs and the originally provided examples, i.e.,
 	the training data. The square of absolute error is selected as the cost function of the optimization problem to assure better estimation performance for the data with higher magnitude.
 The training is performed with the back-propagation method using the RMSProp update rule \cite{murphy2013machine}. 
 We used the Leaky Rectified Linear ($LReLU(x)=max(x,0.001x)$) Activation Function in the hidden layers, and linear activation function at the output layer \cite{ReLu}. The NN weights are given the random Xavier initialization \cite{pmlr-v9-glorot10a}.
If the training process is overextended, typically, the NN performance is only good for the training set; this phenomenon is called overfitting. To avoid this, a validation error is computed on an independent set of samples and used to monitor overfitting \cite{murphy2013machine}. 
 Providing the appropriate training and validation data sets are key steps in designing a NN. 
	The optimal NN weights are obtained as those minimizing the error on the validation set. 
	 The inputs and outputs are standardized before introduction to the NN, by being centered around their mean and normalized by their standard deviation, to improve training performance \cite{murphy2013machine}. 
 
\subsection{Sample Selection}
Enough samples must be provided to capture the problem complexity, while 
 limiting the computational cost of the training. 
The flame-variable modeling is challenging because it is multi-scale; e.g., 
PVRR varies from $-284$~\si{\kilogram \per \meter\cubed \per\second} to $1.28e5$~\si{\kilogram \per \meter\cubed\per\second}. Combustion instability occurs from the coupling of high HRR and pressure; thus, the NN must accurately predict points with high HRR. However, the population of these points is relatively smaller than other points, requiring a nonuniform sample selection approach.
 The samples should be selected to provide a rich data set from high HRR points, with enough samples that should be selected from low HRR points to avoid biasing the NN towards uniformly high HRR values.
 Based on the designed table resolution,
 there exist around 10 million cells in the table, which are used for selecting training and validation data sets. Initially, a data set is selected uniformly from the available data in the table, then enriched with more points from higher energy release zones.
 At this stage, around 500,000 data points are selected for training, and around 200,000 data points are selected for validation representing 5\% and 2\% of the available points, respectively. After several iterations, when the learning progress starts to slow down, the partially trained NN is tested on additional data from the table.
 Then, points with errors higher than a specified bound (e.g., 5\%) are added to the training set and points with errors higher than a lower bound (e.g., 2\%)
 are added to the validation set. 
 Here, different NNs are trained for each output of the table; so, after this step, the training and validation sets might be different among outputs.
 A maximum limit of 1.5 million points per set was established after the modification steps.
 
\subsection{Network Structure}
A general feed-forward NN comprises of layers of different numbers of neurons, which communicate through the weighted links. 
The focus of this work is to explore the effectiveness of NNs as closure models in turbulent combustion; optimizing the NN structure was beyond the scope of the work. 
	We arrived at the proposed structures in trial-and-error fashion by increasing the number of neurons in each layer and the number of layers to a point that good accuracy on the training and validation sets was achieved.
Our observation is that increasing the number of NN layers leads to better accuracy than increasing the number of neurons in a single layer.
 Hence, NNs with more layers and fewer neurons on each layer are selected. 
 In particular, two NN structures are introduced here.
 In the first NN, based on the physical behavior of flamelet data, the outputs are grouped as $G_1$: $\widetilde{\dot{\omega}}_C$, $G_2$: $\widetilde{e}_f$, $\widetilde{R}$, $G_3$: $\widetilde{T_f}$, $\widetilde{\lambda}$, $G_4$: $\widetilde{\gamma}$ and $G_5$: $\widetilde{a}_\gamma$. ($G_i$ denotes a group), and five different NNs (6-layer) are designed, one for each of these sets of outputs, each having 5 hidden layers with 15, 20, 25, 20, and 15 neurons, respectively. The input layer has 4, and the output layer has 1 or 2 neurons depending on the output group. 
A single output NN with the above structure requires 1771 floating point operations (flop) for data retrieval, while a double-output NN requires 1787 flop. This NN model, referred as $NN_a$, requires 8887 flop for overall flame modeling.
Another set of NNs, referred as $NN_b$, improves the accuracy by increasing the number of layers and using
 7 single-output NNs (8-layer) each with 7 hidden layers with 15, 20, 25, 30, 25, 20, and 15 neurons, respectively. Data retrieval in this NN requires 3326 flop for a single variable, and 23,282 flop for all variables. 
 While the structure proposed in $NN_b$ increases the computational cost of retrieving flame data by 162\% over $NN_a$, it gives an approximately 40\% reduction in the reconstruction Mean Square Error (MSE).

\section{Results }\label{Res}
The closure model for flame variables must be implemented inside the CFD. Both the single-output and the multi-output NNs discussed in Sec. \ref{CMNN} are implemented to replace the flamelet table.
The designed NNs are tested in a stand-alone phase and an in-situ phase. In the stand-alone phase (offline), the NN is tested on all the available table data regardless of whether they are used or not for the training and validation, giving error bounds for each NN output compared
to the corresponding table output. 
The second (online) phase tests whether the NN model can satisfactorily model combustion instability in the single injector LPRE case. 
 Particularly, the highly coupled dynamics governing the LPREs makes combustion modeling very sensitive and complicated. Through the online test, NNs
 are successfully implemented within the CFD, and the are compared with the table-based simulations.
In the following, the offline and online test results for $NN_a$ and $NN_b$ are provided.

\subsection{Offline test}
The NNs are tested on all table data, while at most 15\% of the data is used for the training process.
\tablename{~\ref{tabOFFres}} presents the relative MSE for each variable calculated from Eq.~\eqref{mseoff}; 
 where $x_i$ is an element from the $n$ elements in the flamelet table, and $y_i$ is the NN-estimated value for that element.

 \begin{equation}
 \label{mseoff}
 e(\%)=100\times\frac{\sqrt{\sum_i^n{(y_i-x_i)^2}}}{\sqrt{\sum_i^n{x_i^2}}}
 \end{equation}
 
\begin{table}[hbt!]
	\centering
	\caption{Comparison of relative MSE (\%) results for $NN_a$ and $NN_b$ on all data in the flamelet table-based CFD} \label{tabOFFres}
	\begin{tabular}{|c|c|c|c|c|c|c|c|}
		\hline
		Model &$\widetilde{R}$ & $\widetilde{e}_f$ & $\widetilde{T_f}$ &$\widetilde{\dot{\omega}}_C$ & $\widetilde{\lambda}$ & $\widetilde{\gamma}$ & $\widetilde{a}_\gamma$ \\ \hline
$NN_a$ & 0.0889 & 0.14 & 0.99 & 5.96 & 1.04 & 0.14 & 1.58 \\ \hline		
		$NN_b$ & 0.0506 & 0.0852 & 0.63 & 3.59 & 0.68 & 0.0778 & 0.86 \\ \hline
	\end{tabular}
\end{table}

\begin{figure}[hbt!]
	\centering		
			\begin{subfigure}{.49\textwidth}
		\centering
		\includegraphics[width=0.98\textwidth]{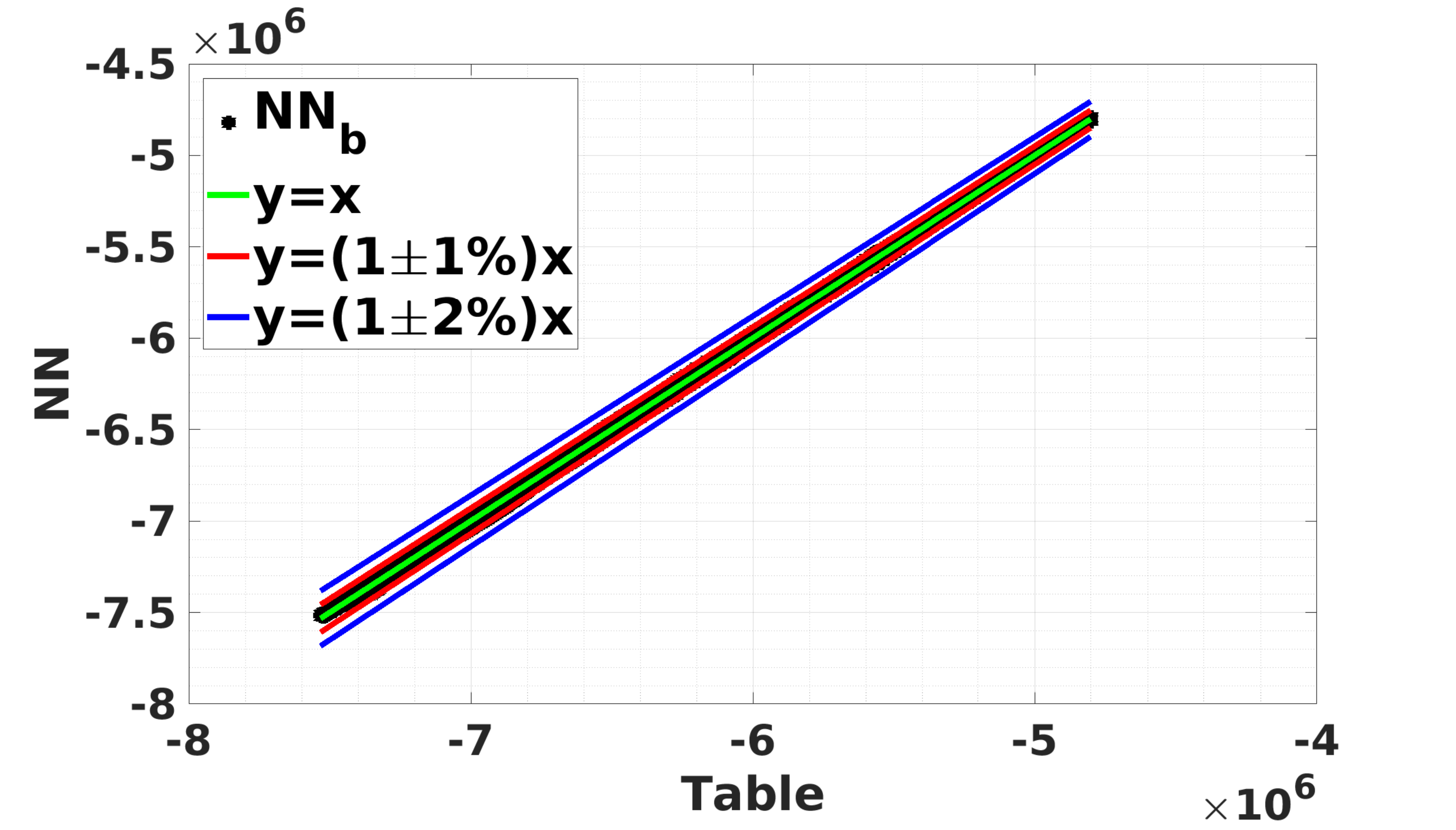}
		\caption{$\widetilde{e}_f$ (\si{\meter\squared\per\second\squared}), $NN_b$}\label{9lrsIEfNN}				
			\end{subfigure}				
			\begin{subfigure}{.49\textwidth}
				\centering		
				\includegraphics[width=0.98\textwidth]{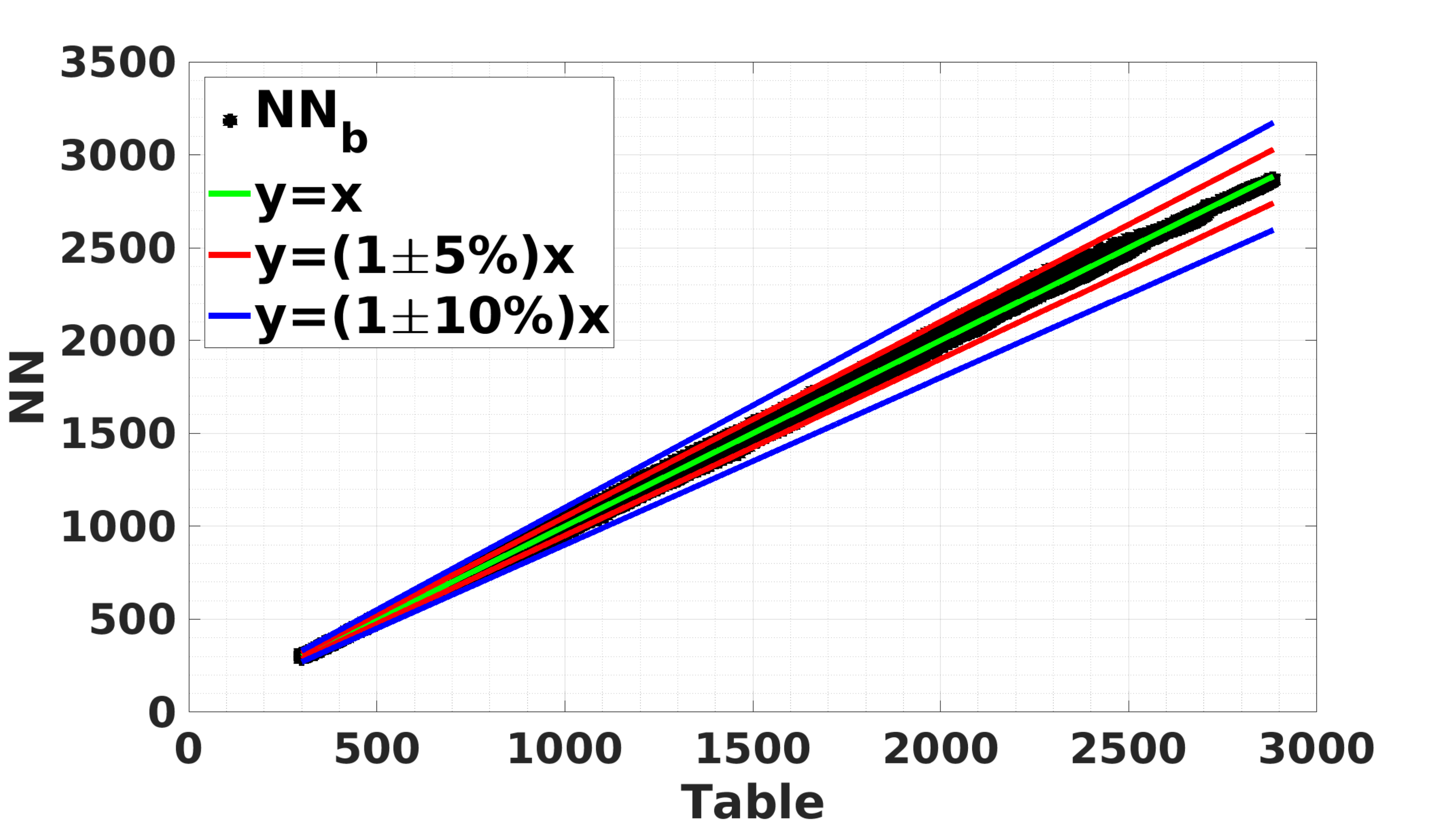}
				\caption{$\widetilde{T_f}$ (\si{\kelvin}), $NN_b$}\label{9lrsTfNN}			
			\end{subfigure}					
			\begin{subfigure}{.49\textwidth}		
				\centering
				\includegraphics[width=0.95\textwidth]{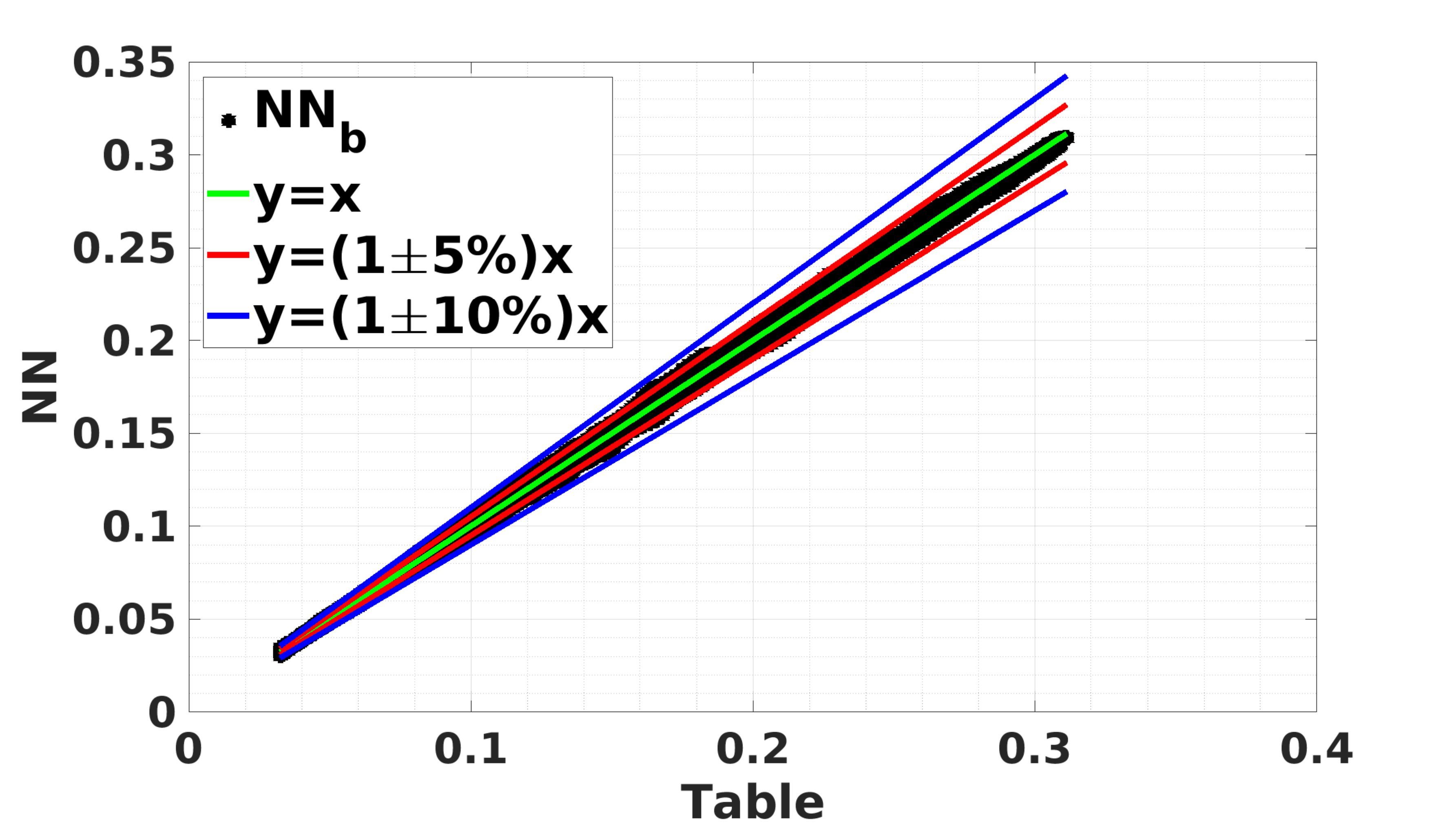}
				\caption{$\widetilde{\lambda}$ (\si{\watt\per\meter\per\kelvin}), $NN_b$}\label{9lrslambdaNN}			
			\end{subfigure}								
			\begin{subfigure}{.49\textwidth}
					\centering
				\includegraphics[width=0.95\textwidth]{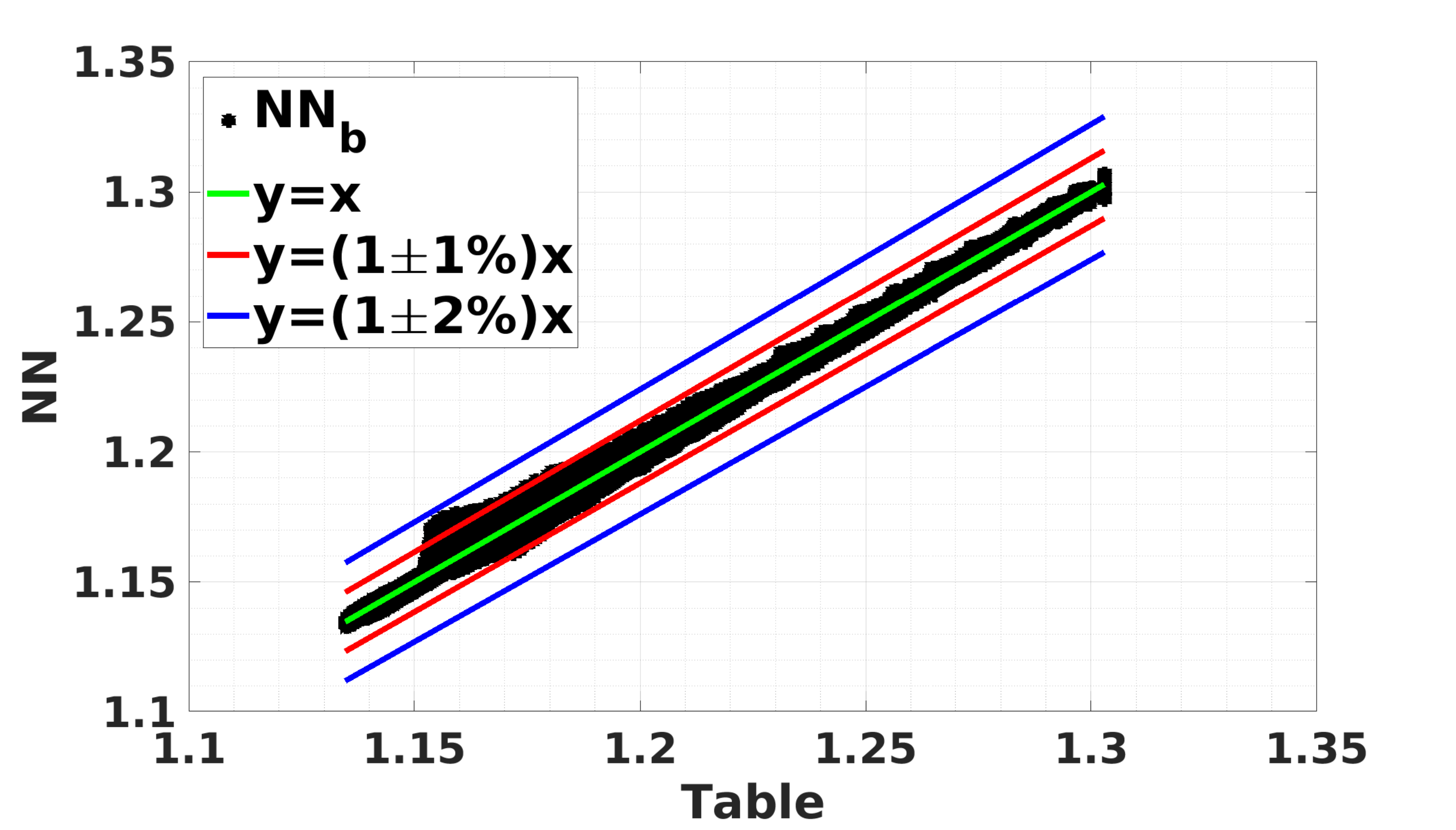}
				\caption{$\widetilde{\gamma}$, $NN_b$}\label{9lrsgammaNN}						
			\end{subfigure}					
			\begin{subfigure}{.49\textwidth}			
				\centering
				\includegraphics[width=0.95\textwidth]{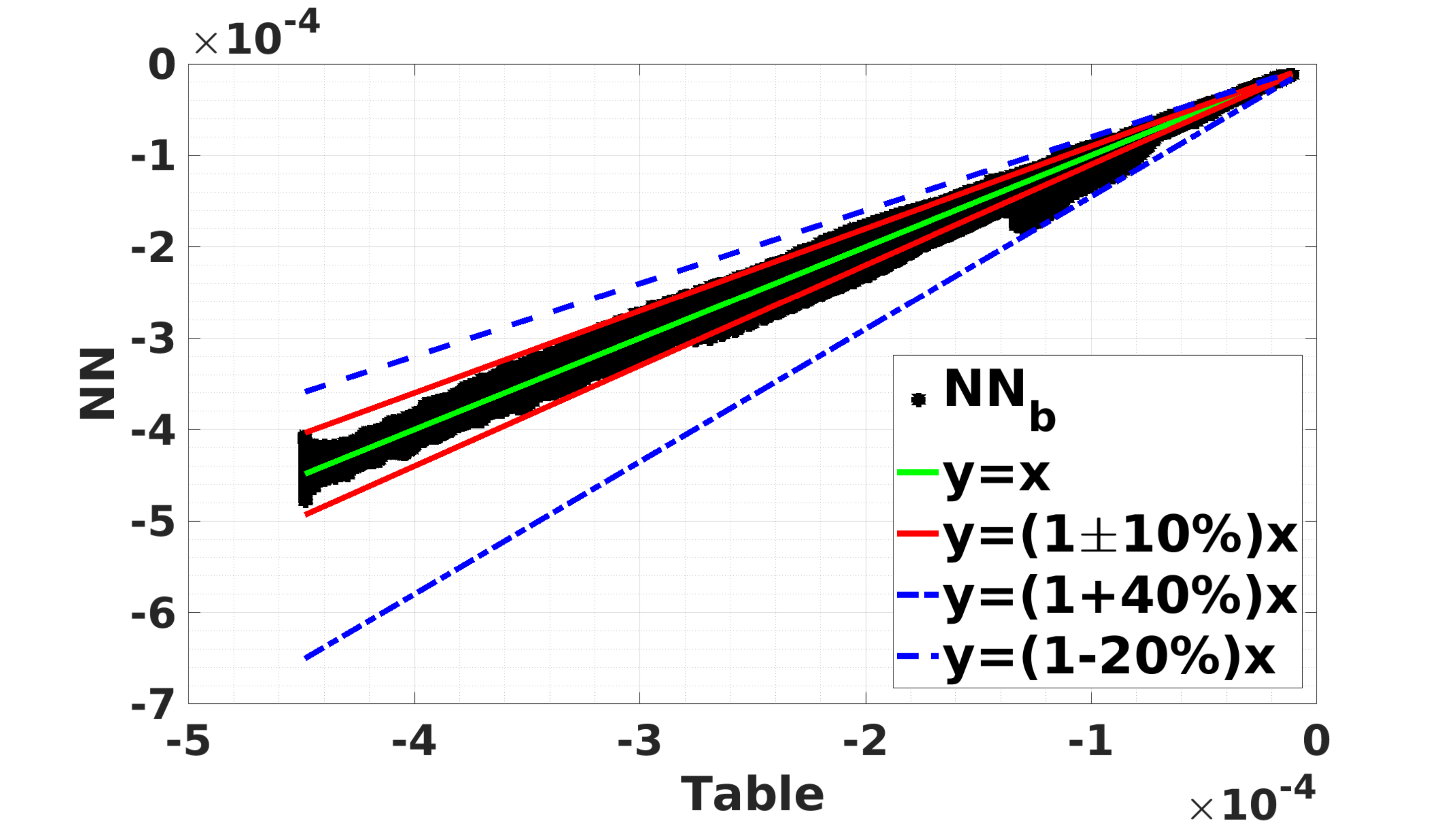}
				\caption{$\widetilde{a}_\gamma$ (\si{\per\kelvin}), $NN_b$}\label{9lrsagammaNN}				
			\end{subfigure}											
			\begin{subfigure}{.49\textwidth}								
				\centering
				\includegraphics[width=0.98\textwidth]{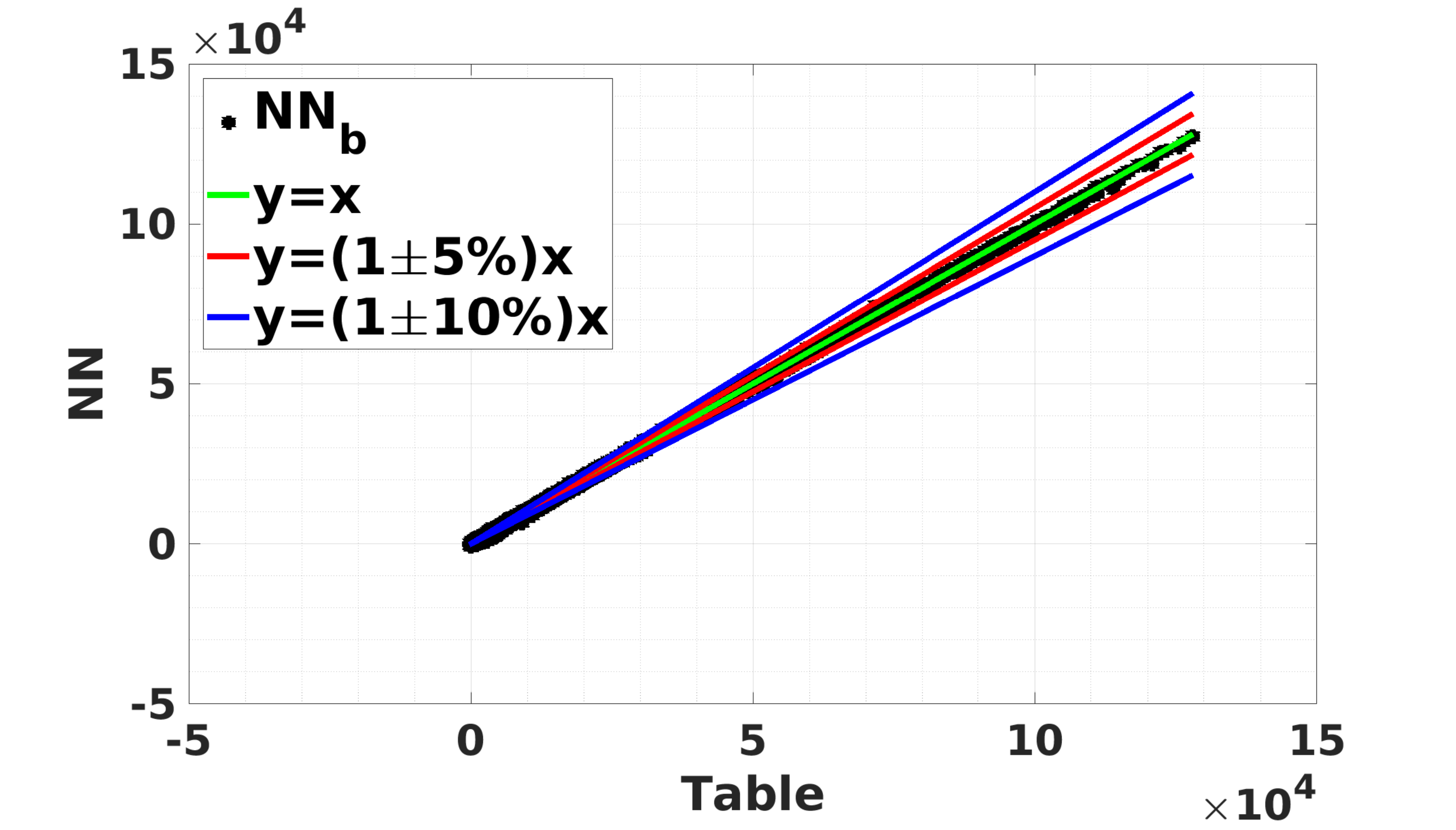}
				\caption{$\widetilde{\dot{\omega}}_C$ (\si{\kilogram\per\meter\cubed\per\second}), $NN_b$}\label{9lrsProdCNN}		
			\end{subfigure}		
	\caption{Offline result comparison for internal energy, 
%		gas constant, 
		flame temperature,
		 thermal conductivity, ratio of heat capacity,
		 ratio of heat capacity coefficient,
		 and PVRR }\label{G2G5offlineNN}				
		\end{figure}

\tablename{~\ref{tabOFFres}} shows better performance for $NN_b$ than $NN_a$ by 36\% to 44\%.
 Having more layers and neurons, besides separate NNs for each variables in $NN_b$ provides more degrees of freedom for the model while, on the downside, it increases the computational cost relative to $NN_a$ by around 2.4 times.
As $NN_b$ provides better estimation, the data from the flamelet table is compared with its counterpart calculated from $NN_b$ in \figurename{~\ref{G2G5offlineNN}} for $\widetilde{e}_f$, 
%$\widetilde{R}$, 
$\widetilde{T_f}$, 
$\widetilde{\lambda}$, $\widetilde{\gamma}$, $\widetilde{a}_\gamma$,
 and PVRR.
 In the graphs, the x-axis shows the value of an output variable in the table, and the y-axis shows the value of the same variable estimated from NNs. Perfect estimation implies that all points lie on the $y=x$ line in each graph.
On the graphs the lines $y=(1\pm 1\%) x$, $y=(1\pm 2\%) x$, $y=(1\pm 5\%) x$, and $y=(1\pm 10\%) x$ are provided as guidelines for evaluating the NN performance in estimating the value of each point. Variable $\widetilde{R}$ was estimated with the lowest error (less than 1\%) and not plotted here.
\tablename{~\ref{tabOFFres}} also shows that the performance of similar NN structures in estimating distinct variables is different. Some variables are estimated much better than others. One of the factors in determining NN performance is the variable range relative to its average value.
 $\widetilde{R}$ and $\widetilde{\gamma}$ have relatively short ranges, and they are estimated with the lowest errors. PVRR, on the other hand, varies by five orders of magnitudes, thus it is estimated with the worst error.
 Figure~{\ref{9lrsProdCNN}} shows that the worst relative accuracy in PVRR estimation occurs for low values,
 since higher values are dominating the training. Although each variable is estimated with different errors, our studies showed that required levels of accuracy are different for each variable for successfully implementing the model in CFD. The studies involved a sensitivity analysis of CFD to perturbations in the outputs of the flamelet table. Different simulations are conducted, and in each of them, one of the outputs of the flamelet table is perturbed with random Gaussian noise with bounded amplitudes. Then, we measured how that affects the pressure signal.
 For instance, our modeling error sensitivity analysis in CFD showed that estimating $\widetilde{e}_f$ accurately is more important than others. 
		
\subsection{Online Test}\label{online}
Next, $NN_a$ and $NN_b$ are introduced in the CFD simulations for replacing the flamelet table. 
 According to \cite{Tuan1}, the 14-cm oxidizer post is a CVRC configuration with a very high level of instability.
In the 14-cm oxidizer post, for each of the NN sets, two different simulations have been conducted starting from two distinct initial conditions. The two initial conditions are selected at a time before the emerging instability and at a time after the instability has been fully developed. The first is called the "transient case," and the second is called the "dynamic equilibrium case." 
 The pressure signals represent the time waveform of pressure at each of the spatial grid points over the combustor geometry.
To validate the NN-based flame models, the results from each
 simulation are evaluated by investigating the similarity between pressure signals from NN-based and table-based CFD simulations.
 To this end, first, we looked at the overall relative error between pressure signals calculated from each CFD. Assume $\boldsymbol{x}$ is the pressure signal from table-based CFD, and $\boldsymbol{y}$ is the pressure signal from the NN-based CFDs. The relative error of signal $\boldsymbol{y}$ with respect to $\boldsymbol{x}$ is calculated based on Eq.~\eqref{mseoff}, where the index i runs over the time signals. $n$ is the number of time snapshots, and $x_i$ and $y_i$ are the values of the signals at the i-th time sample. This gives an overall measure of the distance between the two signals relative to the magnitude of the target signal. As shown later in \tablename{~\ref{errsumTAB}}, the third column, the pressure signals in the dynamic equilibrium case are estimated on average with 7.18\% and 6.23\% error by the $NN_a$-based and the $NN_b$-based simulations, respectively. For the transient case, the average error resulted from both simulations is 5.02\%. This close results from both NN-based models raise the question of whether the relative error is a comprehensive measure for the purpose of modeling combustion instability. 
Equation \eqref{mseoff} normalizes the distance of two signals by the magnitude of the reference. For oscillating signals with high mean values, such a measure ignores the fluctuations errors. In the combustion instability study, it is of high significance for a model to capture the phase and amplitude of the pressure fluctuation.
 In that regard, the root mean square (rms) of the pressure signals and their correlations are compared. 

 The similarity between the phases of two signals ($\boldsymbol{x}$ and $\boldsymbol{y}$) can be measured through their correlation (Pearson's linear correlation coefficient), calculated as:
 
 \begin{equation}
 \label{Corrdef}
 \kappa(\boldsymbol{x},\boldsymbol{y})=\frac{\sum_{i=1}^n(x_i-\bar{x}_i)(y_i-\bar{y}_i)}{\sqrt{\sum_{i=1}^n(x_i-\bar{x}_i)^2\sum_{i=1}^n(y_i-\bar{y}_i)^2}}
 \end{equation}
 
 where $\bar{x}$ and $\bar{y}$ are their empirical mean values. 
 The similarity between the fluctuation amplitudes of the two signal is measured by comparing their rms. The rms ratio of the fluctuations of $\boldsymbol{y}$ with respect to $\boldsymbol{x}$, is defined as:
 
\begin{equation}
\label{rmsratdef}
\eta(\boldsymbol{x},\boldsymbol{y})=\frac{rms(\boldsymbol{y})}{rms(\boldsymbol{x})}=\frac{\sqrt{\frac{1}{n}\sum_{i=1}^n(y_i-\bar{y}_i)^2}}{\sqrt{\frac{1}{n}\sum_{i=1}^n(x_i-\bar{x}_i)^2}}
\end{equation}		

In Eq.~\eqref{Corrdef} and Eq.~\eqref{rmsratdef}, the time averages of the signals are subtracted from the main signal to calculate the fluctuation signals. Hence, the similarity between their mean values is also measured. The relative errors of the mean values ($e_m$) of the signals are calculated through Eq.~\eqref{mseoff} by replacing $x$ and $y$ by
% the mean values, 
 $\bar{x}$ and $\bar{y}$. When signal amplitude is not constant over time (transient signal), the $\bar{x}_i$ and $\bar{y}_i$ are smoothed versions of the actual signals. 
 Otherwise, the actual means $\bar{x}_i$ and $\bar{y}_i$ are equal for all indexes (dynamic equilibrium).

\tablename{~\ref{errsumTAB}} summarizes the values of the above criteria for each of the simulations for dynamic equilibrium and transient cases. Each criterion is measured at each grid points on the CVRC geometry. The mean and standard deviation of all the measurements for each criterion is presented in \tablename{~\ref{errsumTAB}}. The first and second columns identify the case and NN model used in each case. The third main column presents the overall error statistics by their mean and standard deviation in the sub-columns. Similarly, the fourth, fifth, and sixth columns present the correlation, rms ratio, and error in mean values. Note that the desired value for correlation and the rms ratio is 100\%.

\begin{table}[hbt!]
	\centering
	\caption{Error analysis summary for the 14-cm oxidizer post configuration test base}\label{errsumTAB}
	\begin{tabular}{|c|c|c|c|c|c|c|c|c|c|}
		\hline
		\textbf{14-cm oxidizer-} & \multirow{2}{*}{\textbf{Model}} & \multicolumn{2}{c|}{\textbf{$e$ (\%):} Eq.~\eqref{mseoff}} & \multicolumn{2}{c|}{\textbf{ $\kappa$ (\%):} Eq.~\eqref{Corrdef}}& \multicolumn{2}{c|}{\textbf{$\eta$(\%):} Eq.~\eqref{rmsratdef}} & \multicolumn{2}{c|}{\textbf{$e_m$(\%):} Eq.~\eqref{mseoff}}\\ \cline{3-10} 
		\textbf{post cases}& & \textbf{mean} & \textbf{std} & \textbf{mean} & \textbf{std} & \textbf{mean} & \textbf{std} & \textbf{mean} & \textbf{std} \\ \hline
		\multirow{2}{*}{\textbf{Dynamic equilibrium}} & \textit{\textbf{$NN_a$}} & 7.18 & 2.47 & 78.5 & 12.69 & 93.71 & 4.58 & 1.41 & 0.31 \\ \cline{2-10} 
		& \textit{\textbf{$NN_b$}} & 6.23 & 2.34 & 82.1 & 13.71 & 93.91 & 2.85 & 0.64 & 0.19 \\ \hline
		\multirow{2}{*}{\textbf{Transient}} & \textit{\textbf{$NN_a$}} & 5.02 & 1.32 & 76.66 & 18.74 & 75.84 & 2.35 & 1.27 & 0.26 \\ \cline{2-10} 
		& \textit{\textbf{$NN_b$}} & 5.02 & 1.4 & 79.61 & 14.89 & 103.63 & 2.89 & 1.16 & 0.22 \\ \hline
	\end{tabular}
					\footnotesize{STD=standard deviation}
\end{table}

 In the dynamic equilibrium case, $NN_b$ has a slightly better performance in all criteria.
 However,
 $NN_a$ is selected as the representing flamelet model considering its computational cost, which is around 40\% of the $NN_b$-based simulation computational cost. The results from the $NN_a$-based simulations and the table-based simulations are discussed in more detail in Sec.~\ref{DynEqcase}. In the transient case, both NN-based models perform similarly with respect to the overall error, mean, and correlation criteria. But for the pressure fluctuation, the rms ratio captured by the $NN_b$-based and the $NN_a$-based simulations are 103.63\% and 75.84\%, respectively. Therefore, $NN_b$ is selected for representing the flamelet model. The results from the $NN_b$-based and the table-based simulations are discussed in more detail in 
Sec.~\ref{Transcase}. 
 $NN_b$, which shows higher capability in capturing flame dynamics, is also implemented in the 9-cm oxidizer post configuration, which, according to \cite{Tuan1}
 is recognized as a stable configuration. This is discussed later in Sec.~\ref{9cmcase}.
 
\subsubsection{ Dynamic Equilibrium Case} \label{DynEqcase}
\begin{figure}[hbt!]
	\begin{subfigure}{.5\textwidth}
		\centering
		\includegraphics[width=0.95\textwidth]{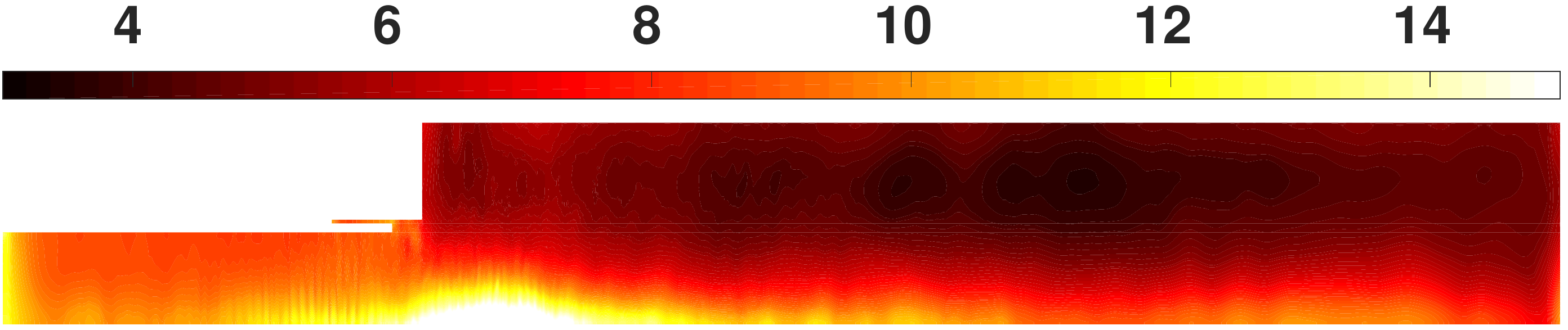}
		\caption{Overall relative error (\%)}\label{efullcont6}		
	\end{subfigure}
	\begin{subfigure}{.5\textwidth}
		\centering
		\includegraphics[width=0.95\textwidth]{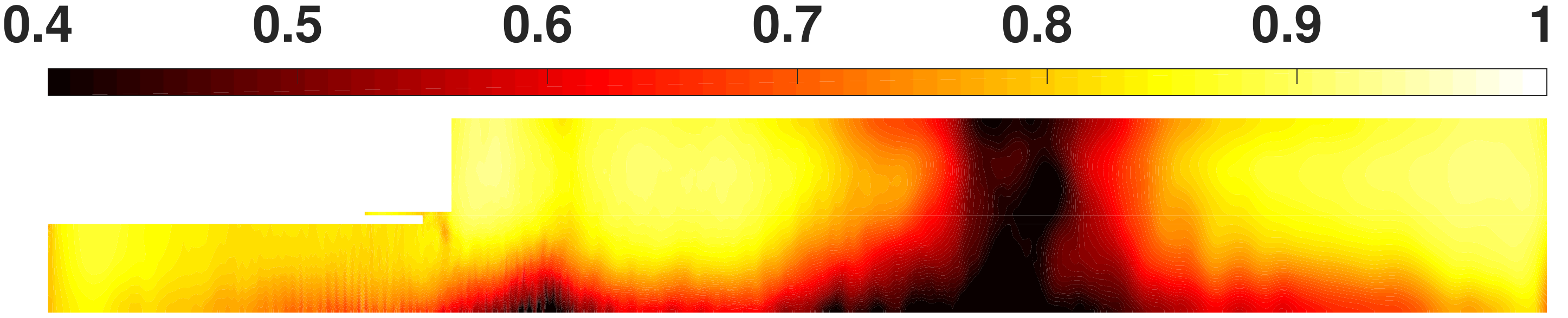}
		\caption{Fluctuation correlation}\label{corcont6}
	\end{subfigure}
	\caption{14-cm, dynamic equilibrium: distribution of relative error (\%) and fluctuation correlation between pressure signals calculated from $NN_a$-based and table-based simulations}\label{corcomp2}
	\begin{subfigure}{.5\textwidth}
		\centering
		\includegraphics[width=0.95\textwidth]{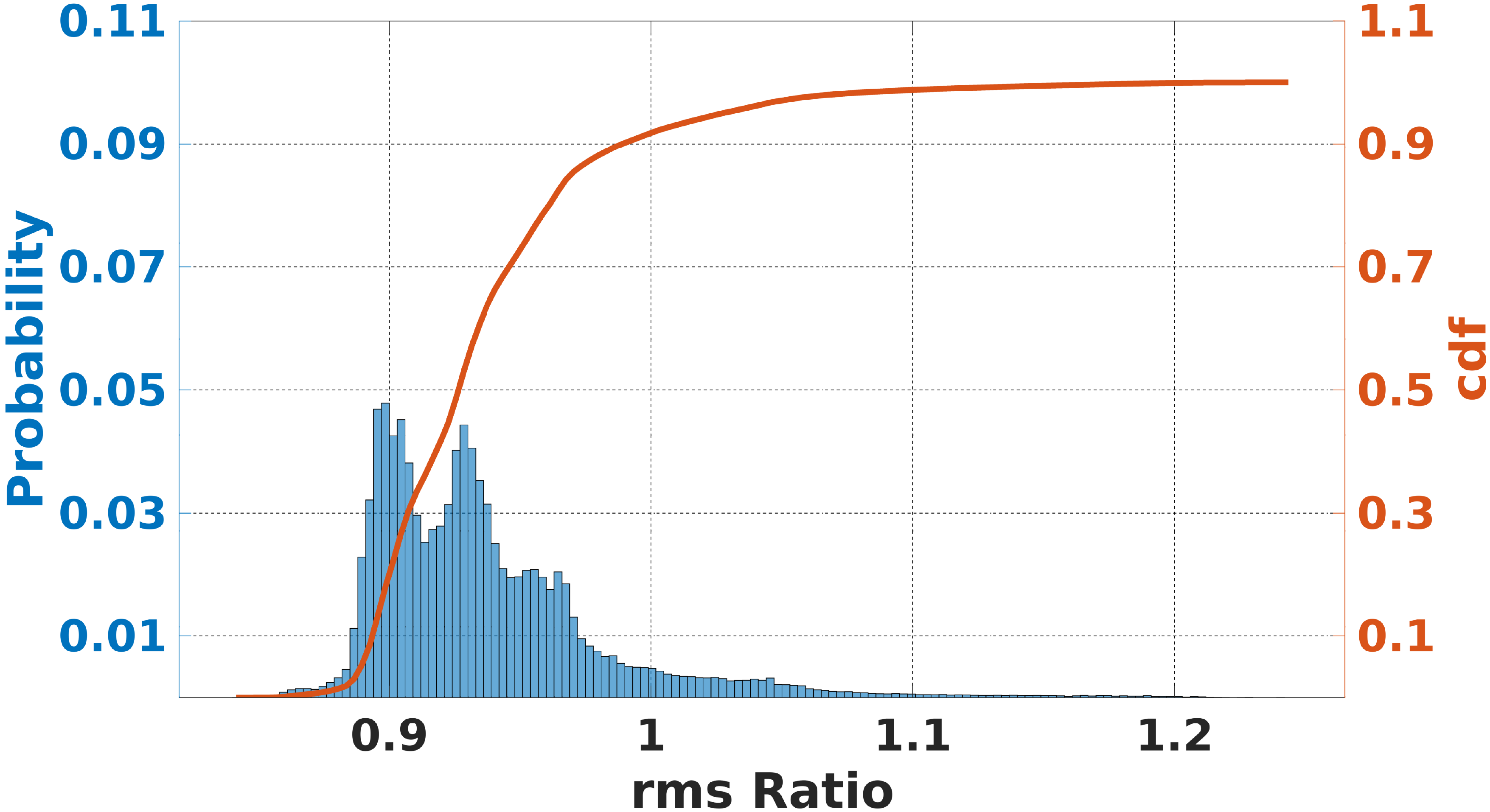}
		\caption{rms ratio distribution}\label{rmsrathist6}		
	\end{subfigure}
	\begin{subfigure}{.5\textwidth}
		\centering
		\includegraphics[width=0.95\textwidth]{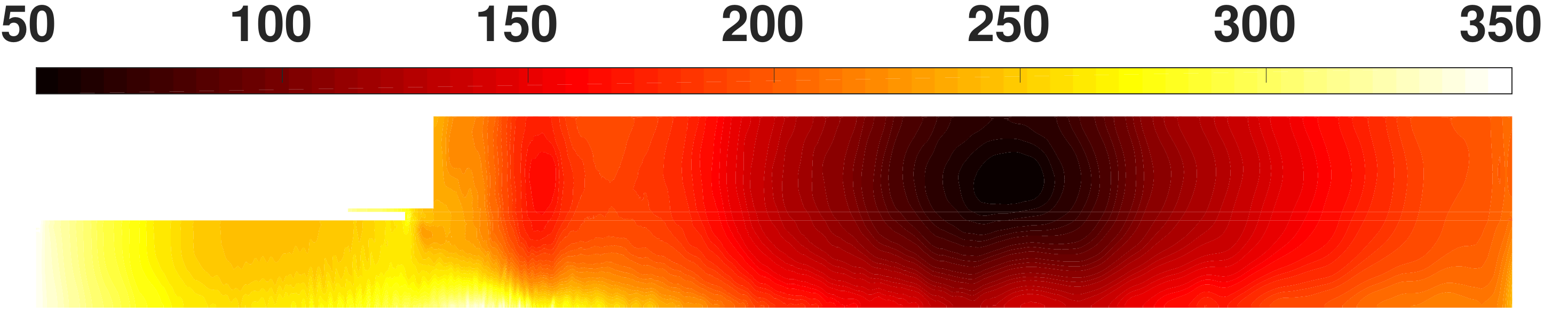}
		\caption{Table-based }\label{dynrmspowTAB}		
		\centering
		\includegraphics[width=0.95\textwidth]{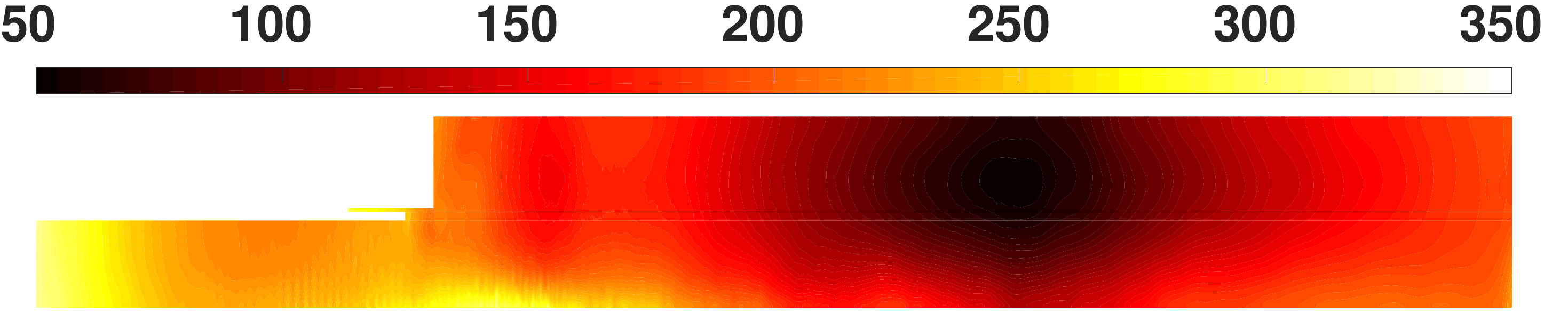}
		\caption{$NN_a$-based}\label{dynrmspowNN6}
	\end{subfigure}
	\caption{ 14-cm, dynamic equilibrium: The distribution of $\kappa$ calculated from $NN_a$-based (\ref{dynrmspowNN6}) to the one calculated from table-based (\ref{dynrmspowTAB}) simulations}\label{rmsratcomp3}
\end{figure}

In combustion instability studies, the accuracy in simulating pressure is perhaps the most significant for validating the model.
 The relative error between pressure signals from the $NN_a$-based and the table-based simulations, calculated by Eq.~\eqref{mseoff}, is shown in \figurename{~\ref{efullcont6}}.
The highest error happens near the centerline. In the axisymmetric configuration, the centerline ($r=0$) can be considered as a singularity and is prone to numerical errors. Accordingly, near the centerline, the signals are noisier and can be modeled with less quality.
The correlation of the fluctuation of the pressure signal at each grid point with the pressure signal at the same point from the original table-based simulation is calculated and shown in the contour plot in \figurename{~\ref{corcont6}}. The points with lower than 80\% correlation are located in the pressure node, where the amplitudes of pressure harmonics are almost equal and smaller relative to other locations; therefore, the fluctuation signals there can be considered as noise. 
Moreover, \figurename{~\ref{rmsratcomp3}}, compares the ratio of pressure fluctuation rms calculated from the $NN_a$-based simulation to the rms calculated from the table-based simulation; the rms of fluctuations of the two simulations are shown in \figurename{~\ref{dynrmspowNN6}} and \figurename{~\ref{dynrmspowTAB}}, respectively. The parts of the combustor in which the rms of pressure fluctuations is lower is consistent with the regions where correlation is lower. These errors that are calculated from Eq.\eqref{mseoff} to Eq.\eqref{rmsratdef} are demonstrated by comparing pressure signals calculated from the table-based and the NN-based simulations at few points in the following.

\begin{figure}[hbt!]
	\begin{subfigure}{.49\textwidth}
		\centering
		\includegraphics[width=0.9\textwidth]{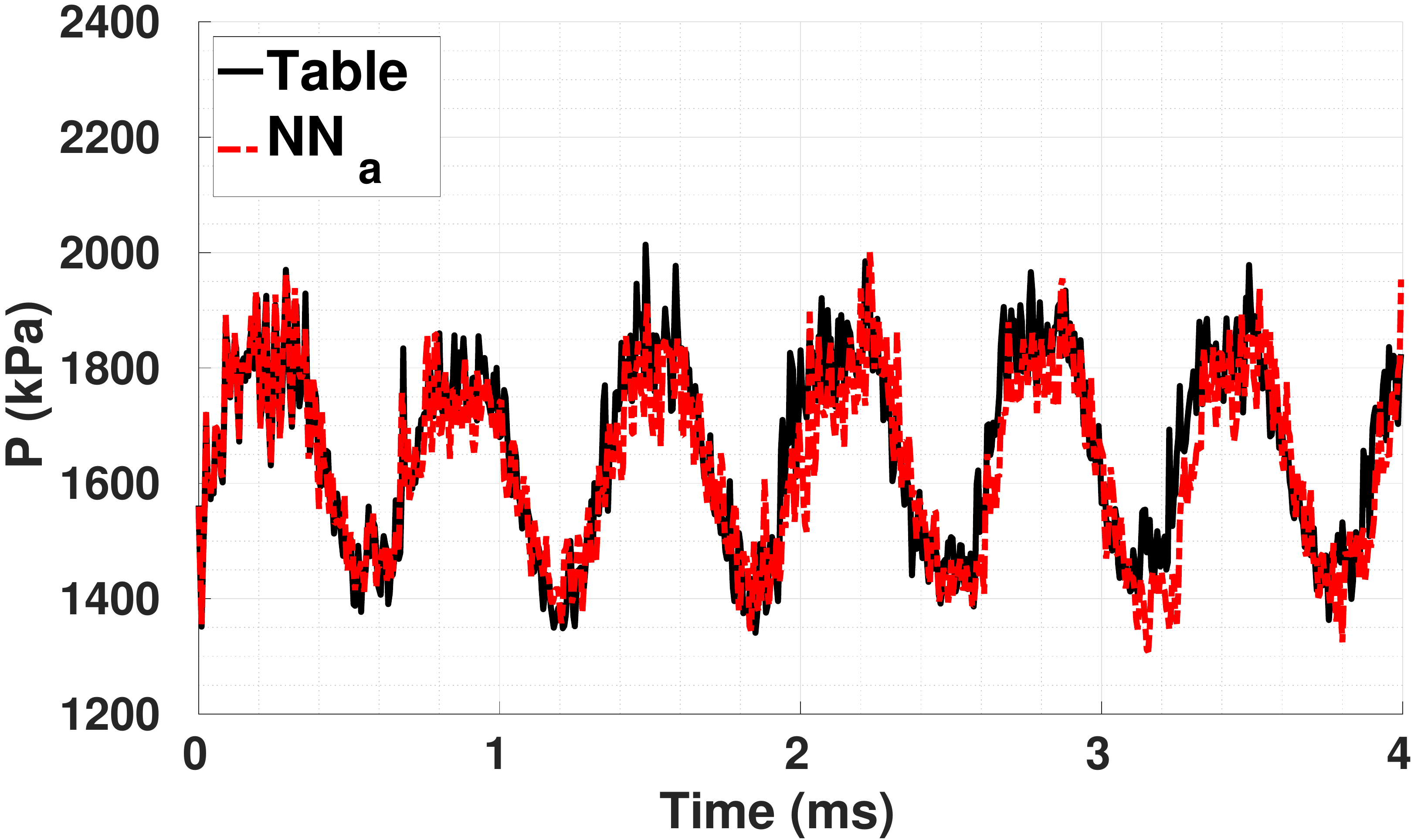}
		\caption{$x=10$ \si{\centi\meter}, top wall}\label{dyn643_404}		
	\end{subfigure}
	\begin{subfigure}{.49\textwidth}
		\centering
		\includegraphics[width=0.9\textwidth]{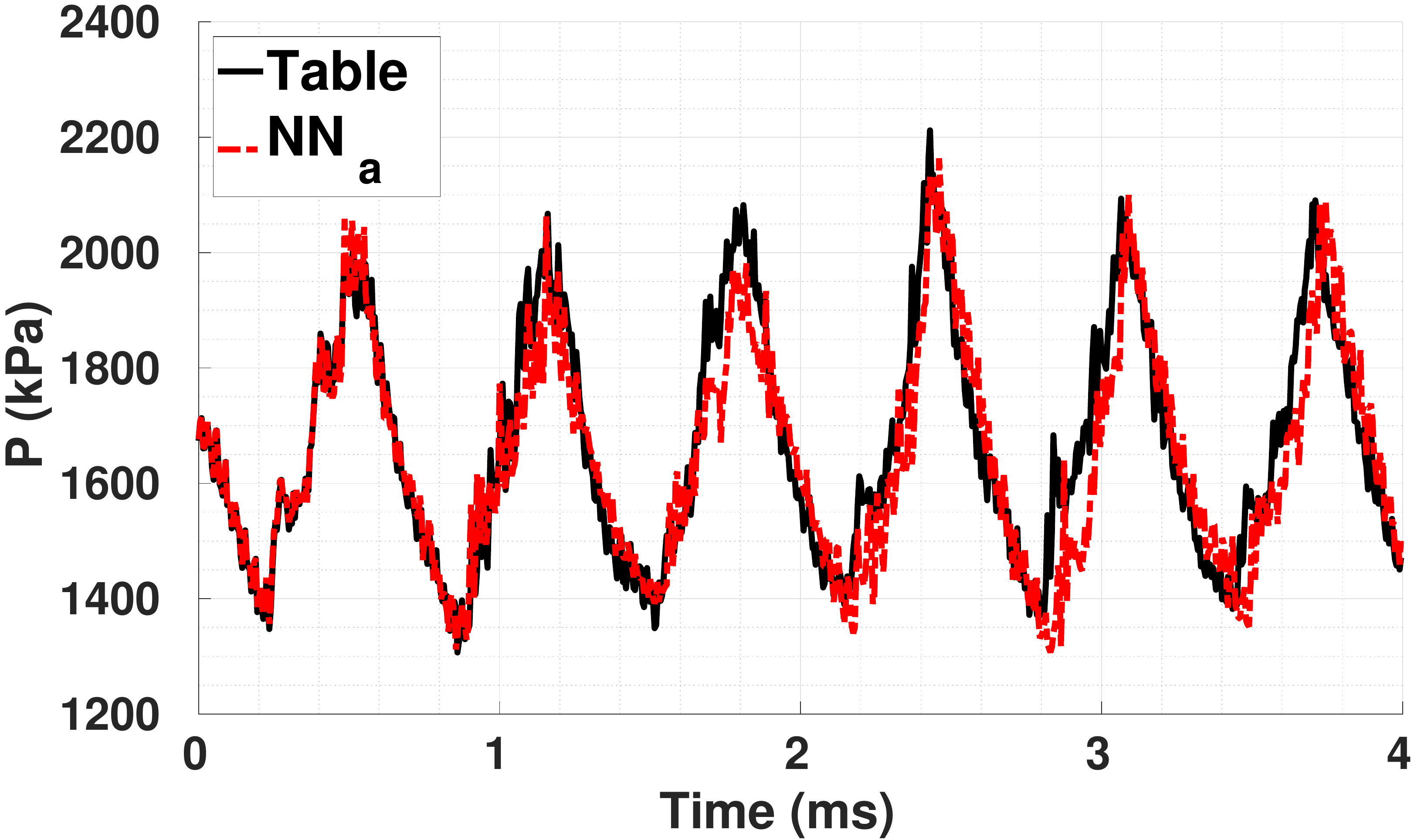}
		\caption{ $x=37$ \si{\centi\meter}, top wall}\label{dyn643_645}			
	\end{subfigure}			
	\begin{subfigure}{.49\textwidth}
		\centering
		\includegraphics[width=0.9\textwidth]{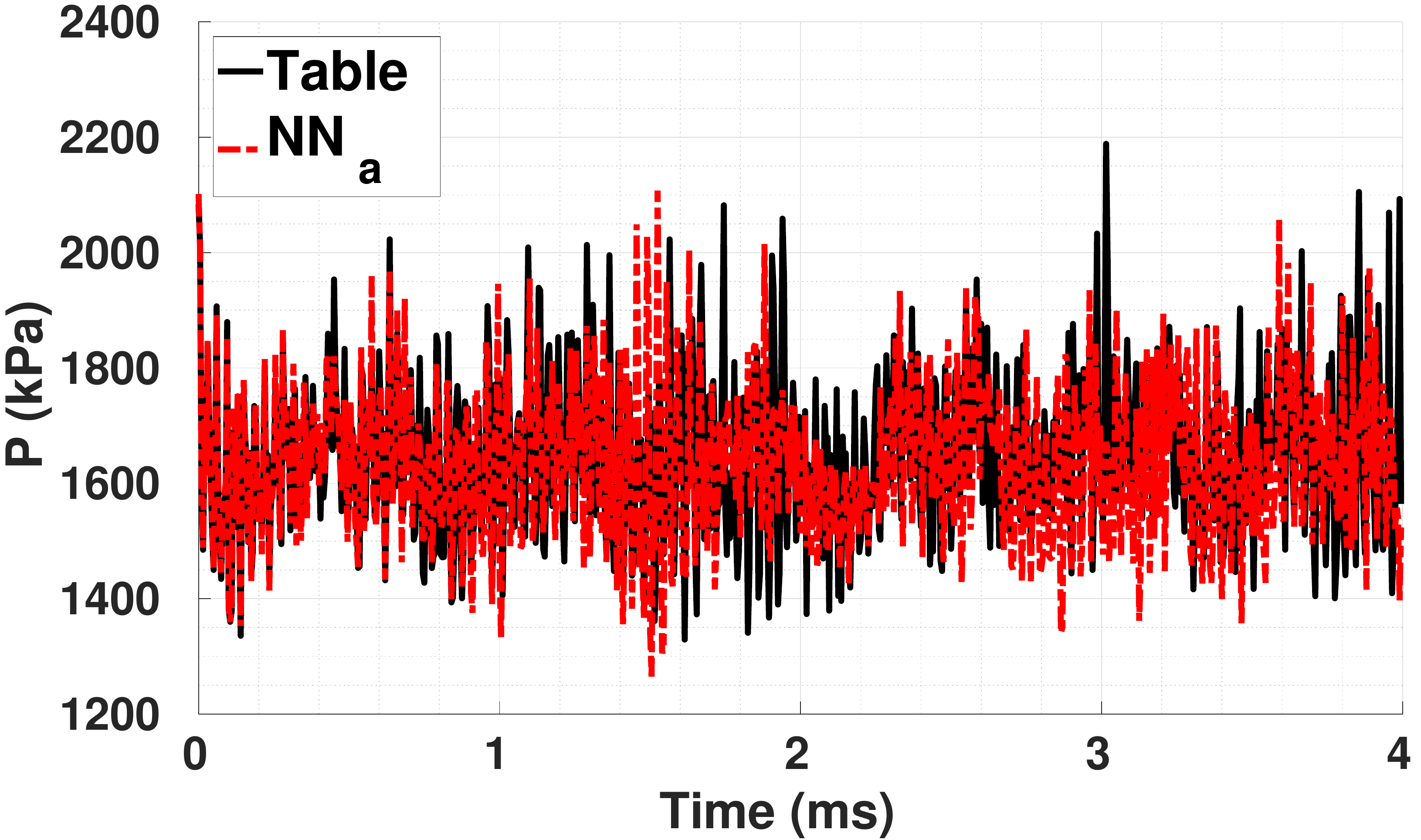}
		\caption{$x=21.8$ \si{\centi\meter}, centerline}\label{dyn61_847}		
	\end{subfigure}
	\begin{subfigure}{.49\textwidth}
		\centering
		\includegraphics[width=0.9\textwidth]{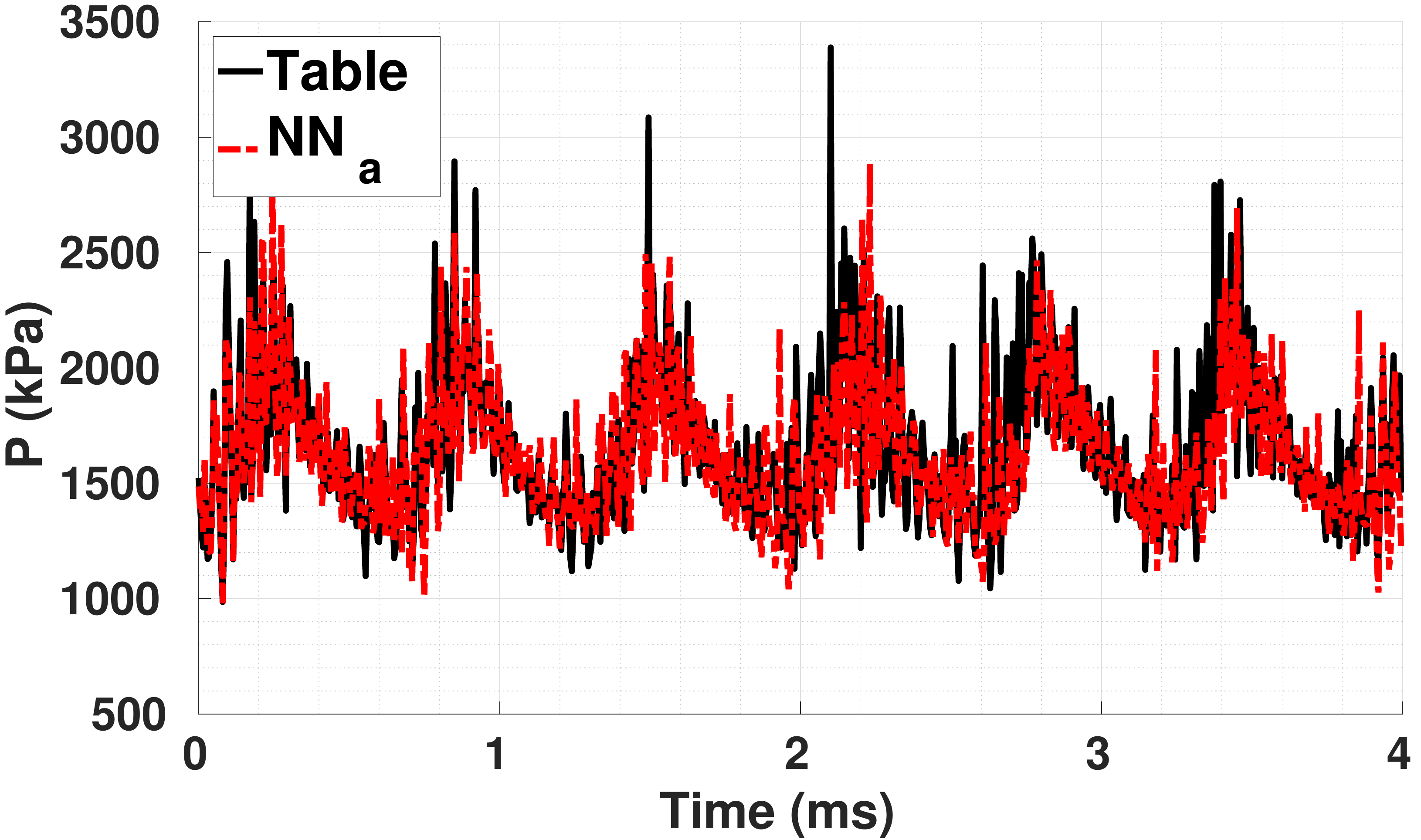}
		\caption{$x=1.14$ \si{\centi\meter}, centerline}\label{dyn61_353}		
	\end{subfigure}			
	\caption{ 14-cm, dynamic equilibrium: comparison of pressure signals 
		 at different points on top wall and centerline between the $NN_a$-based and the table-based simulations} \label{Psigcomp2}		
\end{figure}

Figure{ \ref{dyn643_404}} and \figurename{~\ref{dyn643_645}} compare the pressure signals
 at the antinode (10 \si{\centi\meter}), and near the nozzle at the pressure tab (37 \si{\centi\meter}) on the top wall. At the antinode, the correlation is 87.73\%, and near the nozzle, the correlation is 91.21\%. At these points, the overall relative errors are 4.91\% and 4.98\%, respectively. The rms ratio is 95\%, and the mean value is estimated with a 1.4\% error at $x=10$ \si{\centi\meter}. At $x=37$ \si{\centi\meter}, the rms ratio is 94.74\%, and the mean value is estimated with a 1.14\% error. The first and second modes of oscillations are captured with high accuracy in these signals.
 The antinode and pressure tab are the common points that are investigated for each test because of their geometrical importance and to provide reference points among different cases.
Figure{ \ref{dyn61_847}} and \figurename{~\ref{dyn61_353}} compare pressure signals
 at points with lower correlations and higher error values. One of the points with a lower correlation is located on the centerline at 21.8 \si{\centi\meter}, i.e., the pressure node vicinity. Figure{ \ref{dyn61_847}} compares the pressure signal at this point between the NN-based and the table-based simulations. The correlation at this point is 40.07\%. The overall error at this point is 9\%, the rms ratio is 96.72\%, and the mean value error is 0.9\%. 
At 21.8 \si{\centi\meter}, looking at the pressure signal in the frequency domain, the power of the signal is distributed almost evenly among different modes, and the first mode is weaker relative to other locations. Essentially the higher frequencies are more significant at this point; however, the rms of the signal is captured to a good extent. The other point is located on the centerline at 1.14 \si{\centi\meter}, which is right after the dump plane near the mixing layer. Figure{ \ref{dyn61_353}} compares the pressure signal at this point with 53.64\% correlation. The overall error at this point is 17.72\%, the rms ratio is 89.03\%, and the mean value error is 2.15\%. 
This point is located near the interface of the oxidizer post and combustor, at $r=0$; a singular point with a very high density mesh.
The spikes are associated with quantitative excursions of pressure 
	at and beyond the limits used for the construction of the table (30 atm). Spikes do occur in close-neighboring physical locations with the NN as well. The spikes fortunately occur only in 
	regions where little reaction occurs---near the centerline and near the 
	entrance for the propellant flow into the main chamber. Thereby, with no 
	reaction locally, these spikes have no consequence on the Rayleigh Index 
	and the stability of the combustion chamber. Note that the physical behavior is chaotic, and therefore each realization can differ in detail. Spikes cannot be expected to occur always at the same location when the small quantitative differences (i.e., numerical errors) in constraints can appear between NN and the table.

\begin{figure}[hbt!]
	\begin{subfigure}{.33\textwidth}
		\centering
		\includegraphics[width=1\textwidth]{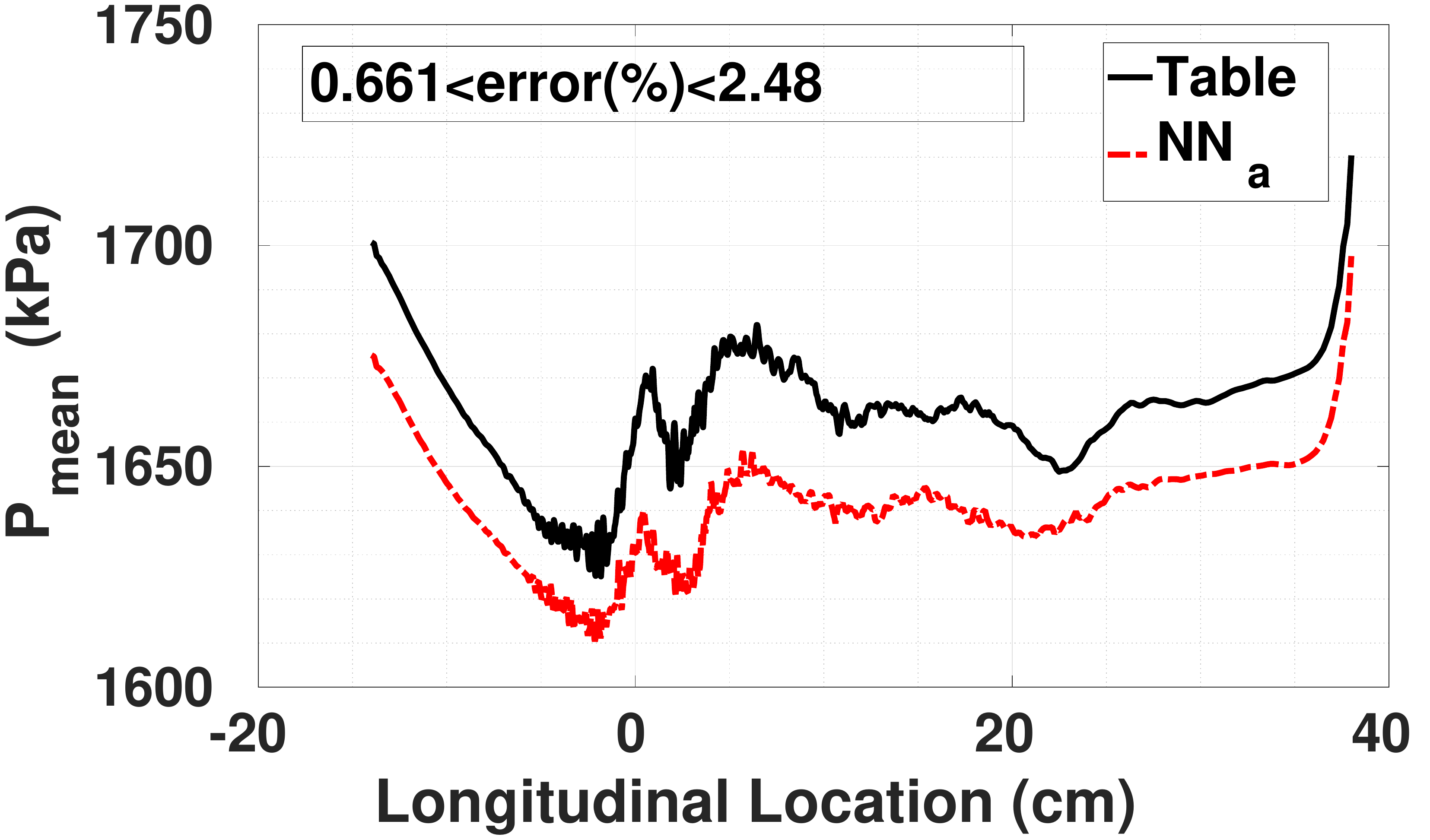}
		\caption{Mean pressure (\si{\kilo\pascal})}\label{PMSmean6}
			\end{subfigure}		
\begin{subfigure}{.33\textwidth}			
\centering
\includegraphics[width=1\textwidth]{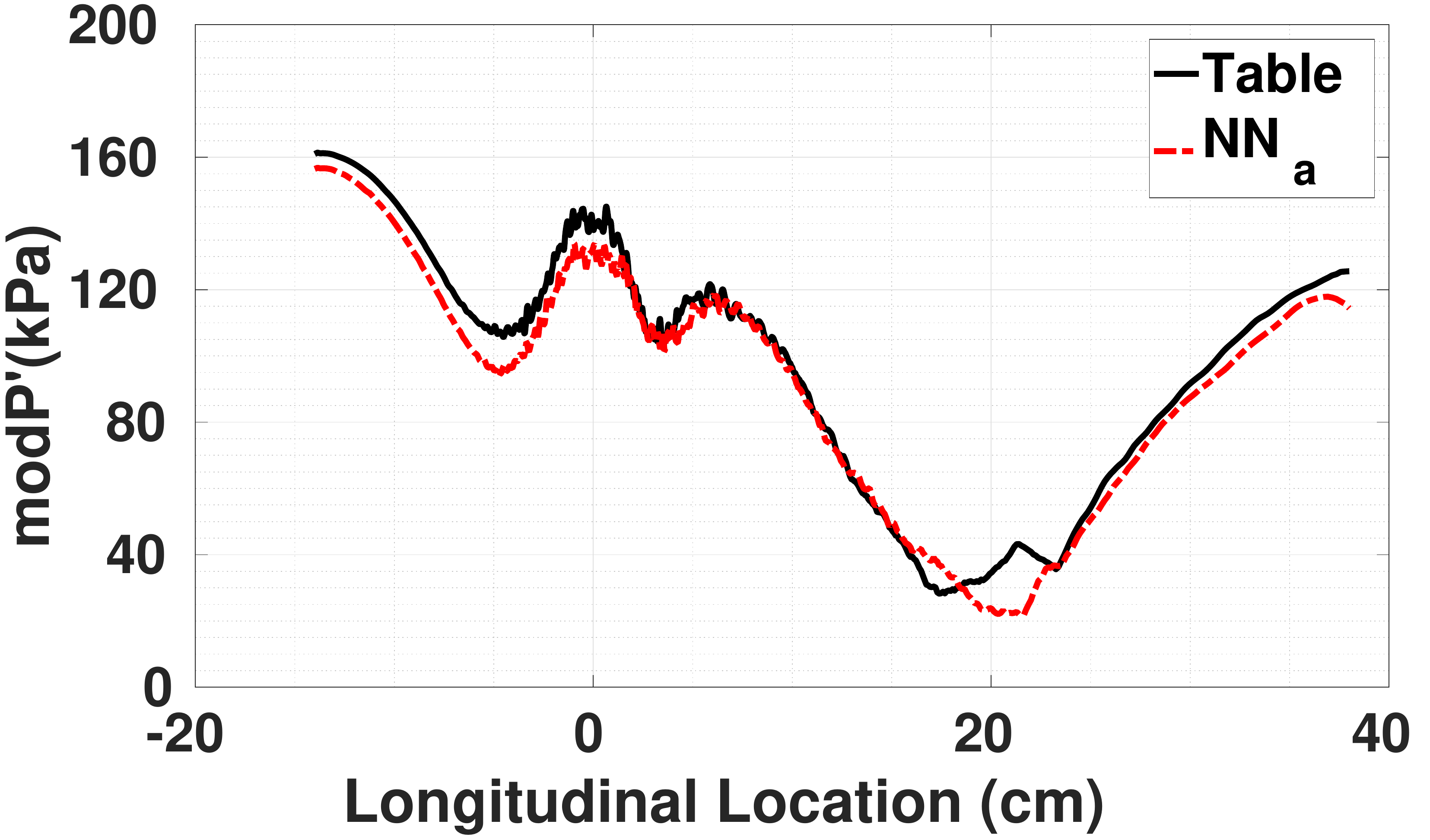}
\caption{First mode shape (\si{\kilo\pascal}) }\label{PMSfirst6}	
			\end{subfigure}		
			\begin{subfigure}{.33\textwidth}	
		\centering
		\includegraphics[width=1\textwidth]{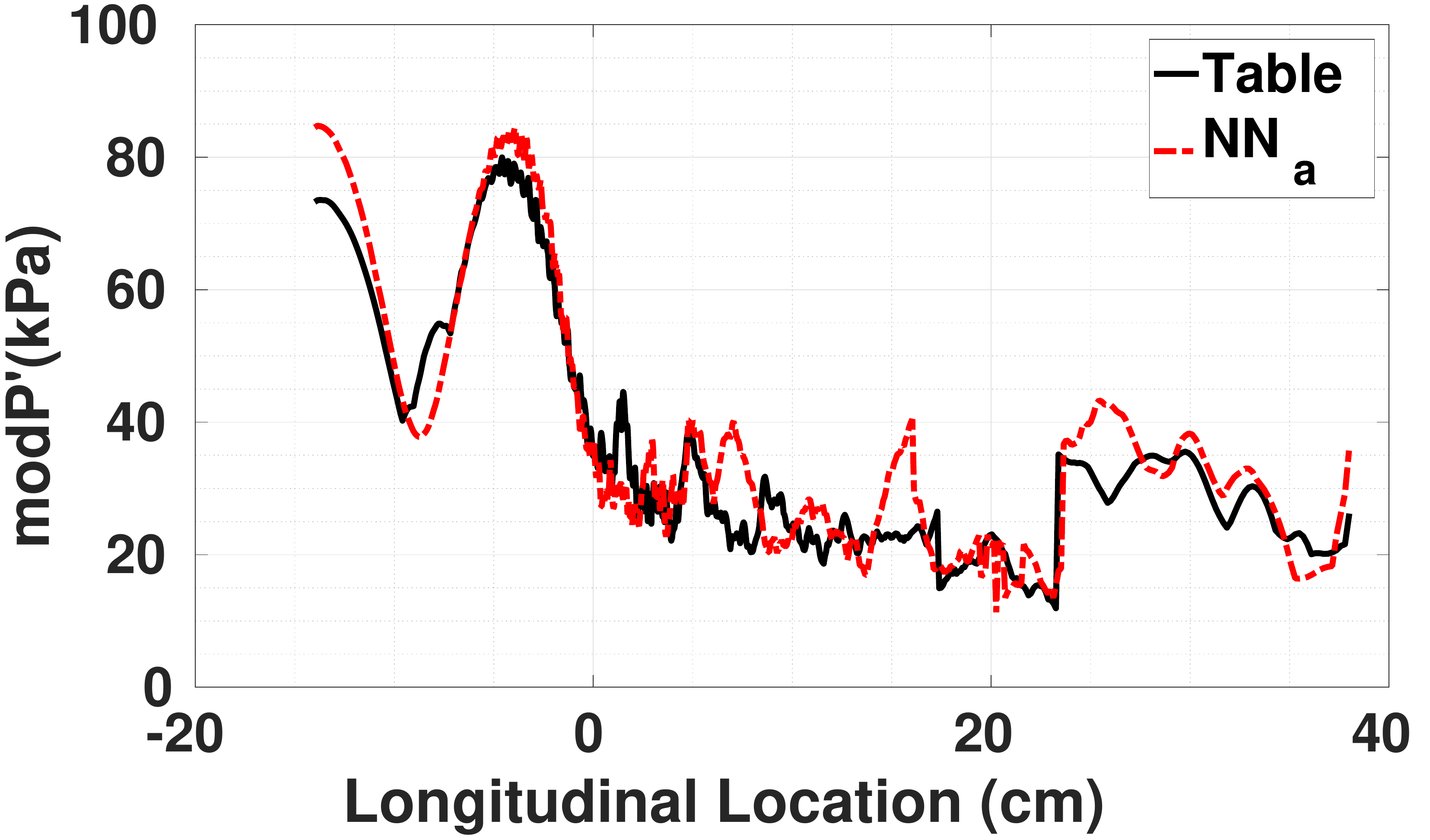}
		\caption{Second mode shape (\si{\kilo\pascal})}\label{PMSsec6}		
	\end{subfigure}
	\caption{14-cm, dynamic equilibrium: comparison of pressure mean, the first, and the second mode shapes between $NN_a$-based and table-based simulations}\label{modshcompnew}
\end{figure}

Also, to compare the frequency content of pressure signals, the mode shapes of the signals over the centerline are compared. Pressure mode shapes are demonstrated by plotting the modulus of the Fourier spectrum peaks at each grid point along the centerline. First, the time-average of the signals, which are plotted in \figurename{~\ref{PMSmean6}}, are subtracted from it. Then, the Fourier spectrum is calculated for each of the centered signals. The first and the second modes are plotted in \figurename{~\ref{PMSfirst6}} and \figurename{~\ref{PMSsec6}}. The $NN_a$-based simulation gives a good estimate of the mean ($e_m$ ranges between 0.66\% to 2.46\%), the first, and the second modes; it starts to deviate as for the third 
mode; however, the effect of the third mode in the overall results is considered negligible. 

\begin{figure}[hbt!]
	\begin{subfigure}{.5\textwidth}
			\centering
			\includegraphics[width=0.95\textwidth]{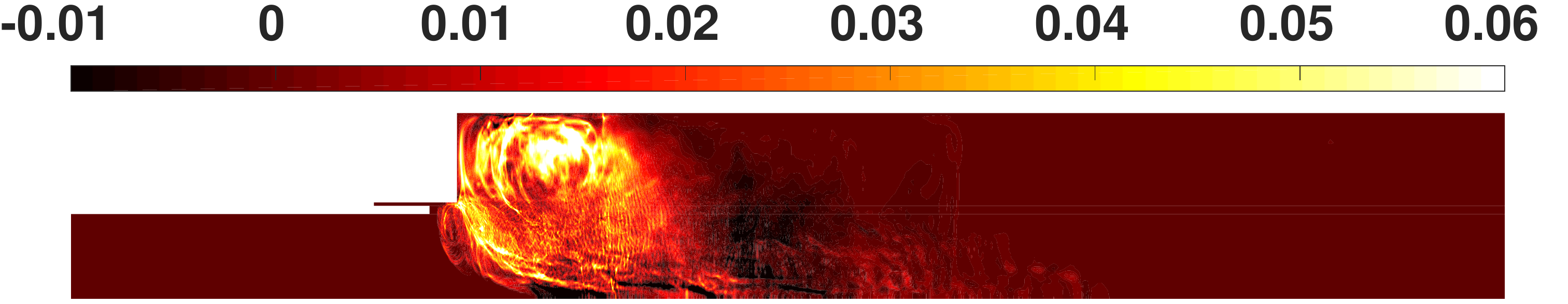}
			\caption{ Table-based CFD: \textit{mRI}}\label{WCRIrr}			
		\centering
		\includegraphics[width=0.95\textwidth]{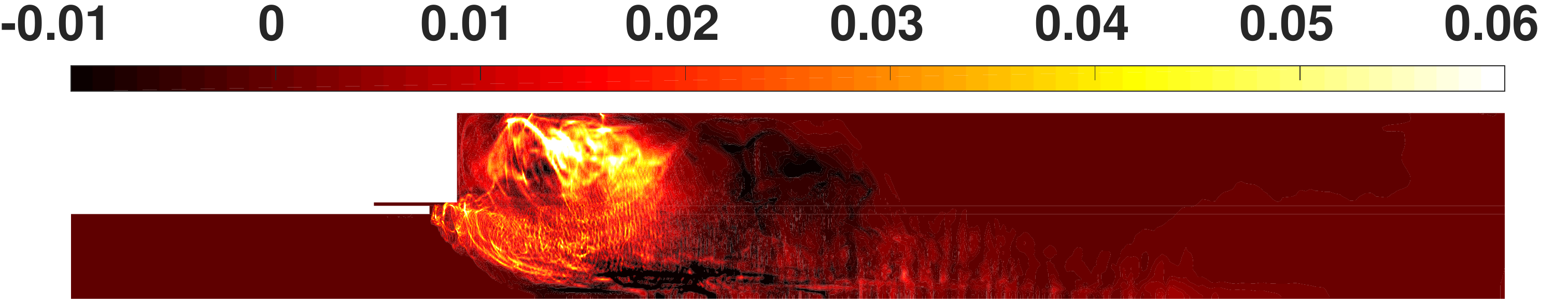}
		\caption{ $NN_a$-based CFD: \textit{mRI} }\label{WCRInna}							
	\end{subfigure}		
		\begin{subfigure}{.5\textwidth}
			\centering	
			\includegraphics[width=0.95\textwidth]{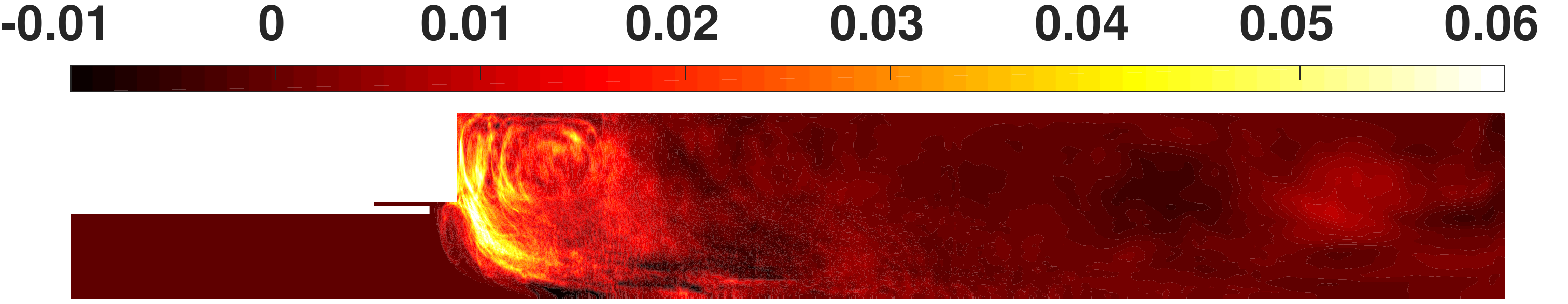}
			\caption{ Table-based CFD: \textit{RI}}\label{HRRRIrr}				
		\centering	
		\includegraphics[width=0.95\textwidth]{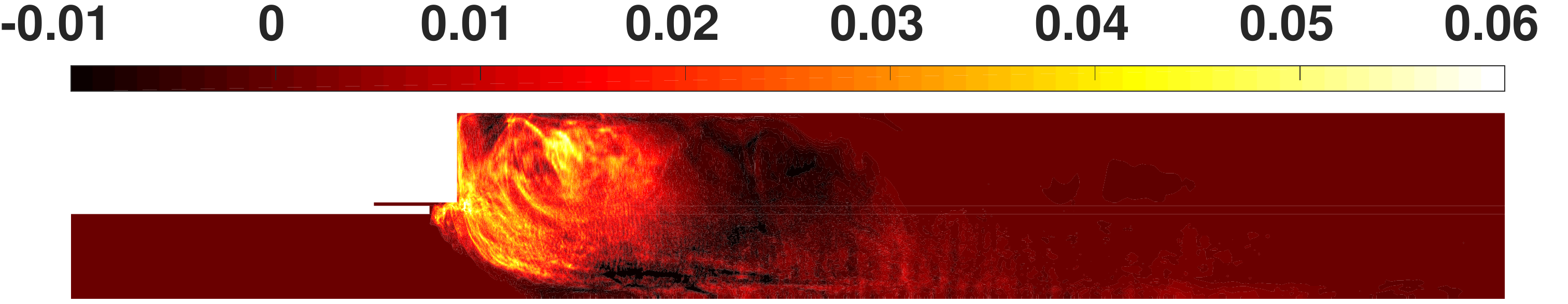}
		\caption{$NN_a$-based CFD: \textit{RI} }\label{HRRRInna}								
	\end{subfigure}		
		\caption{ 14-cm, dynamic equilibrium: comparison of \textit{RI} and \textit{mRI} from the $NN_a$- and the table-based simulations} \label{RIlabcomp1}		
\end{figure}

As discussed before, it is important for a model to provide a good estimation of \textit{RI} to be useful in combustion stability research. The local \textit{mRI} and \textit{RI} are compared for the NN-based and the table-based simulations in \figurename{~\ref{RIlabcomp1}}. Comparing \figurename{~\ref{WCRInna}} with \figurename{~\ref{WCRIrr}} and comparing \figurename{~\ref{HRRRInna}} with \figurename{~\ref{HRRRIrr}} show that the $NN_a$-based simulation captures the location of the flame that drives instability, yet underestimates the flame near the corner. Particularly,
 the $NN_a$-based simulation underestimates \textit{mRI} from the table-based simulation near the corner and dump plane. HRR is calculated after the CFD simulation was conducted, only for the purpose of calculating \textit{RI}. For \textit{RI}, there exists a
 good correlation in the flame zone, 
 but discrepancies downstream. The slight underestimation of \textit{RI} and \textit{mRI} are consistent with the slight underestimation of the rms value of the limit cycle with the NN-based simulation.

\begin{figure}[hbt!]
		\begin{subfigure}{.49\textwidth}
			\centering
			\includegraphics[width=0.95\textwidth]{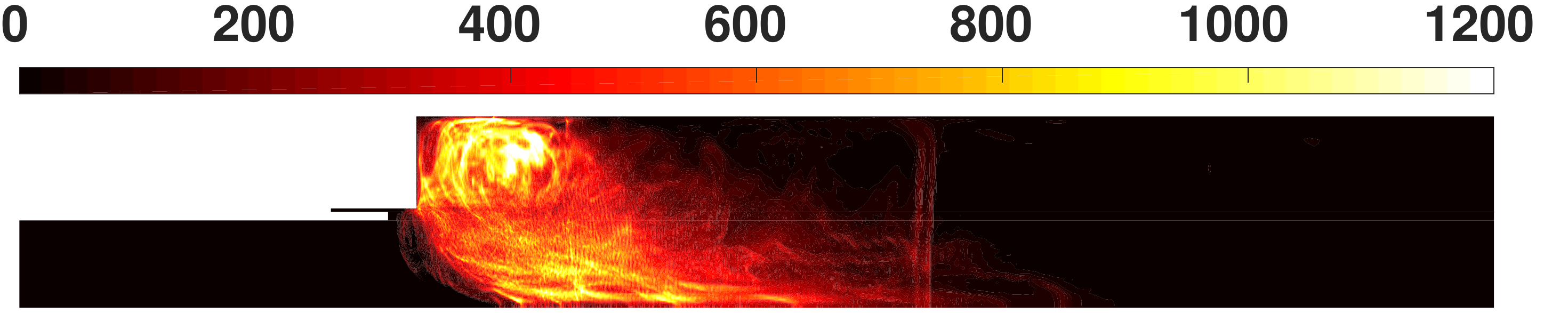}
			\caption{$\widehat{\dot{\omega}}_C$ (\si{\kilogram\per\meter\cubed\per\second}), table}\label{WProdCsigcotimeavgdyn8}	
			\centering
			\includegraphics[width=0.95\textwidth]{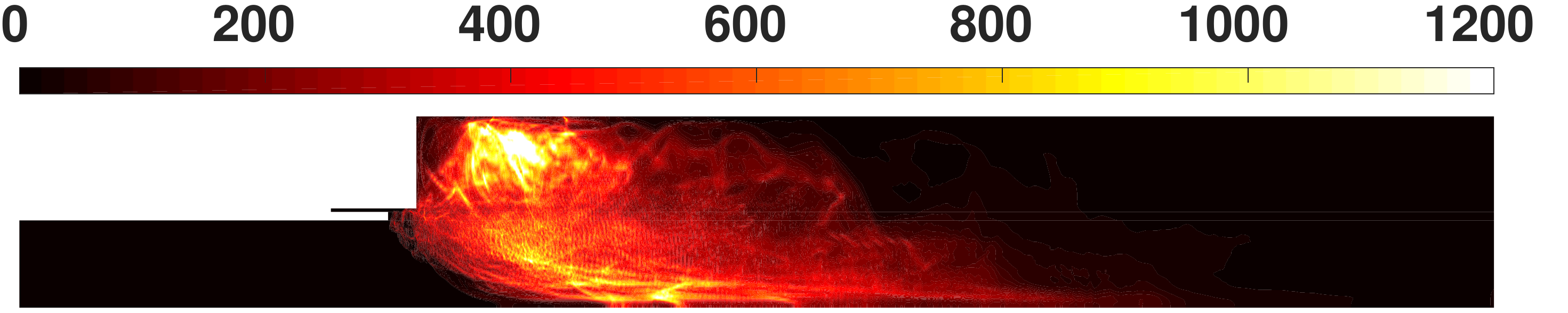}
			\caption{$\widehat{\dot{\omega}}_C$ (\si{\kilogram\per\meter\cubed\per\second}), $NN_a$}\label{NNProdCsigcotimeavgdyn8}					
			\centering		
			\includegraphics[width=0.95\textwidth]{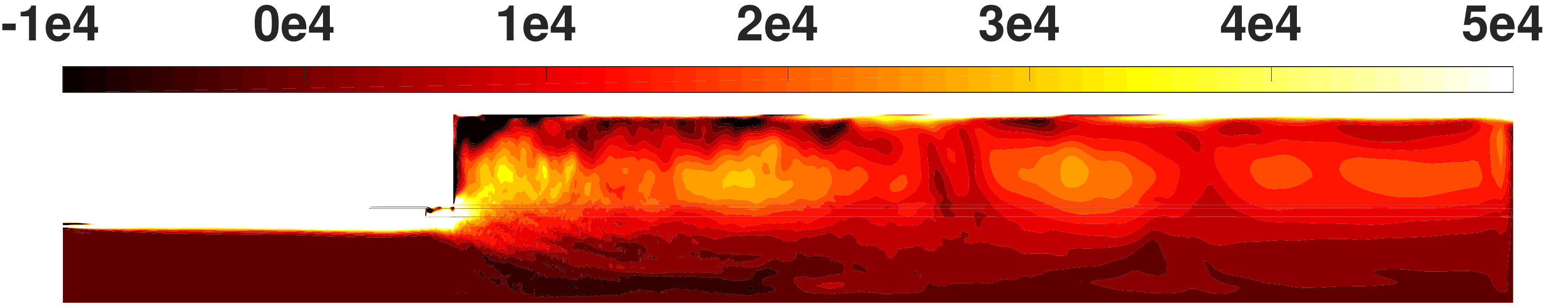}
			\caption{$\widehat{\Omega}$ (\si{\per\second}), table}\label{Wvortsigcotimeavgdyn8}			
			\centering		
			\centering		
			\includegraphics[width=0.95\textwidth]{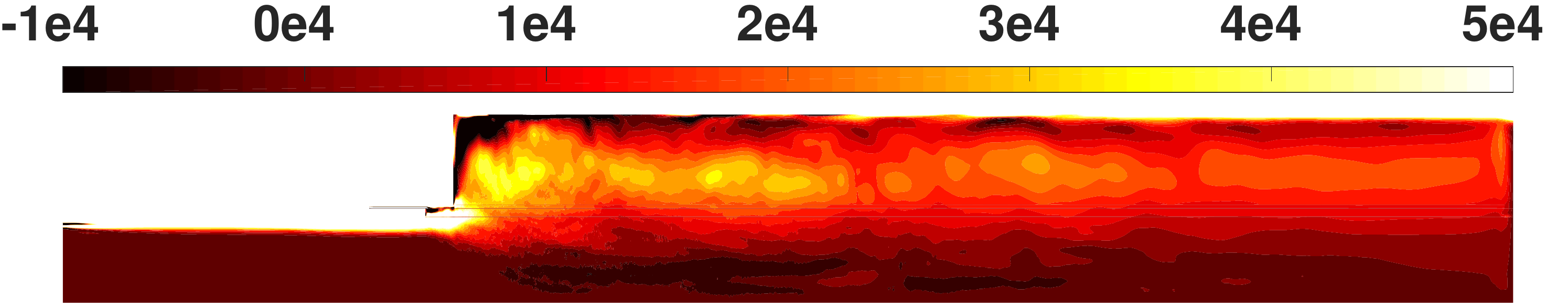}
			\caption{$\widehat{\Omega}$ (\si{\per\second}), $NN_a$}\label{NNvortsigcotimeavgdyn8}				
		\end{subfigure}				
		\begin{subfigure}{.49\textwidth}		
			\centering		
			\includegraphics[width=0.95\textwidth]{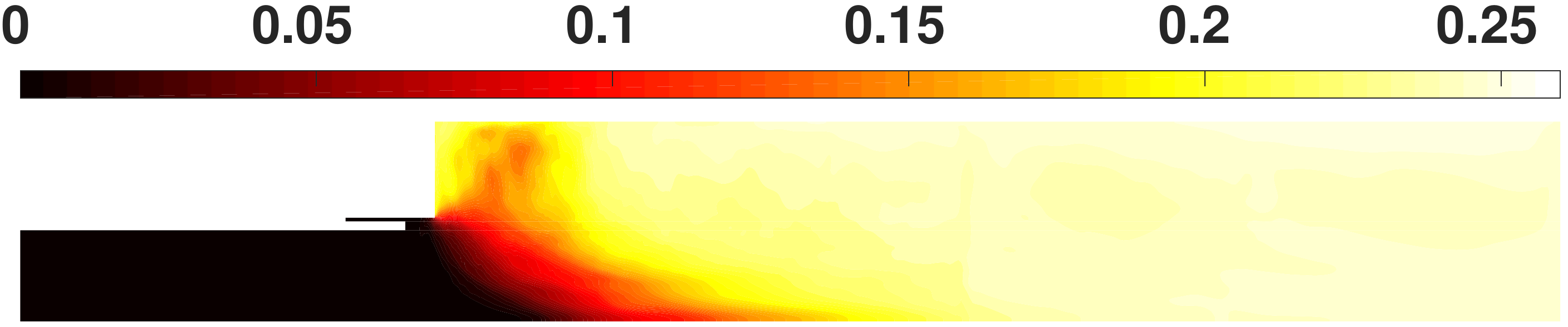}
			\caption{$\widehat{C}$, table }\label{WCsigcotimeavgdyn8}					
			\centering		
			\includegraphics[width=0.95\textwidth]{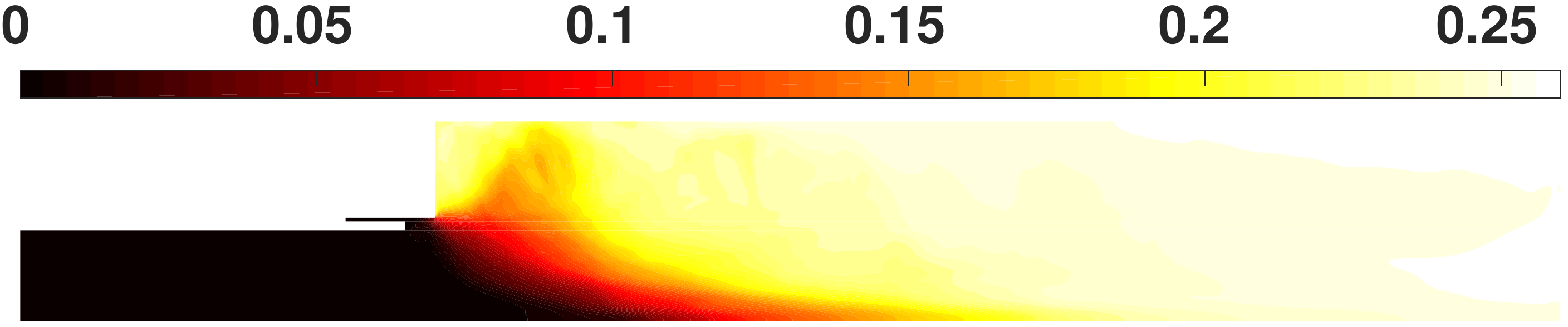}
			\caption{$\widehat{C}$, $NN_a$}\label{NNCsigcotimeavgdyn8}						
			\centering				
			\includegraphics[width=0.95\textwidth]{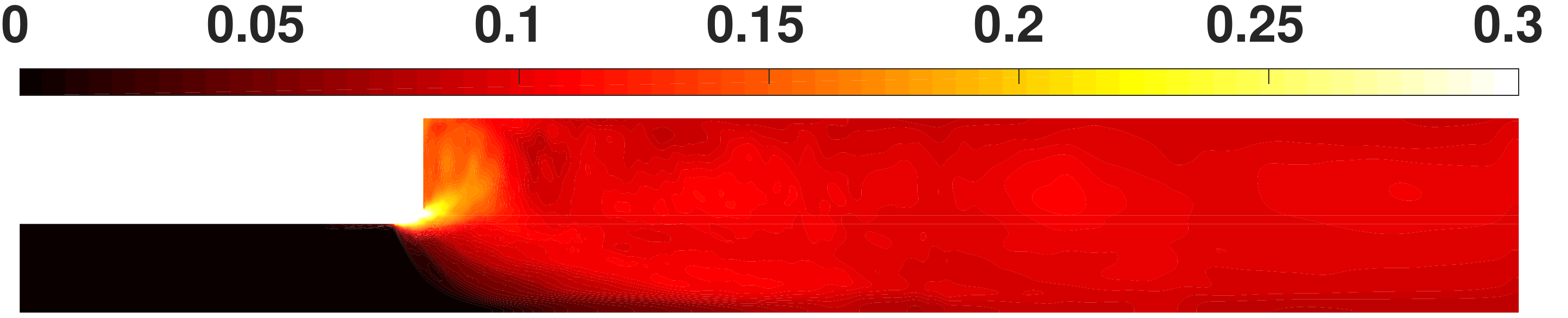}
			\caption{$\widehat{Z}$, table }\label{WZsigcotimeavgdyn8}		
			\centering		
			\includegraphics[width=0.95\textwidth]{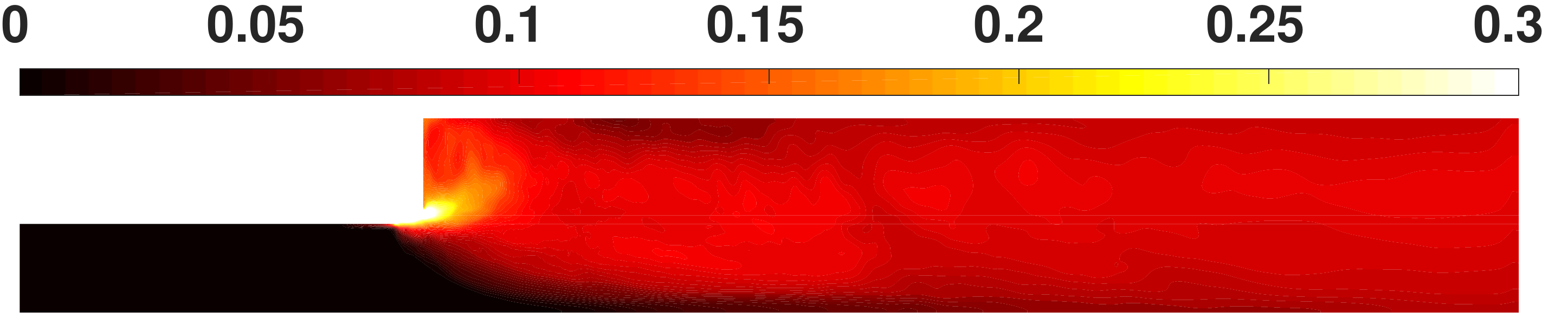}				
			\caption{$\widehat{Z}$, $NN_a$}\label{NNZsigcotimeavgdyn8}									
		\end{subfigure}			
		\caption{ 14-cm, dynamic equilibrium: time-averaged values of: PVRR, progress variable, mixture fraction, vorticity from the $NN_a$-based and the table-based simulations}	\label{timeavgcontcomp}					
	\end{figure}

In the following, other significant variables are briefly compared between the $NN_a$-based and table-based simulations by demonstrating their time-averaged behavior. Figure{ \ref{timeavgcontcomp}} compares the contour plots of time-averaged PVRR (\figurename{~\ref{WProdCsigcotimeavgdyn8}} and \figurename{~\ref{NNProdCsigcotimeavgdyn8}}), time-averaged vorticity (\figurename{~\ref{Wvortsigcotimeavgdyn8}} and \figurename{~\ref{NNvortsigcotimeavgdyn8}}), time-averaged progress variable (\figurename{~\ref{WCsigcotimeavgdyn8}} and \figurename{~\ref{NNCsigcotimeavgdyn8}}), and time-averaged mixture fraction (\figurename{~\ref{WZsigcotimeavgdyn8}} and \figurename{~\ref{NNZsigcotimeavgdyn8}}). In all these figures, although there are differences in the detailed behavior, the overall shapes are very similar between the two simulations. 

\subsubsection{ Transient Case }\label{Transcase}
\begin{figure}[hbt!]
	\begin{subfigure}{.5\textwidth}
		\centering
		\includegraphics[width=0.95\textwidth]{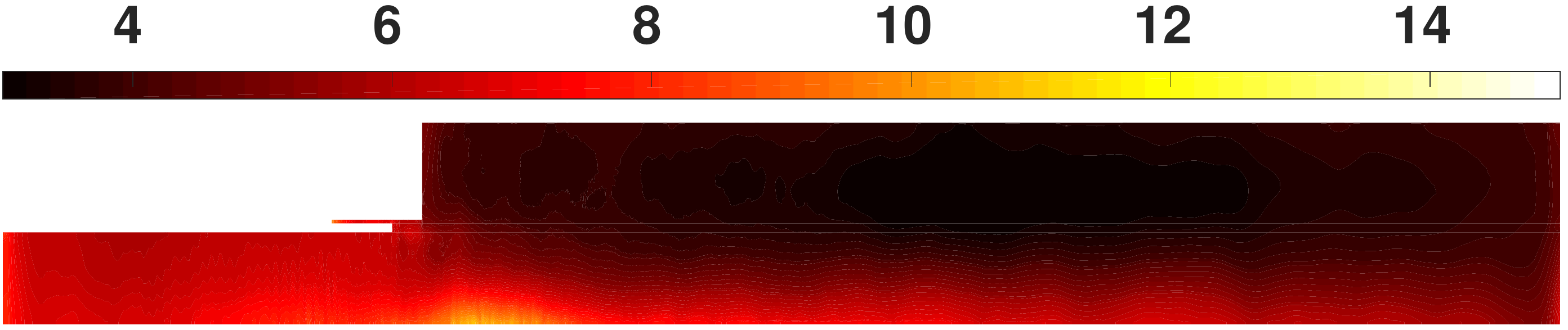}
		\caption{Overall relative error (\%)}\label{trsefullcont8}		
	\end{subfigure}	
		\begin{subfigure}{.5\textwidth}
			\centering
			\includegraphics[width=0.95\textwidth]{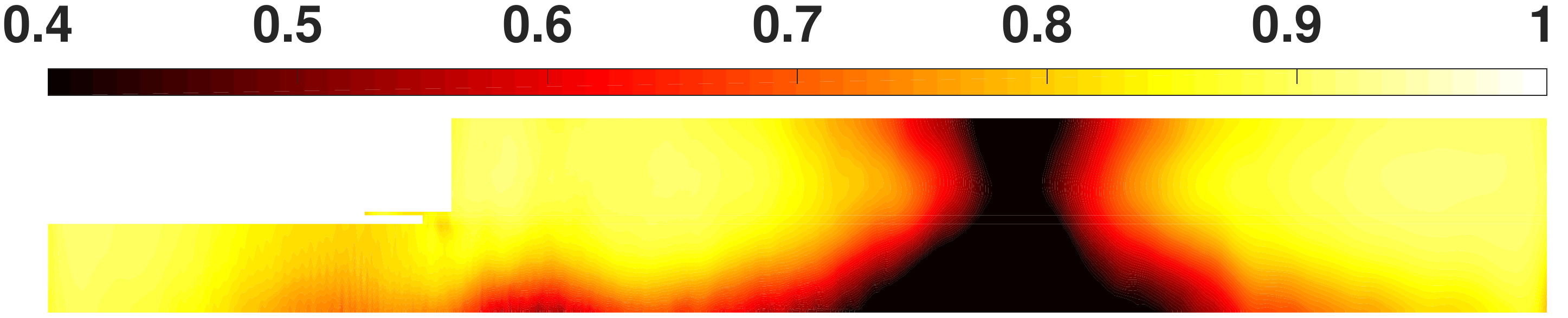}
			\caption{Fluctuation correlation}\label{trscorcont8}			
		\end{subfigure}	
	\caption{ 14-cm, transient: distribution of relative error (\%) and fluctuation correlation between pressure signals calculated from the $NN_b$-based and the table-based simulations}\label{trscorcomp1}
	\begin{subfigure}{.5\textwidth}
		\centering
		\includegraphics[width=0.95\textwidth]{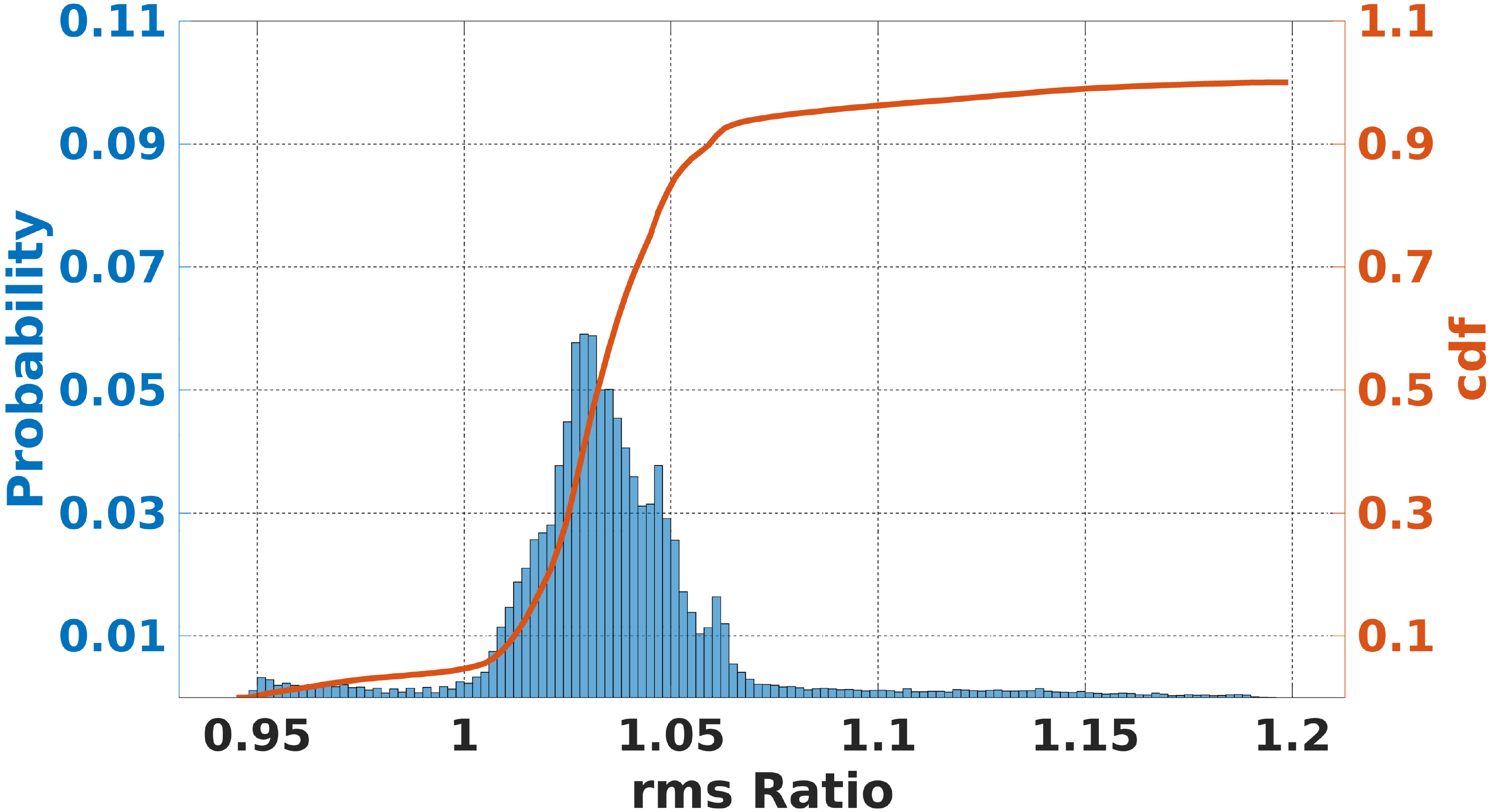}
		\caption{rms ratio distribution}\label{trsrmsrathist8}				
	\end{subfigure}	
	\begin{subfigure}{.5\textwidth}
		\centering
		\includegraphics[width=0.95\textwidth]{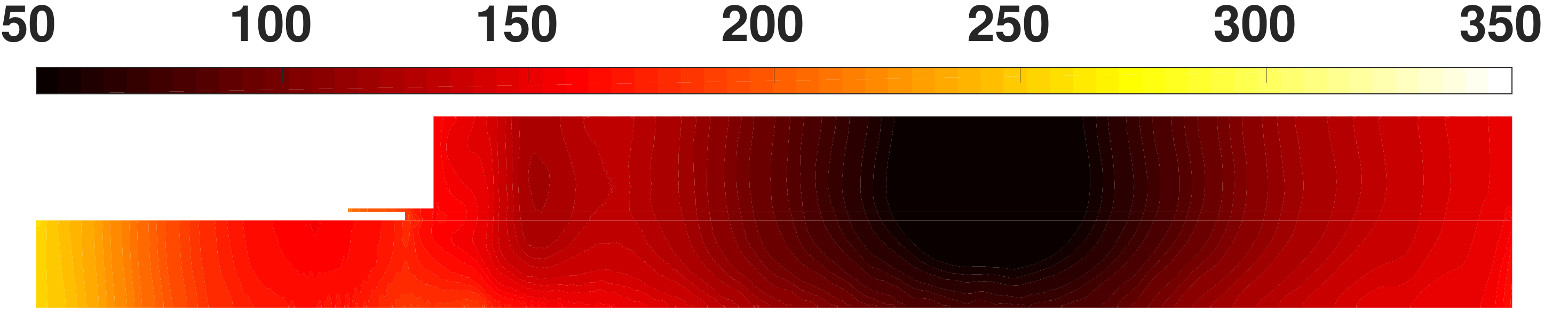}
		\caption{Table-based pressure signal rms (\si{\kilo\pascal})}\label{trsrmspowTAB}		
		\centering
		\includegraphics[width=0.95\textwidth]{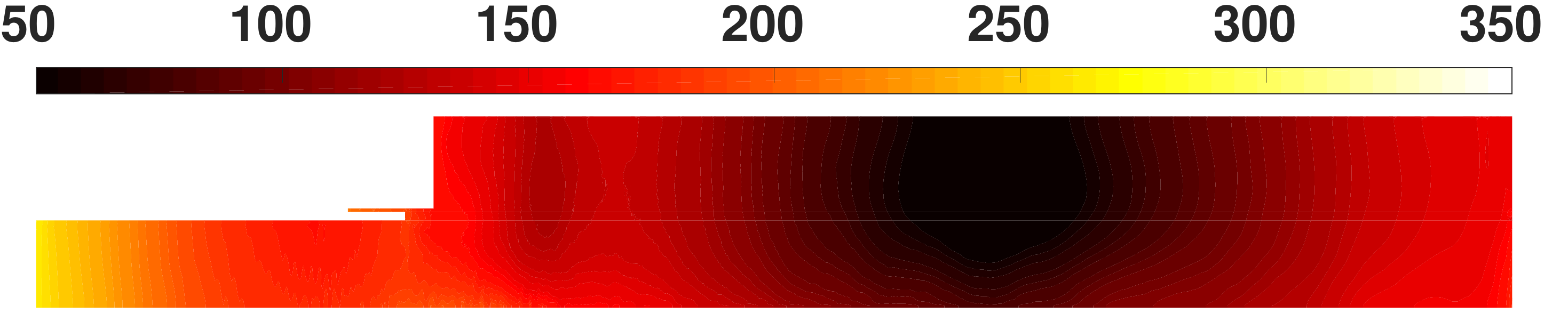}
		\caption{$NN_b$-based pressure signal rms (\si{\kilo\pascal}) }\label{trsrmspowNN8}		
	\end{subfigure}
 \caption{14-cm, transient: The distribution of $\kappa$ calculated from the $NN_b$-based (\ref{trsrmspowNN8}) to the one from the table-based (\ref{trsrmspowTAB}) simulations}\label{trsPpowcomp1}
\end{figure}

As discussed before, the $NN_a$-based simulation underestimates the rms value of pressure fluctuation in the transient case, while the $NN_b$-based simulation has a similar rms with the table-based one.
 Accordingly, $NN_b$ is selected as the flame model for the transient simulation.
 The relative error between the $NN_b$-based and the table-based simulations is shown in \figurename{~\ref{trsefullcont8}}.
Similar to the dynamic equilibrium case, in the axisymmetric configuration, the centerline ($r=0$) can be considered as a singularity and prone to numerical error and noisier signals. 

 Correlation measures how two signals are similar after their mean values have been subtracted. In the transient case, using a global mean value will distort the ability of correlation in measuring the performance of a model. 
To calculate the local mean value, the signals are smoothened using local quadratic regression \cite{Loess}. The mean values are subtracted from each signal at each spatial point to get the fluctuation signals. The correlation of the fluctuation of the pressure signal at each grid point from the $NN_b$-based and the original table-based simulations are calculated and shown in the contour plot in \figurename{~\ref{trscorcont8}}. 
		In addition, the pressure fluctuation rms values calculated from the $NN_b$-based simulation (\figurename{~\ref{trsrmspowNN8}}) are compared to those calculated from the table-based simulation (\figurename{~\ref{trsrmspowTAB}}) through their ratio at each grid point. The rms ratio is shown in \figurename{~\ref{trsrmsrathist8}}.	
		
\begin{figure}[hbt!]
			\begin{subfigure}{.5\textwidth}
				\centering
				\includegraphics[width=0.9\textwidth]{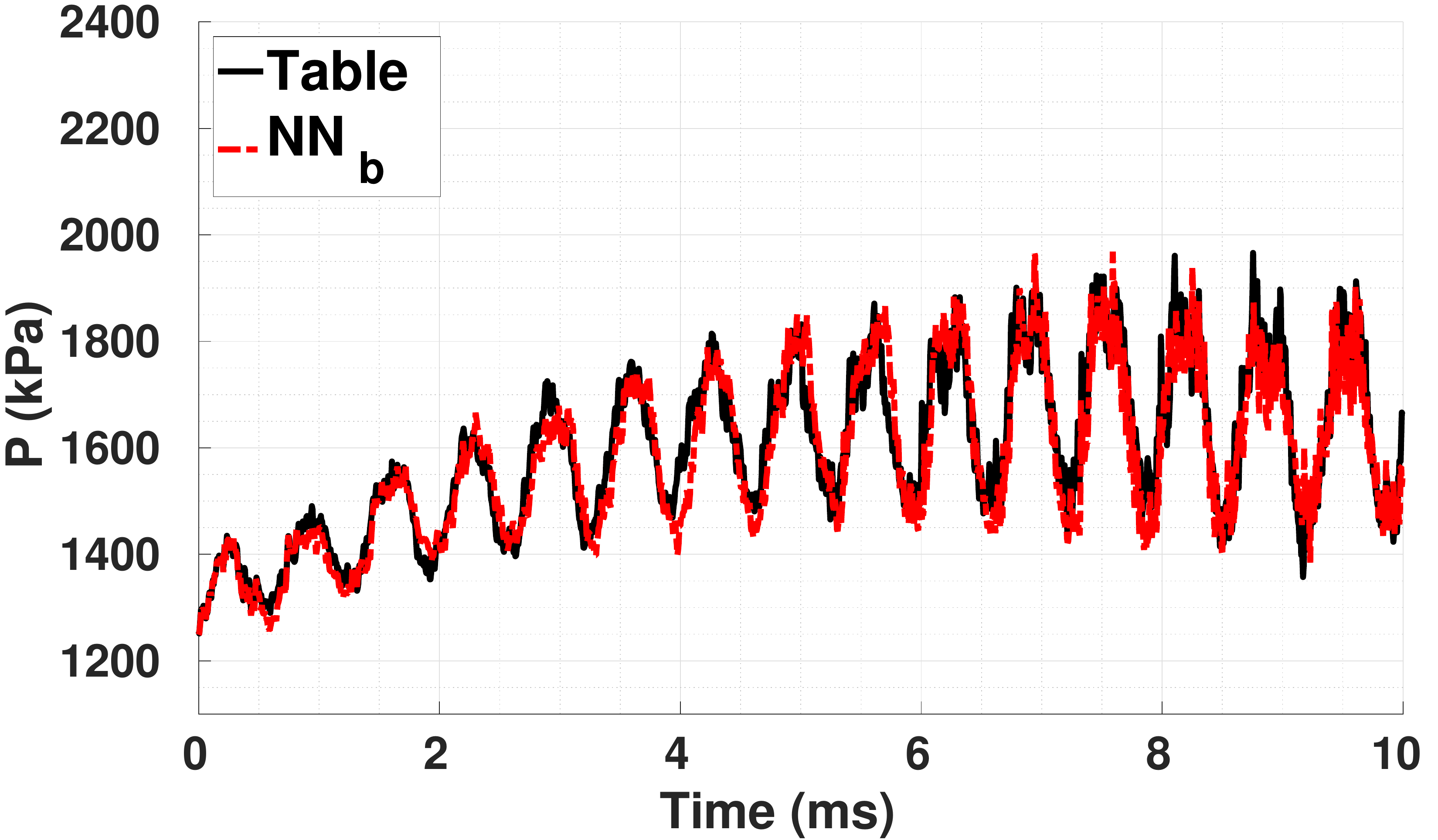}
				\caption{ $x=10$ \si{\centi\meter}, top wall}\label{trsdyn843_404}
			\end{subfigure}		
			\begin{subfigure}{.5\textwidth}
				\centering
				\includegraphics[width=0.9\textwidth]{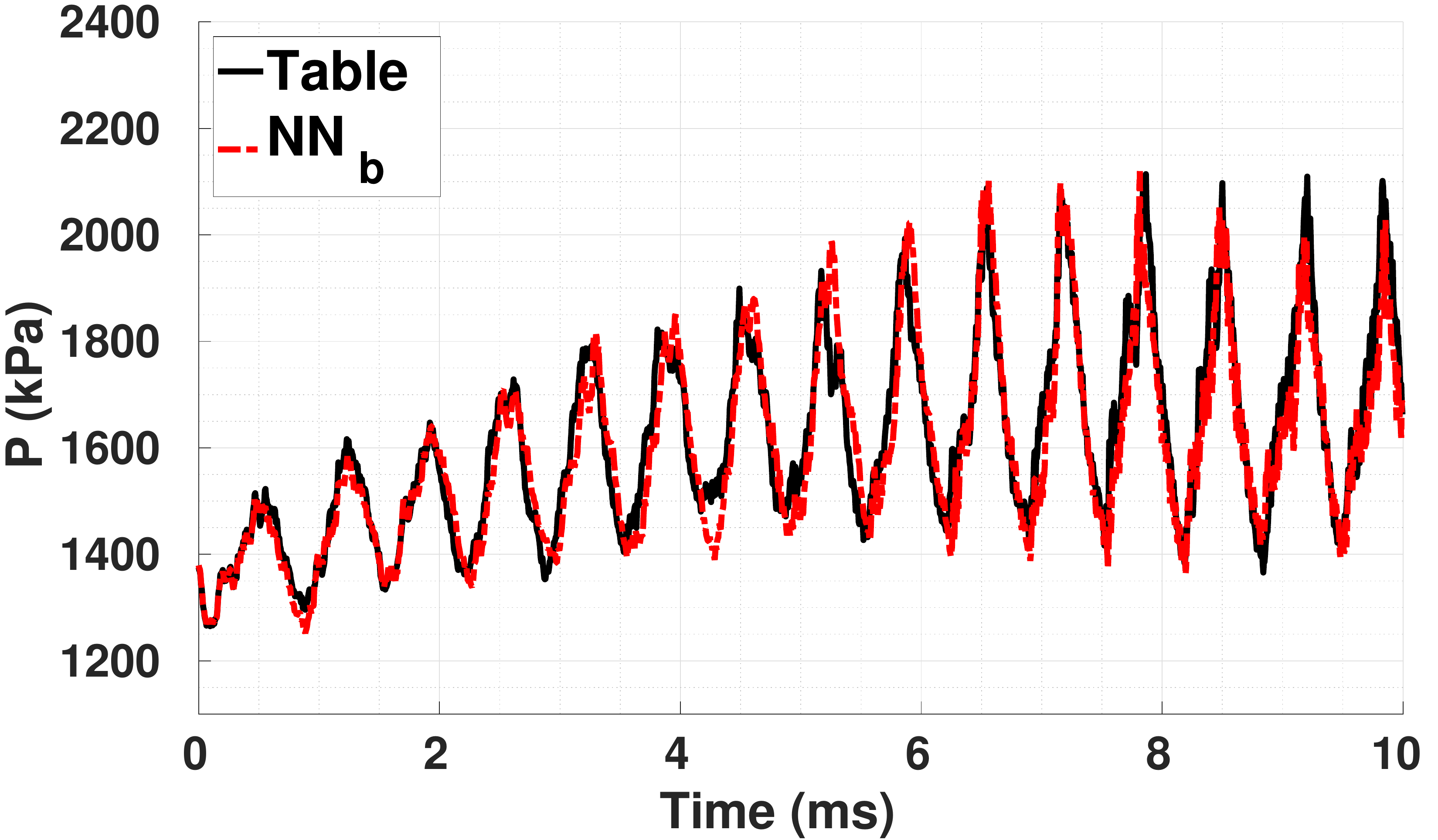}
				\caption{$x=37$ \si{\centi\meter}, top wall}\label{trsdyn843_645} 		
			\end{subfigure}				
			\begin{subfigure}{.5\textwidth}
				\centering
				\includegraphics[width=0.9\textwidth]{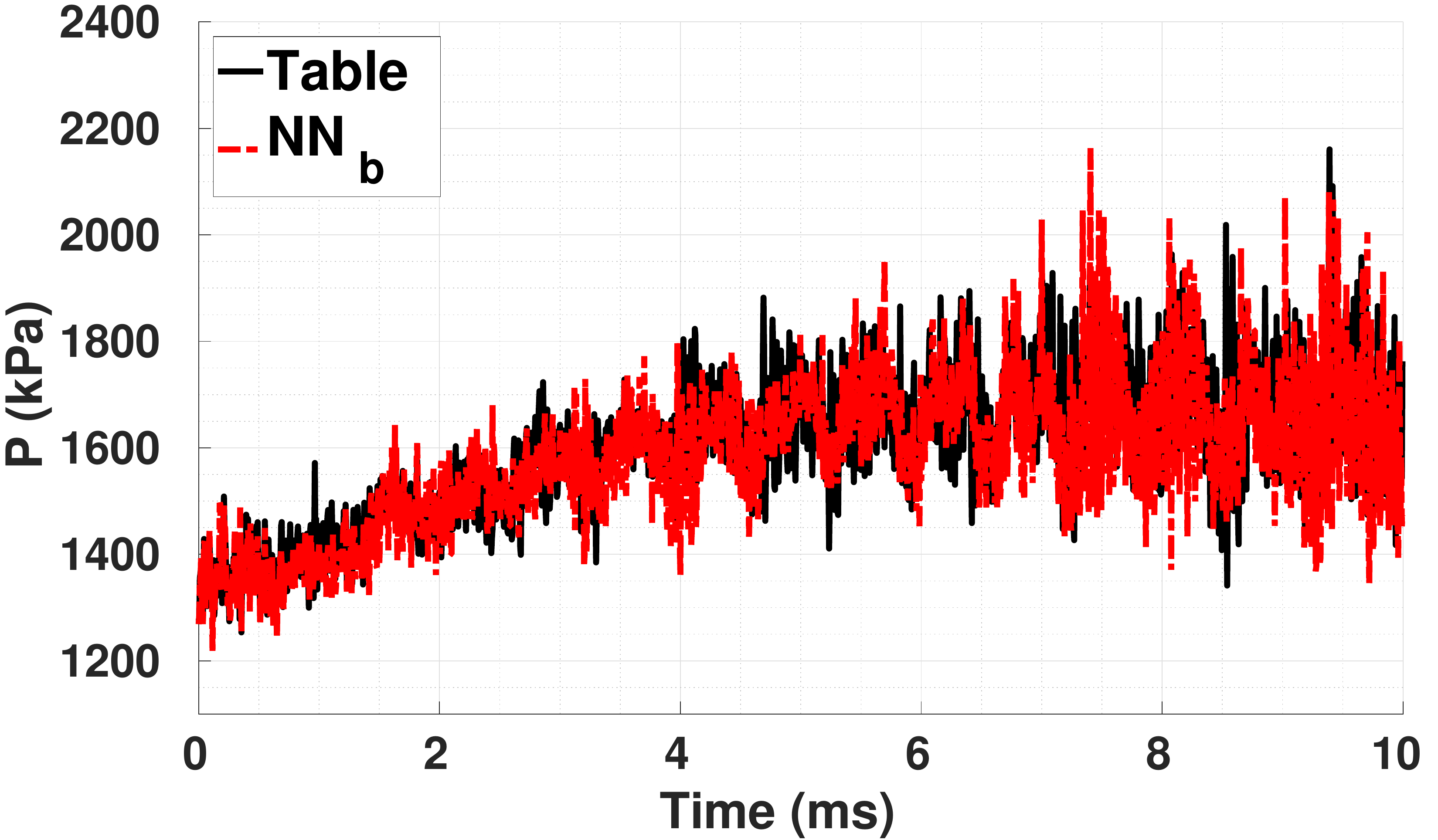}
				\caption{ $x=17.11$ \si{\centi\meter}, centerline}\label{trsdyn81_802}
			\end{subfigure}		
			\begin{subfigure}{.5\textwidth}
				\centering
				\includegraphics[width=0.9\textwidth]{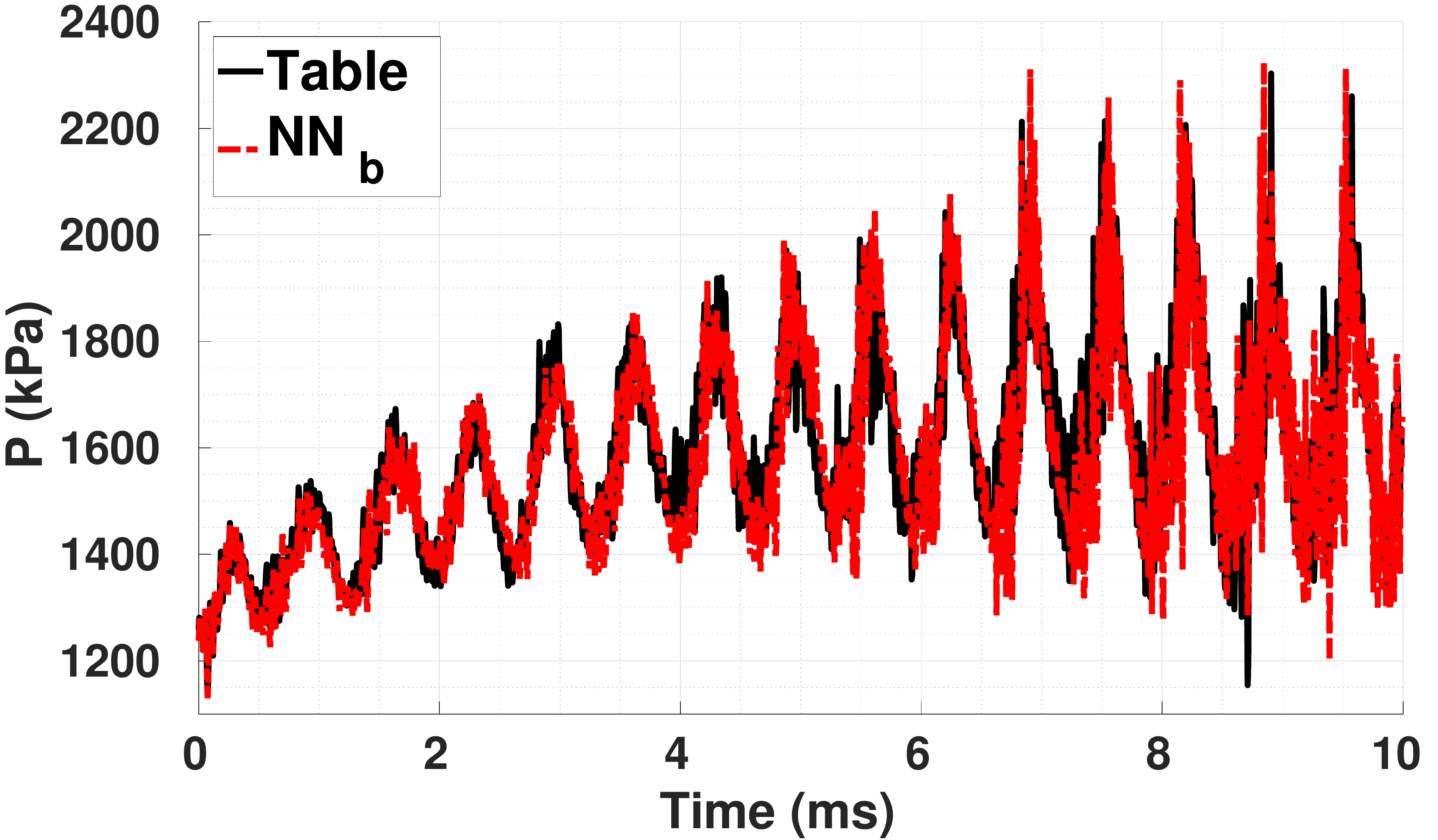}
				\caption{ $x=2.58$ \si{\centi\meter}, $r=0.37$ \si{\centi\meter}}\label{trsdyn81_353} 		
			\end{subfigure}				
			\caption{ 14-cm, transient: the $NN_b$-based and the table-based simulations comparison of pressure signal
at different points}
\label{trsPsigcomp2}		
		\end{figure}

 Figure{ \ref{trsdyn843_404}} and \figurename{~\ref{trsdyn843_645}} compare the pressure signals 
 at the antinode (10 \si{\centi\meter}) and near the nozzle (37 \si{\centi\meter}) on the top wall for the transient case.
 At the antinode, the correlation is 87.65\%, and near the nozzle, the correlation is 91.46\%. At these points, the overall relative errors are 3.65\% and 4.02\%, respectively. The rms ratio is 104.41\%, and the mean value is estimated with 0.95\% error at $x=10$ \si{\centi\meter}. At $x=37$ \si{\centi\meter}, the rms ratio is 100.42\%, and the mean value is estimated with a 1.43\% error.
 Figure{ \ref{trsdyn81_802}} compares the pressure signal located on the centerline near the pressure node (17.1 \si{\centi\meter}) 
 between the NN-based simulation and the original one, where the correlation is 12.48\%. The overall error at this point is 7.33\%, the rms ratio is 112.02\%, and the mean value error is 1.08\%. 
Another point with low correlation is located above the centerline at 2.58 \si{\centi\meter}, which is right after the dump plane near the centerline. Figure{ \ref{trsdyn81_353}} compares the pressure signal at this point with a correlation of 71.36\%. The overall error at this point is 7.42\%, the rms ratio is 105.79\%, and the mean value error is 1.15\%. 
 These points are selected to demonstrate the need for different criteria to measure error. In \figurename{~\ref{trsPsigcomp2}}, the overall errors are similar, yet the correlations are drastically different.
 
\begin{figure}[hbt!]
 	\begin{subfigure}{.33\textwidth}
 		\centering
 		\includegraphics[width=0.95\textwidth]{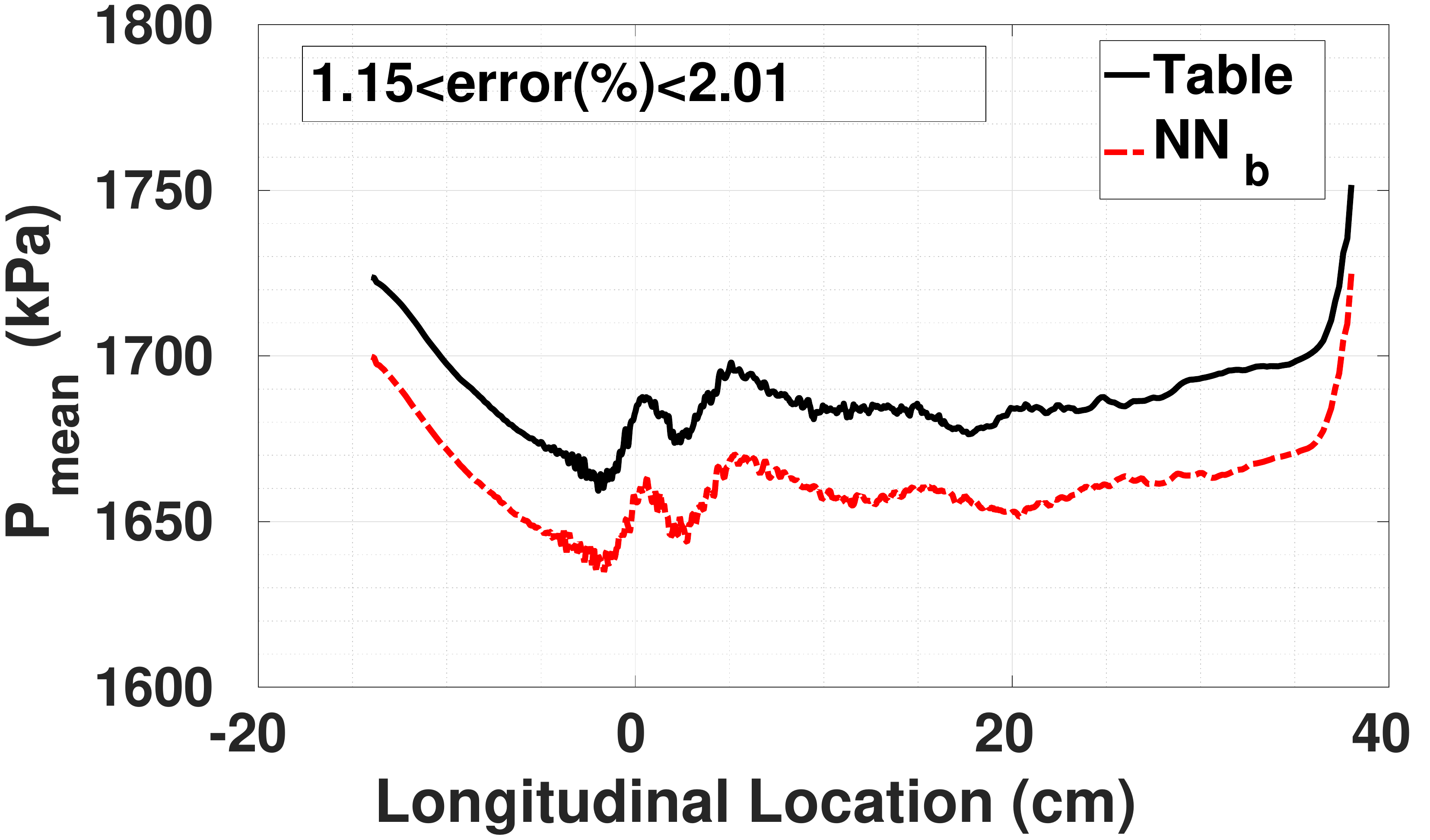}
 		\caption{Mean pressure (\si{\kilo\pascal})}\label{trsPMSmean8}					
 	\end{subfigure}
 	\begin{subfigure}{.33\textwidth}
 		\centering
 		\includegraphics[width=0.95\textwidth]{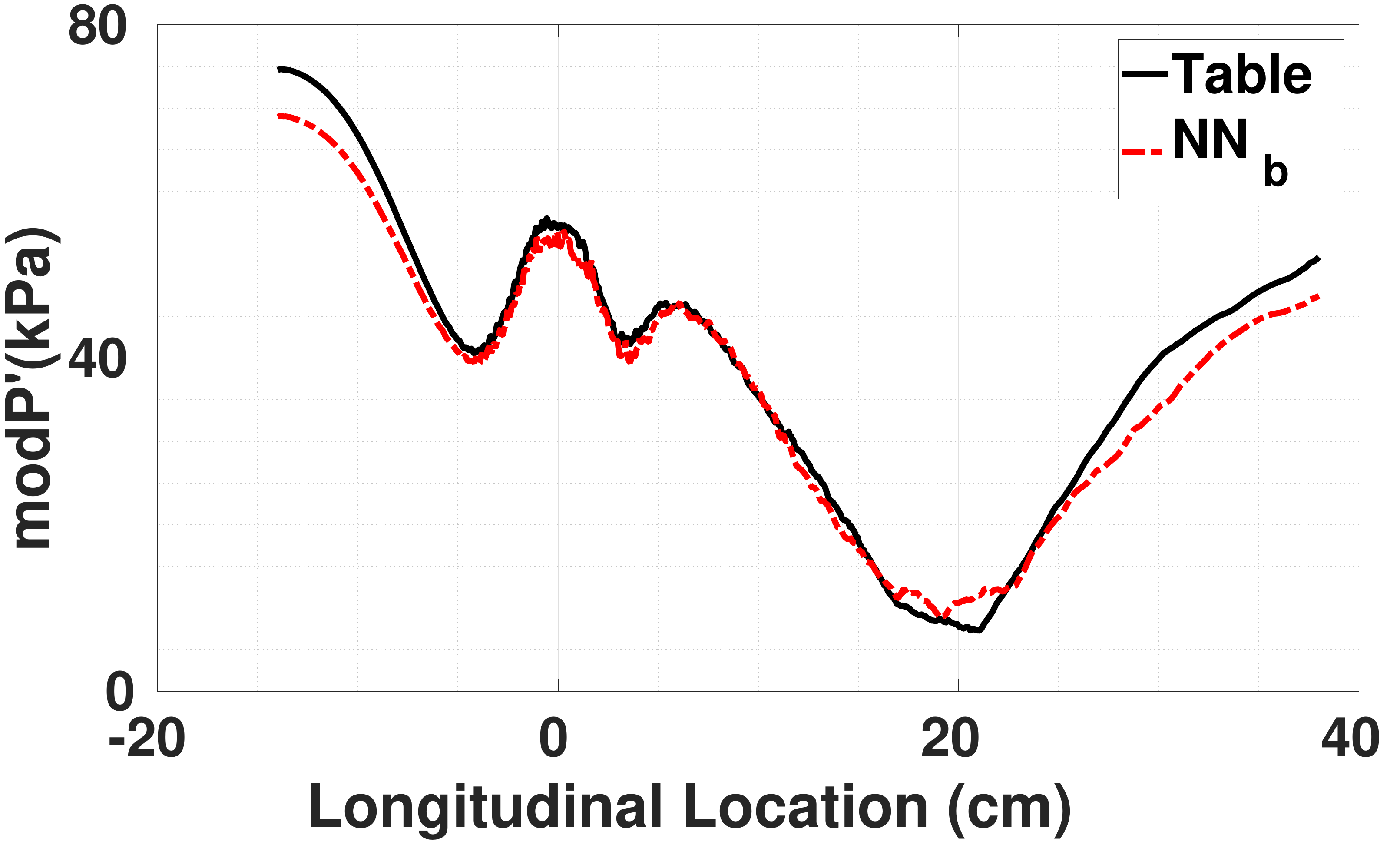}
 		\caption{First mode shape (\si{\kilo\pascal})}\label{trsPMSfirst8}	 
 	\end{subfigure}
 	\begin{subfigure}{.33\textwidth}						
 		\centering
 		\includegraphics[width=0.95\textwidth]{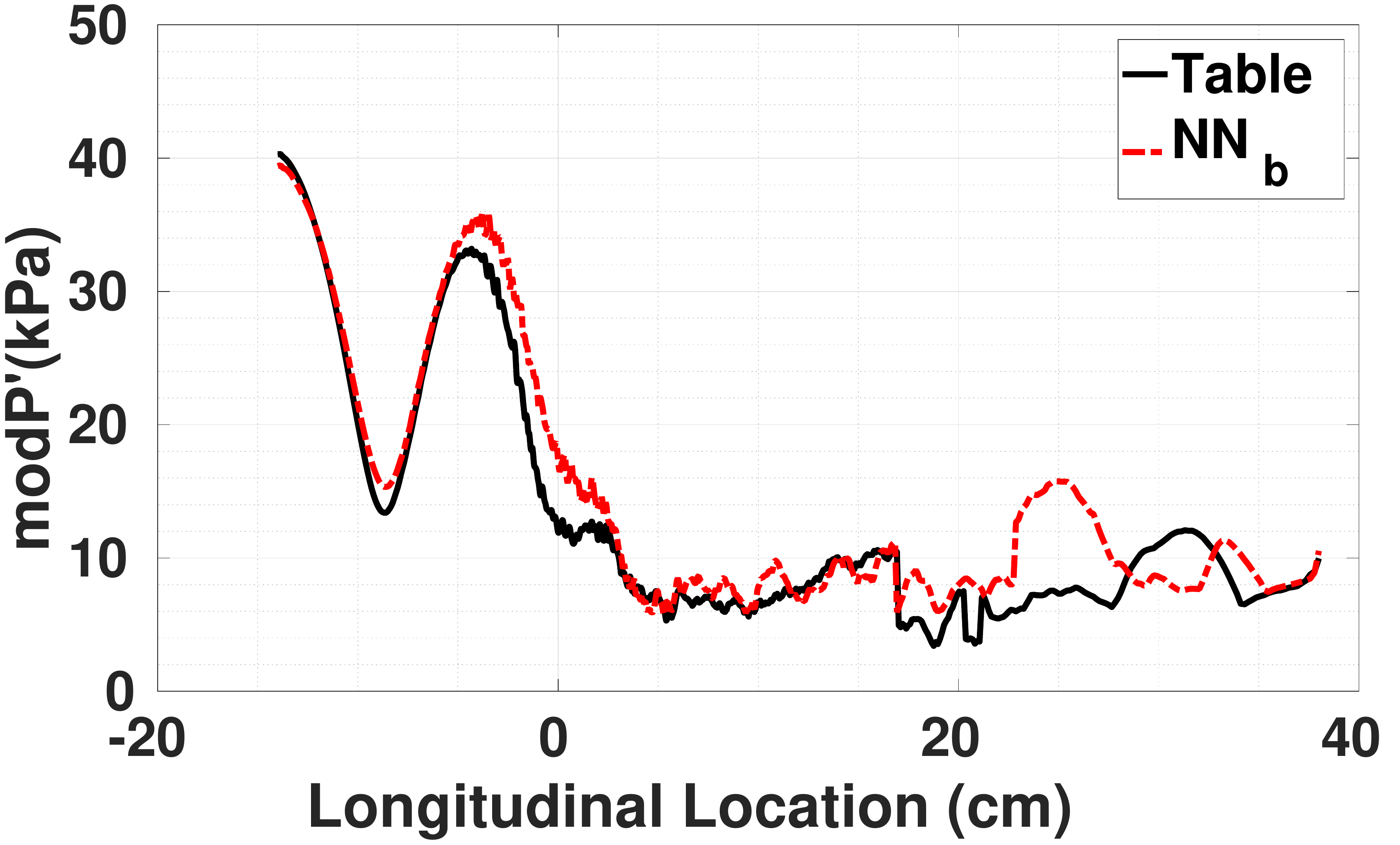}
 		\caption{Second mode shape (\si{\kilo\pascal})}\label{trsPMSsec8}											
 	\end{subfigure}	
 	\caption{ 14-cm, transient: comparison of pressure mean, the first, and the second mode shapes between $NN_b$-based and table-based simulations (last 4 $ms$)}\label{trsmodshcompnew}
 \end{figure}

The frequency content of pressure signals are compared through the mode shapes over the centerline. In the analysis here, only the steady-state parts of the simulations are considered. 
The time-average of each of the signals is plotted in \figurename{~\ref{trsPMSmean8}}. The value of the Fourier spectrum for the first and the second modes are plotted in \figurename{~\ref{trsPMSfirst8}} and \figurename{~\ref{trsPMSsec8}}. The NN-based simulation presents a good estimate of the mean ($e_m$ ranges between 1.15\% to 2.01\%), first and second modes. The discrepancies between the two NN-based and table-based simulations start to grow from the third mode.
 
\begin{figure}[hbt!]
		\begin{subfigure}{.5\textwidth}
				\centering
				\includegraphics[width=0.95\textwidth]{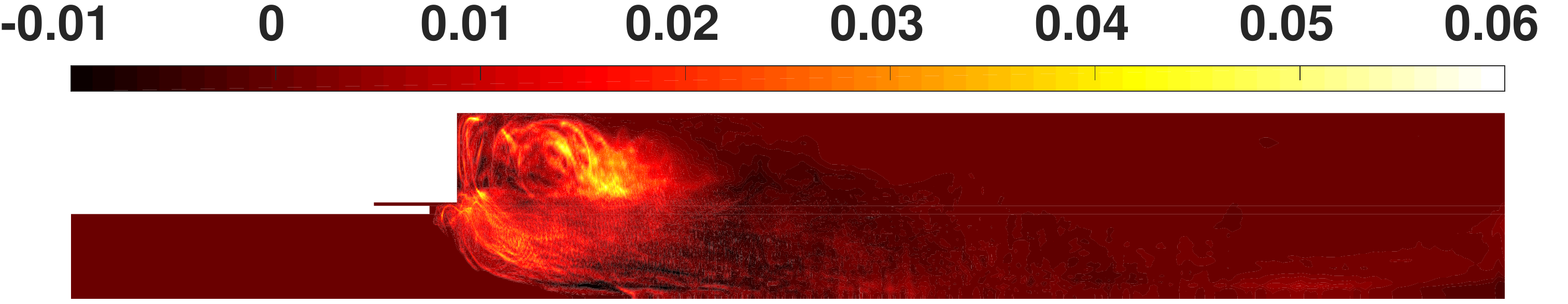}
				\caption{ Table-based CFD: \textit{mRI}}\label{trsWCRIrr}			
			\centering
			\includegraphics[width=0.95\textwidth]{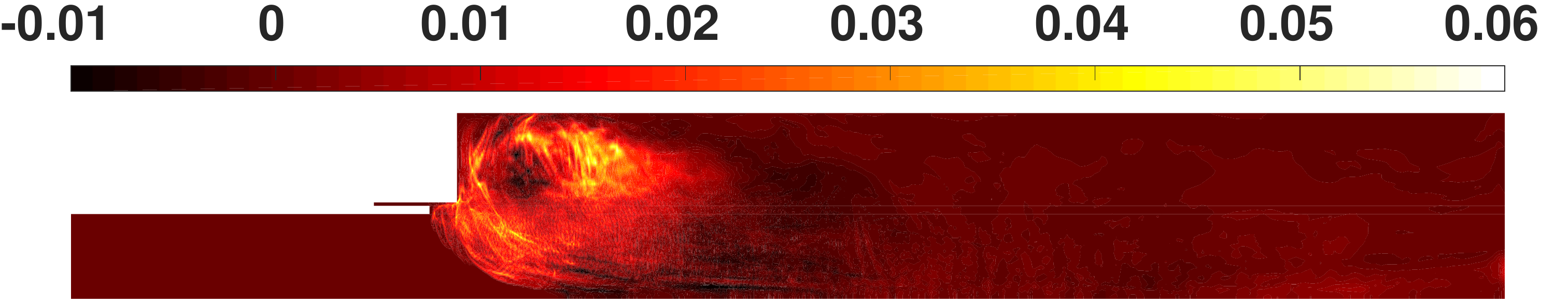}
			\caption{ $NN_b$-based CFD: \textit{mRI}}\label{trsWCRInnb}										
		\end{subfigure}		
		\begin{subfigure}{.5\textwidth}
			\centering
			\includegraphics[width=0.95\textwidth]{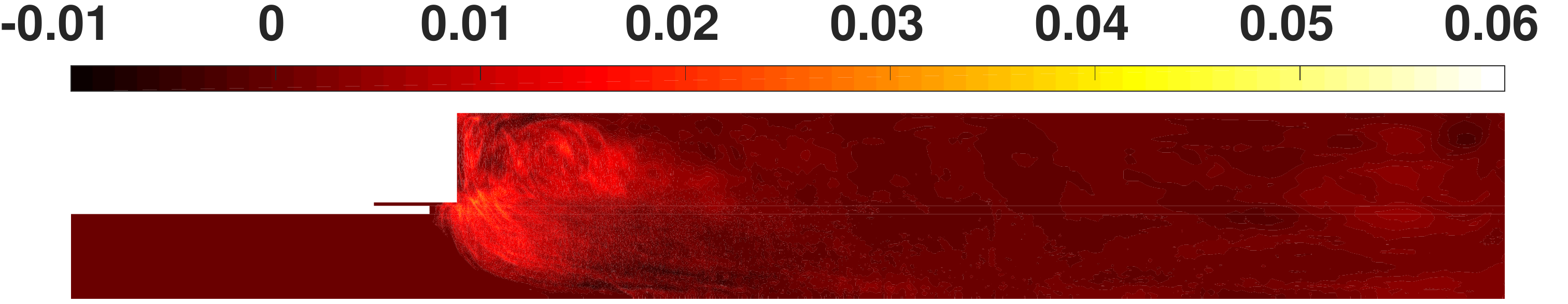}
			\caption{ Table-based CFD: \textit{RI}}\label{trsHRRRIrr}				
			\centering
			\includegraphics[width=0.95\textwidth]{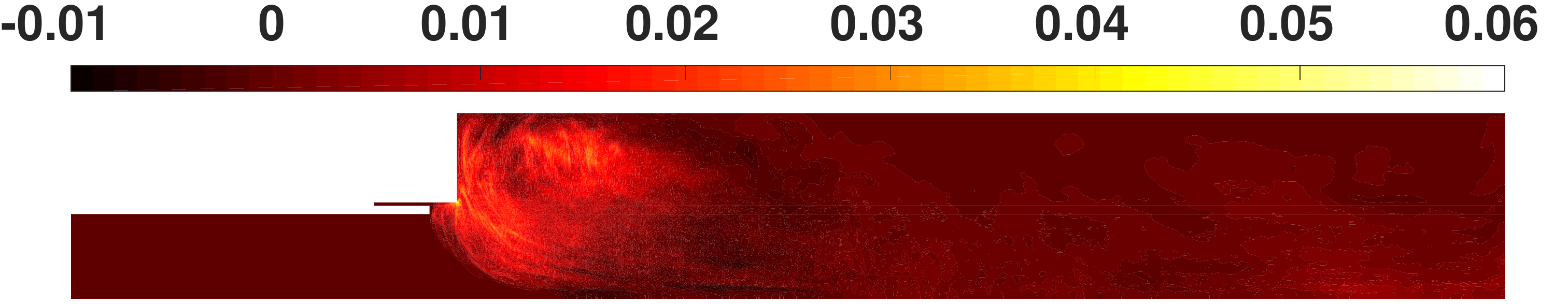}
			\caption{ $NN_b$-based CFD: \textit{RI}}\label{trsHRRRInnb}								
		\end{subfigure}		
		\caption{ 14-cm, transient: comparison of \textit{RI} and \textit{mRI} between the $NN_b$-based and the table-based simulations} \label{trsRIlabcomp1}		
	\end{figure}

 The local \textit{RI} and \textit{mRI} are compared for the NN-based simulation in \figurename{~\ref{trsRIlabcomp1}}. Comparing \figurename{~\ref{trsWCRInnb}} with \figurename{~\ref{trsWCRIrr}}, and comparing \figurename{~\ref{trsHRRRInnb}} with \figurename{~\ref{trsHRRRIrr}}, show a great similarity in flame location and magnitude between the two simulations.
 
\begin{figure}[hbt!]
 	\begin{subfigure}{.5\textwidth}
 		\centering
 		\includegraphics[width=0.98\textwidth]{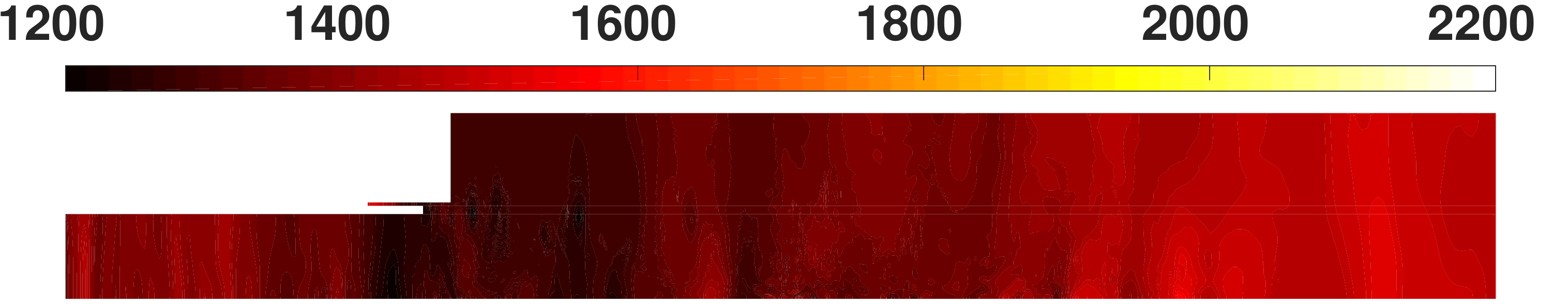}
 		\caption{$\overline{P}$ (\si{\kilo\pascal}), $t_1$, table}\label{trsW8WPsigco_101}	
 		\centering
 		\includegraphics[width=0.98\textwidth]{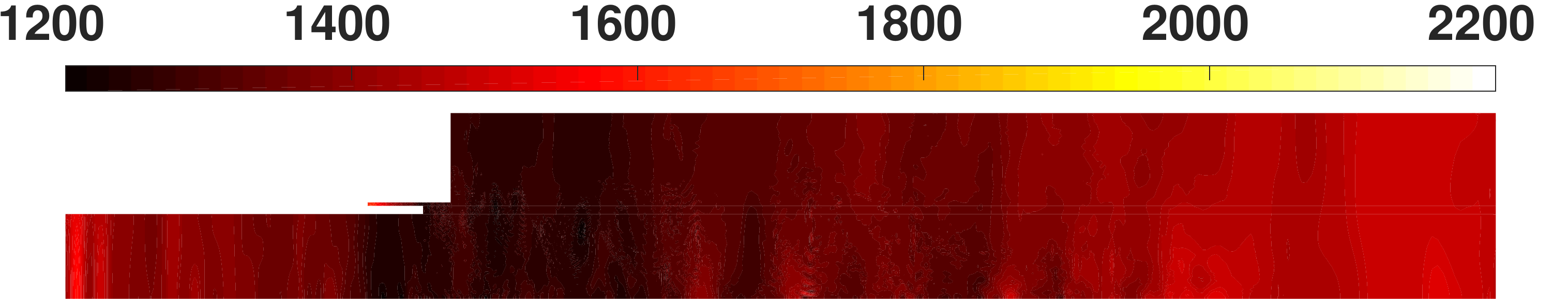}
 		\caption{$\overline{P}$ (\si{\kilo\pascal}), $t_1$, $NN_b$}\label{trsNN8NNPsigco_101}							
 		\centering
 		\includegraphics[width=0.98\textwidth]{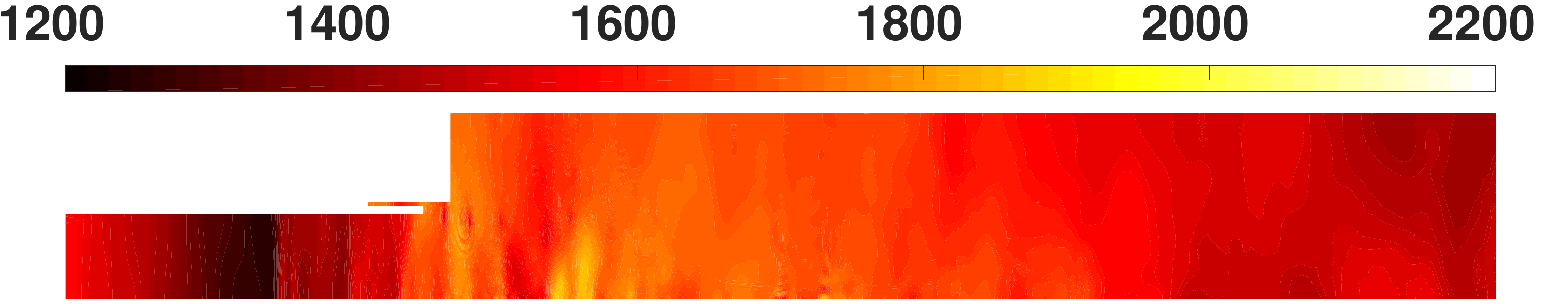}
 		\caption{$\overline{P}$ (\si{\kilo\pascal}), $t_2$, table}\label{trsW8WPsigco_701}	
 		\centering
 		\includegraphics[width=0.98\textwidth]{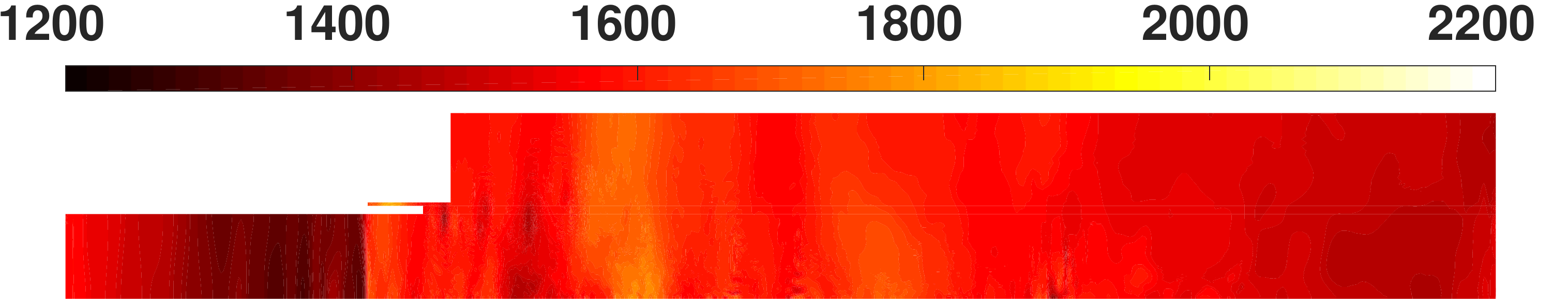}
 		\caption{$\overline{P}$ (\si{\kilo\pascal}), $t_2$, $NN_b$}\label{trsNN8NNPsigco_701}							
 		\centering
 		\includegraphics[width=0.98\textwidth]{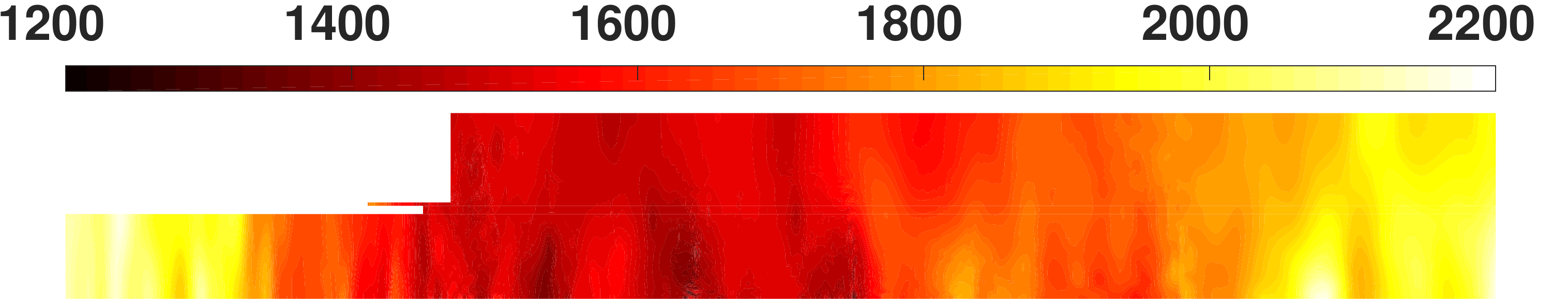}
 		\caption{$\overline{P}$ (\si{\kilo\pascal}), $t_3$, table}\label{trsW8WPsigco_1301}	
 		\centering
 		\includegraphics[width=0.98\textwidth]{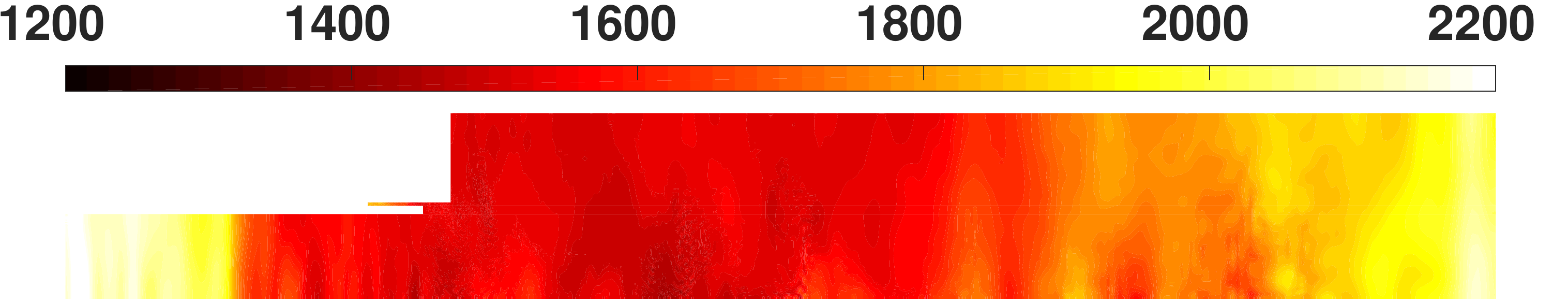}
 		\caption{$\overline{P}$ (\si{\kilo\pascal}), $t_3$, $NN_b$}\label{trsNN8NNPsigco_1301}	
 	\end{subfigure}		
 		\begin{subfigure}{.5\textwidth}
 			\centering
 			\includegraphics[width=0.95\textwidth]{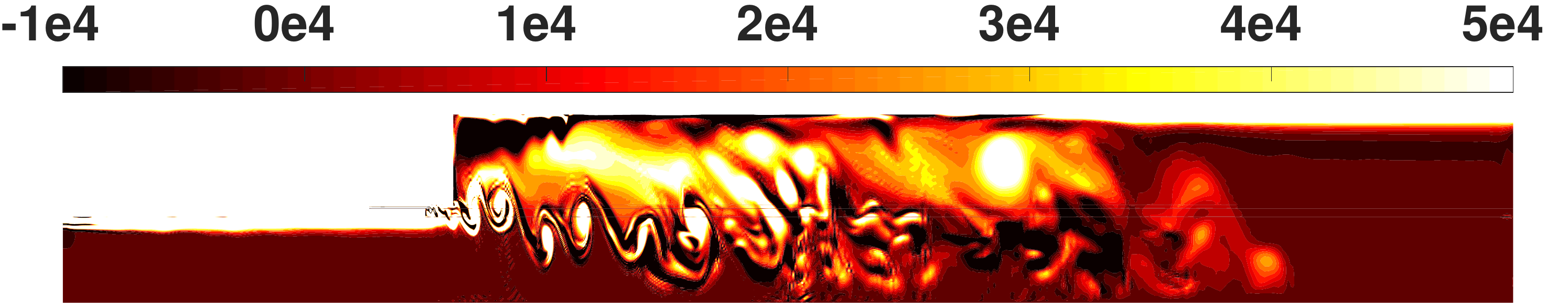}
 			\caption{$\widetilde{\Omega}$ (\si{\per\second}), $t_1$, table}\label{trsW8Wvortsigco_101}	
 			\centering
 			\includegraphics[width=0.95\textwidth]{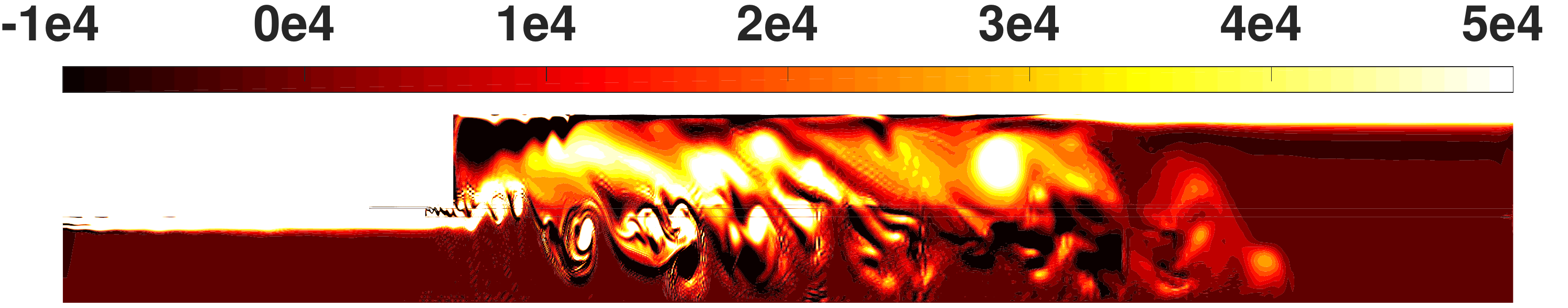}
 			\caption{$\widetilde{\Omega}$ (\si{\per\second}), $t_1$, $NN_b$}\label{trsNN8NNvortsigco_101}						
 			\centering
 			\includegraphics[width=0.95\textwidth]{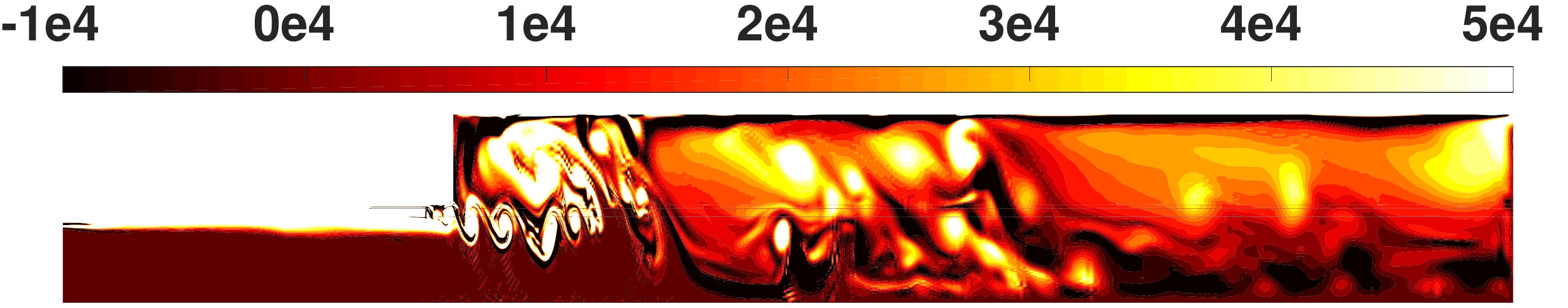}		
 			\caption{$\widetilde{\Omega}$ (\si{\per\second}), $t_2$, table}\label{trsW8Wvortsigco_701}			
 			\centering
 			\includegraphics[width=0.95\textwidth]{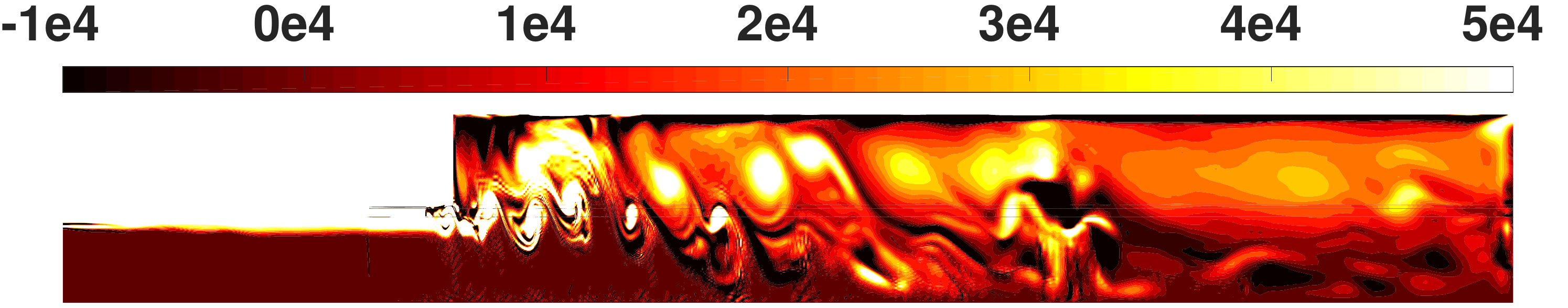}
 			\caption{$\widetilde{\Omega}$ (\si{\per\second}), $t_2$, $NN_b$}\label{trsNN8NNvortsigco_701}				
 			\centering
 			\includegraphics[width=0.95\textwidth]{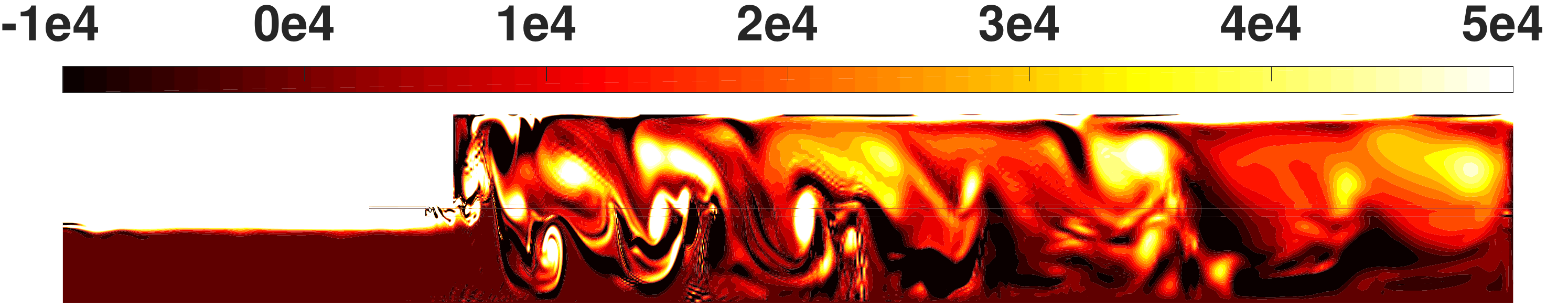}
 			\caption{$\widetilde{\Omega}$ (\si{\per\second}), $t_3$, table}\label{trsW8Wvortsigco_1301}	
 			\centering
 			\includegraphics[width=0.95\textwidth]{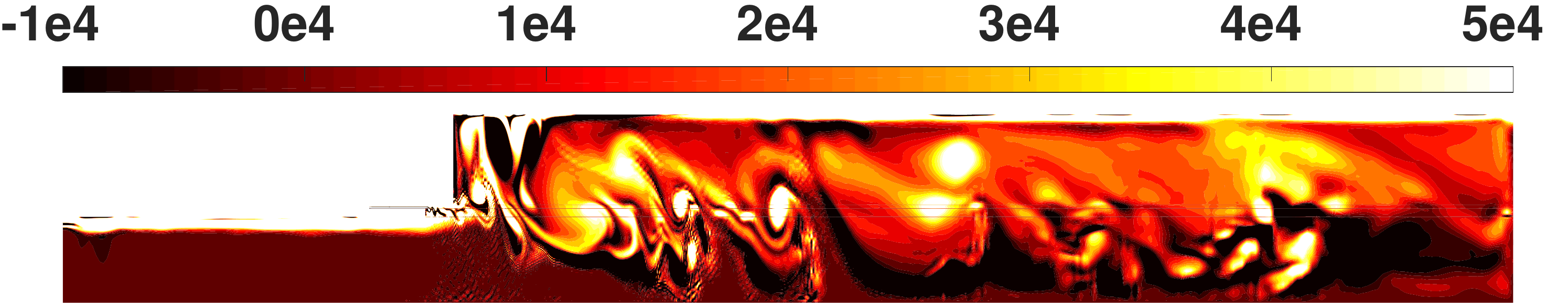}
 			\caption{$\widetilde{\Omega}$ (\si{\per\second}), $t_3$, $NN_b$}\label{trsNN8NNvortsigco_1301}								
 		\end{subfigure}			
 	\caption{ 14-cm, transient: pressure and vorticity snapshots from the $NN_b$-based and the table-based simulations}\label{trsPTtime}	
 	\end{figure}

\begin{figure}[hbt!]	 	
 \begin{subfigure}{.5\textwidth}
 	\centering
 	\includegraphics[width=0.95\textwidth]{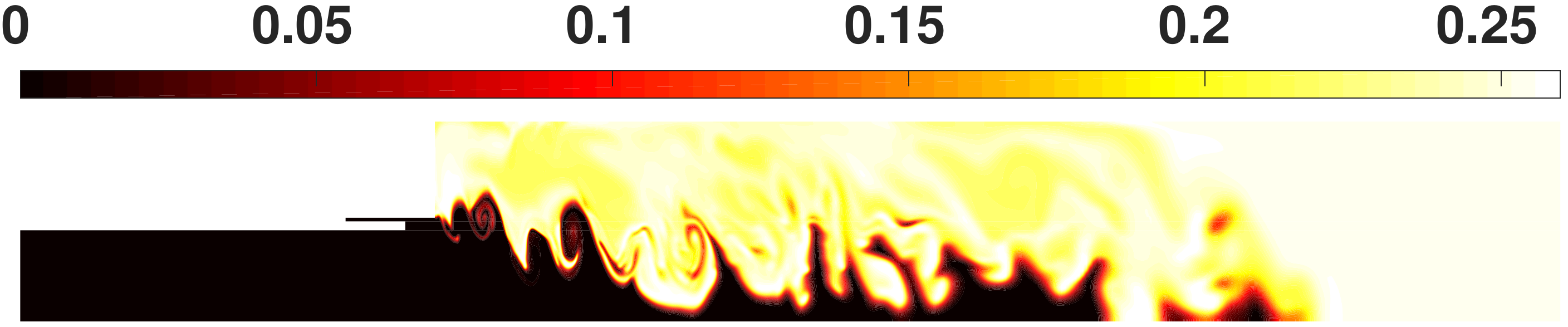}
 	\caption{$\widetilde{C}$, $t_1$, table}\label{trsW8WCsigco_101}
 	\centering
 	\includegraphics[width=0.95\textwidth]{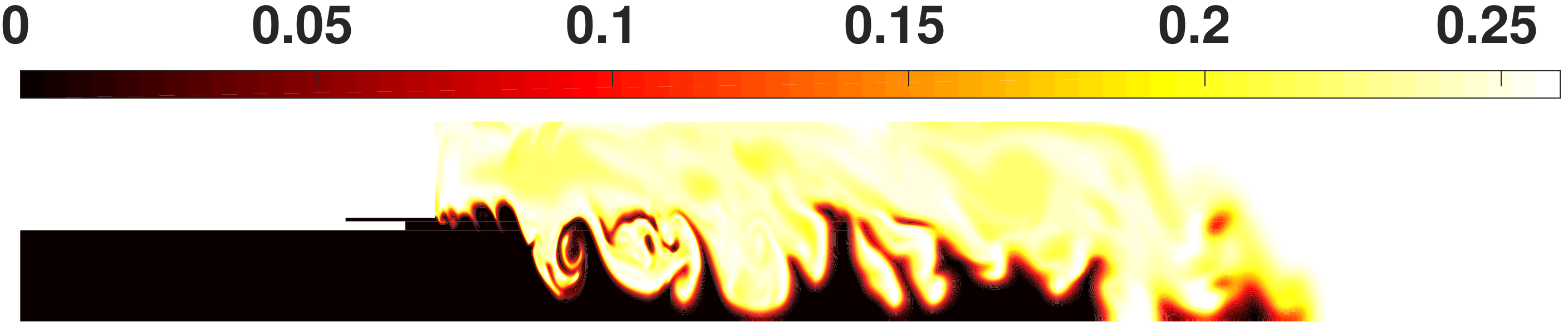}
 	\caption{$\widetilde{C}$, $t_1$, $NN_b$}\label{trsNN8NNCsigco_101}						
 	\centering
 	\includegraphics[width=0.95\textwidth]{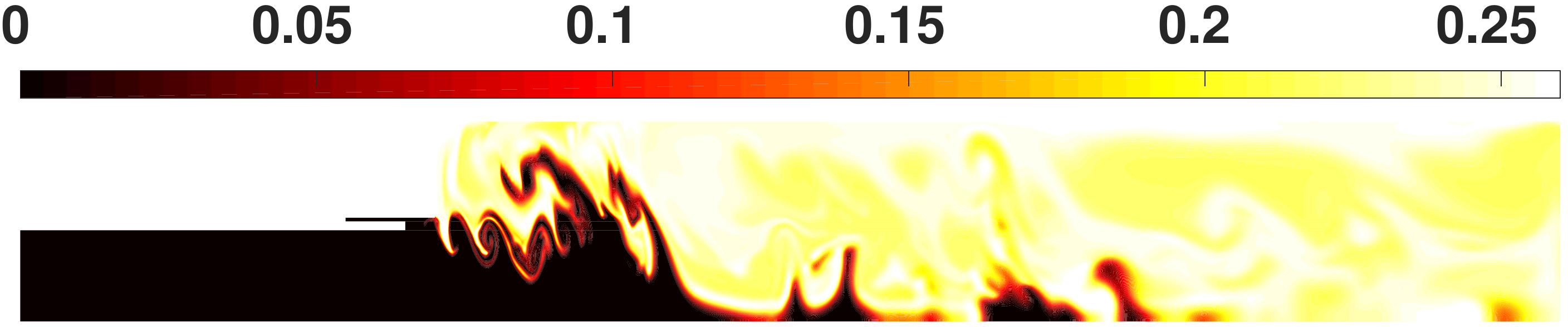}
 	\caption{$\widetilde{C}$, $t_2$, table}\label{trsW8WCsigco_701}
 	\centering
 	\includegraphics[width=0.95\textwidth]{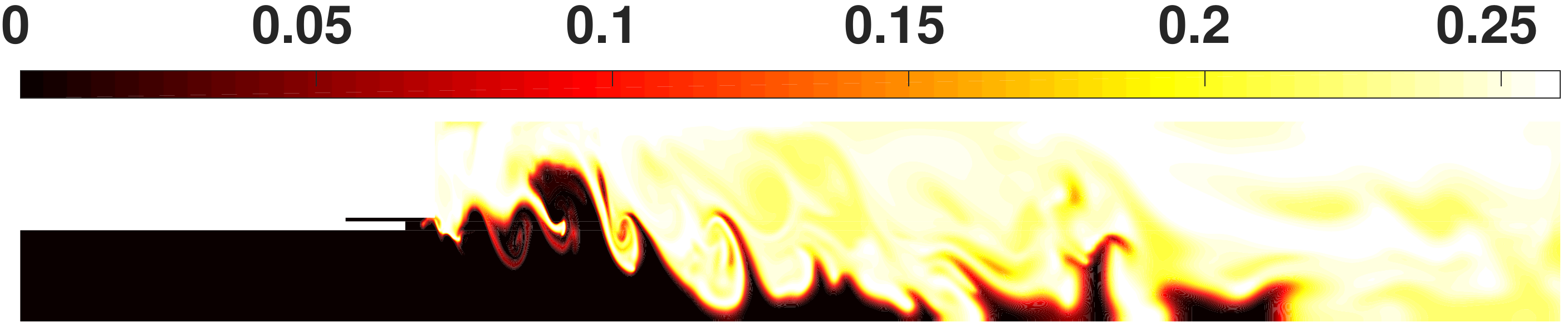}
 	\caption{$\widetilde{C}$, $t_2$, $NN_b$}\label{trsNN8NNCsigco_701}						
 	\centering
 	\includegraphics[width=0.95\textwidth]{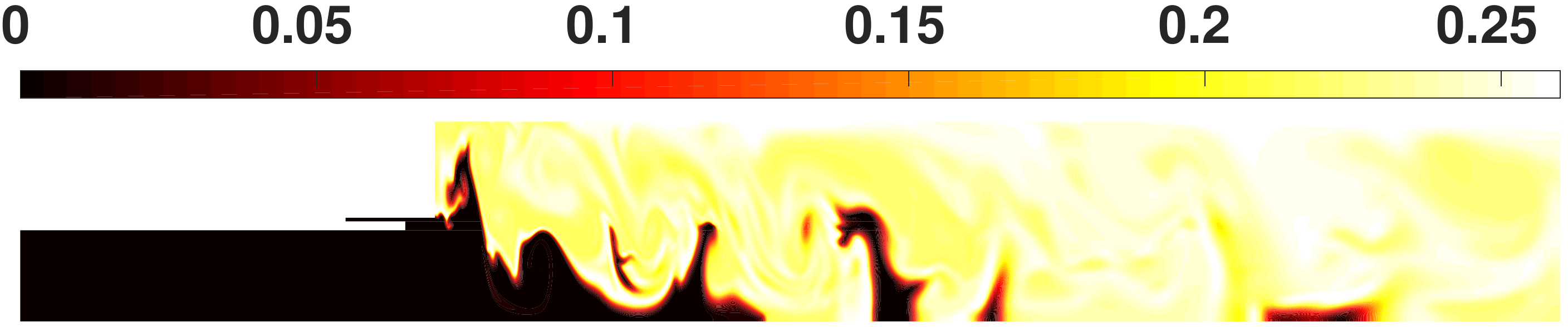}
 	\caption{$\widetilde{C}$, $t_3$, table}\label{trsW8WCsigco_1301}
 	\centering
 	\includegraphics[width=0.95\textwidth]{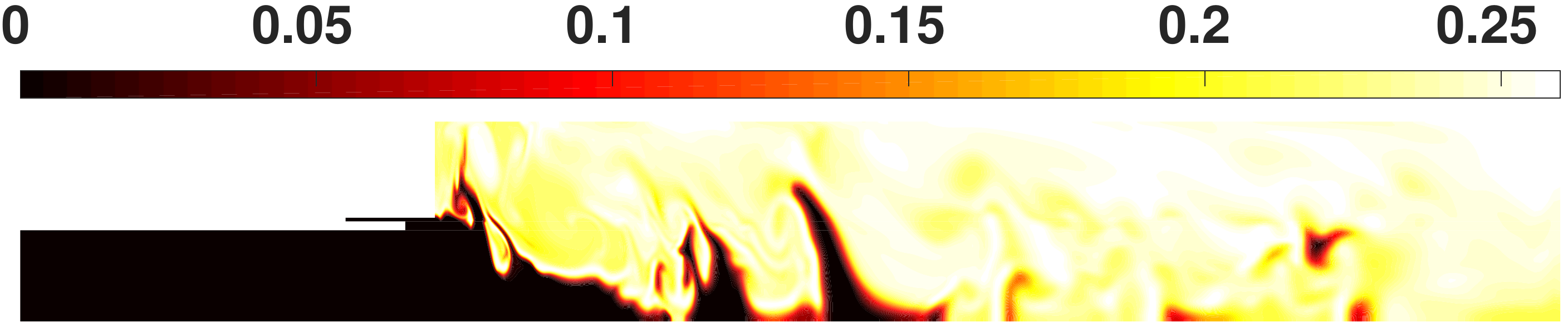}
 	\caption{$\widetilde{C}$, $t_3$, $NN_b$}\label{trsNN8NNCsigco_1301}										
 \end{subfigure}		
 \begin{subfigure}{.5\textwidth}
 	\centering
 	\includegraphics[width=0.95\textwidth]{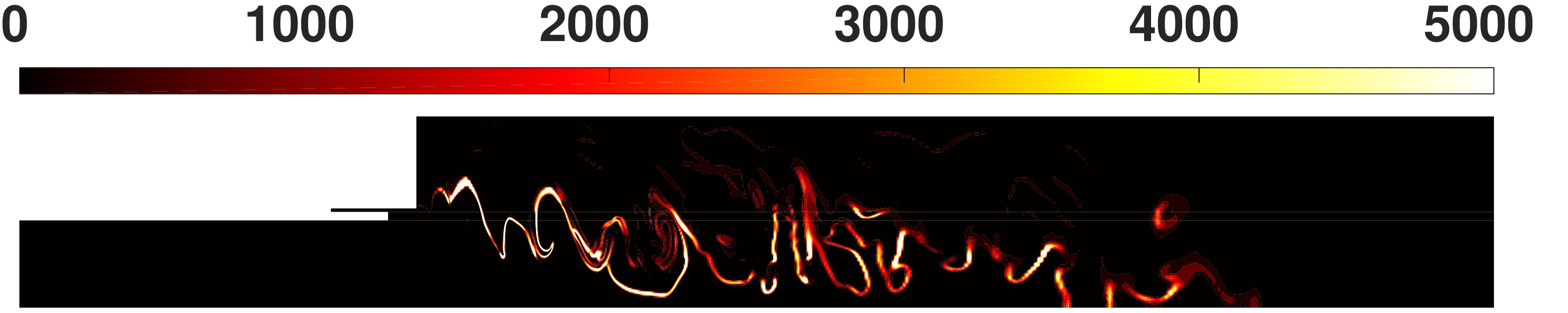}
 	\caption{$\widetilde{\dot{\omega}}_C$ (\si{\kilogram\per\meter\cubed\per\second}), $t_1$, table}\label{trsW8WProdCsigco_101}	
 	\centering
 	\includegraphics[width=0.95\textwidth]{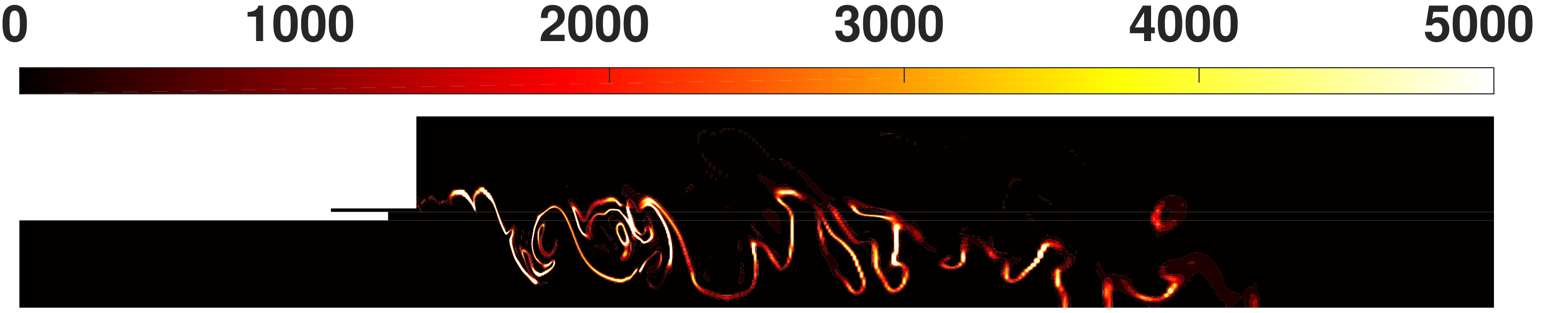}
 	\caption{$\widetilde{\dot{\omega}}_C$ (\si{\kilogram\per\meter\cubed\per\second}), $t_1$, $NN_b$}\label{trsNN8NNProdCsigco_101}					
 	\centering
 	\includegraphics[width=0.95\textwidth]{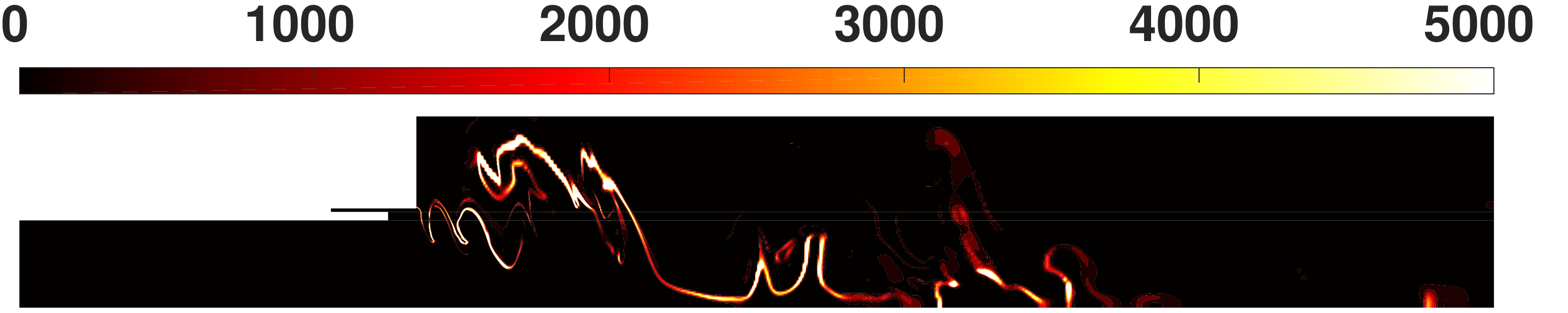}
 	\caption{$\widetilde{\dot{\omega}}_C$ (\si{\kilogram\per\meter\cubed\per\second}), $t_2$, table}\label{trsW8WProdCsigco_701}	
 	\centering
 	\includegraphics[width=0.95\textwidth]{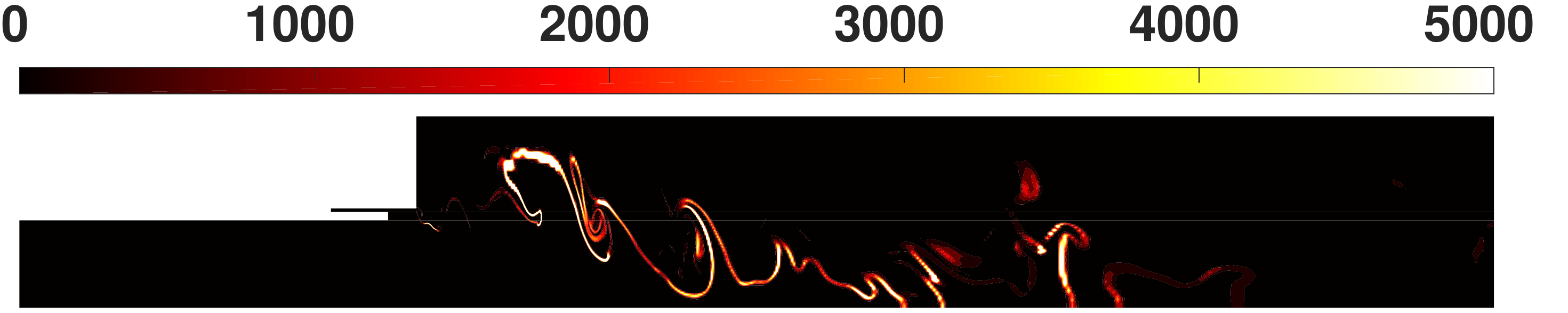}
 	\caption{$\widetilde{\dot{\omega}}_C$ (\si{\kilogram\per\meter\cubed\per\second}), $t_2$, $NN_b$}\label{trsNN8NNProdCsigco_701}							
 	\centering
 	\includegraphics[width=0.95\textwidth]{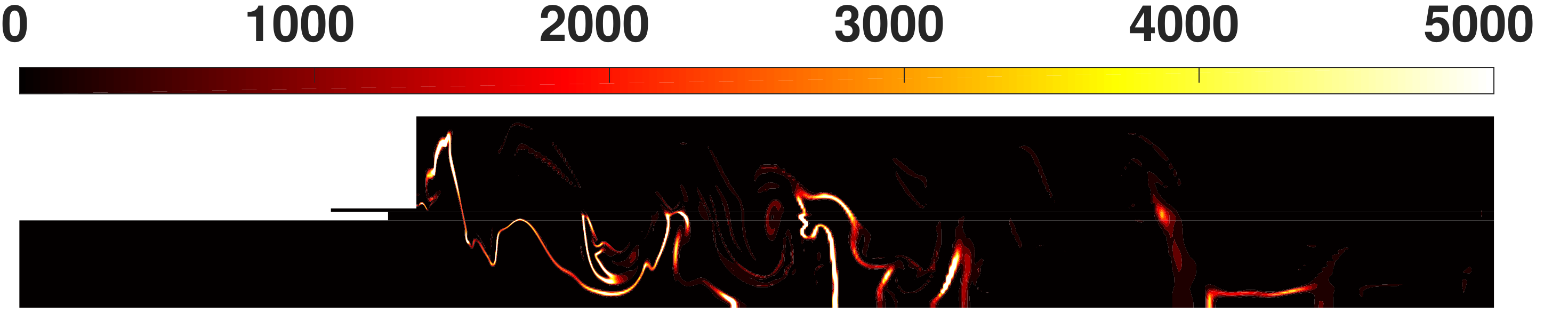}
 	\caption{$\widetilde{\dot{\omega}}_C$ (\si{\kilogram\per\meter\cubed\per\second}), $t_3$, table}\label{trsW8WProdCsigco_1301}	
 	\centering
 	\includegraphics[width=0.95\textwidth]{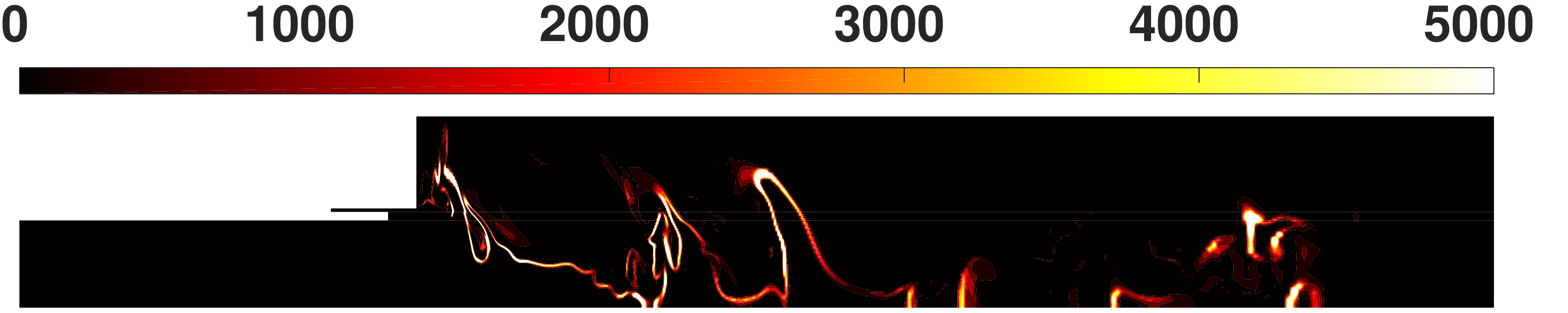}
 	\caption{$\widetilde{\dot{\omega}}_C$ (\si{\kilogram\per\meter\cubed\per\second}), $t_3$, $NN_b$}\label{trsNN8NNProdCsigco_1301}												
 		\end{subfigure}	
 	\caption{ 14-cm, transient: progress variable and PVRR snapshots from the $NN_b$-based and the table-based simulations}\label{trsvorttime}
 \end{figure}

 At three time points, from the transient simulations the snapshots of pressure and vorticity (\figurename{~\ref{trsPTtime}}), 
 progress variable, and PVRR (\figurename{~\ref{trsvorttime}}) are compared between the $NN_b$-based and the table-based simulations. The time snapshots are selected at $t_1=0.5$ \si{\milli\second}, $t_2=3.5$ \si{\milli\second}, and $t_3=6.5$ \si{\milli\second}, where the last time point is associated with the time that signals have reached to the limit cycle, while the other two time points are associated with the growing parts of the pressure signal.
 In the progress variable graphs, the flame front shape is predicted with a great accordance. The flame front is the curve (thin flame) (\figurename{~\ref{trsvorttime}}) that separates the low and high temperature zones.
These quantities are governed by both turbulence and acoustic behavior. The turbulent combustion causes the system to become chaotic, while the acoustic phenomena causes the system to resemble wave forms and modal behavior. 
 Accordingly, for variables that are dominantly affected by turbulence, the performance of the NN-based model can be assessed by the statistical characteristics and consequences of those variables; while for variables where the acoustic phenomena is the significant deriver, the NN-based model can be compared in a point-wise manner. One major statistical consequence for a variable such as PVRR is the modified RI that was shown in \figurename{ \ref{trsWCRInnb}} and \figurename{ \ref{trsWCRIrr}} for the NN-based and table-based simulations, respectively. The $mRI$ calculated from NN-based simulation shows highly consistent behavior to the $mRI$ calculated from table-based Simulation, despite the differences that can be spotted in the detailed behavior of PVRR in the time snapshots (e.g., \figurename{ \ref{trsNN8NNProdCsigco_1301}} vs. \figurename{ \ref{trsW8WProdCsigco_1301}}).
 
\subsubsection{ 9-cm case}\label{9cmcase}
\begin{figure}[hbt!]
	\begin{subfigure}{.5\textwidth}
		\centering
		\includegraphics[width=0.95\textwidth]{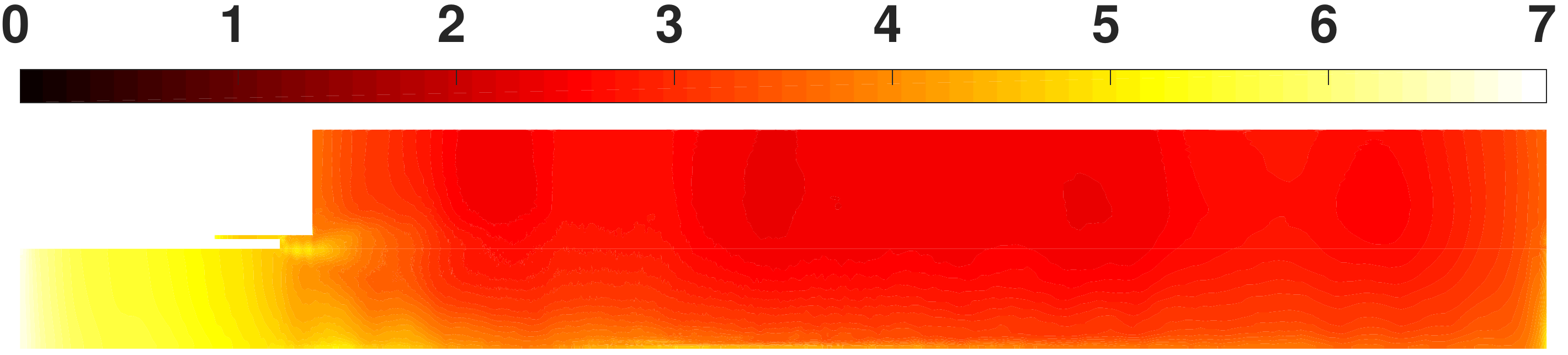}
		\caption{ Overall relative error (\%)}\label{9cmcorcont8}					
	\end{subfigure}
	\begin{subfigure}{.5\textwidth}
		\centering
		\includegraphics[width=0.95\textwidth]{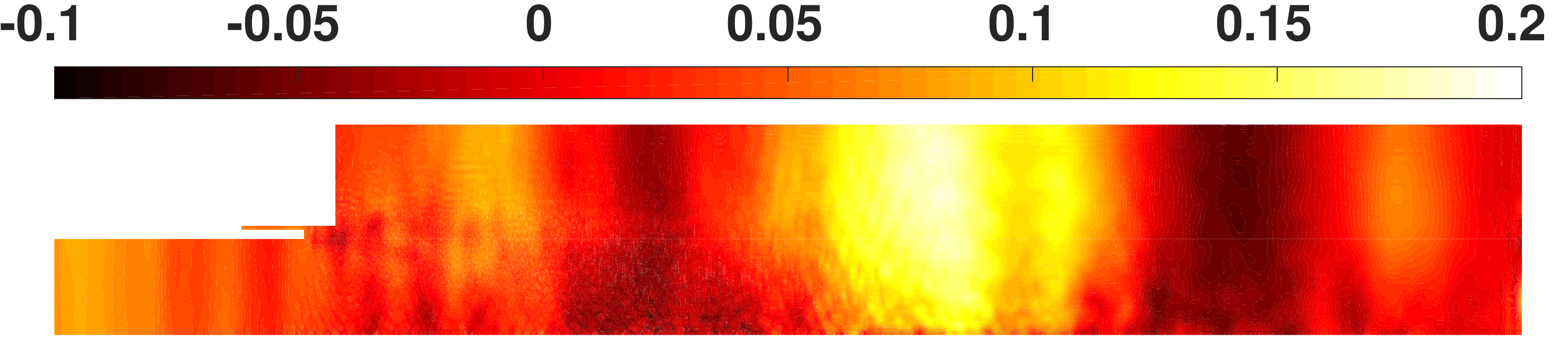}
		\caption{ Fluctuation correlations}\label{9cmcorrampcont8}				
	\end{subfigure}	
	\caption{Case with 9 cm oxidizer post: distribution of relative error (\%) and fluctuation correlation between pressure signals calculated from the $NN_b$-based and the table-based simulations }\label{9cmcorcomp1}
\end{figure}

 NN training was designed mostly for the purpose of reproducing the CFD simulation results of the CVRC experiment under unstable pressure oscillation conditions. This happens when the oxidizer post length is 14 \si{\centi\meter}. If the length is 9 \si{\centi\meter}, the pressure oscillations grow to a limit cycle, which is much smaller than in the 14-cm case. So the 9-cm is considered to be stable. Next, to validate its generality, the $NN_b$ model, which 
 was successfully implemented in the 14-cm transient case, is also implemented in the 9-cm case.
 The relative error between the NN-based and the table-based simulations is shown in \figurename{~\ref{9cmcorcont8}}. 
 The highest error occurs in the oxidizer post, as the NN is not forced to have certain values there. On the other hand, the table values are enforced at the non-reacting zones. In our future work, we will include such constraints in the NN development. 
 The correlation of the fluctuations of the pressure signals from their mean value from the $NN_b$-based and table-based simulations is shown in \figurename{~\ref{9cmcorrampcont8}}. 
 The oscillations are not that well correlated. The reason is that the oscillation amplitude (roughly about 50 \si{\kilo\pascal}) is very small relative to the mean value (roughly about 1600 \si{\kilo\pascal}). Essentially, the maximum oscillation amplitude is around 3\% of the mean value, and it can be considered as noise. 
 
\begin{figure}[hbt!]
	\begin{subfigure}{.5\textwidth}
		\centering
		\includegraphics[width=0.95\textwidth]{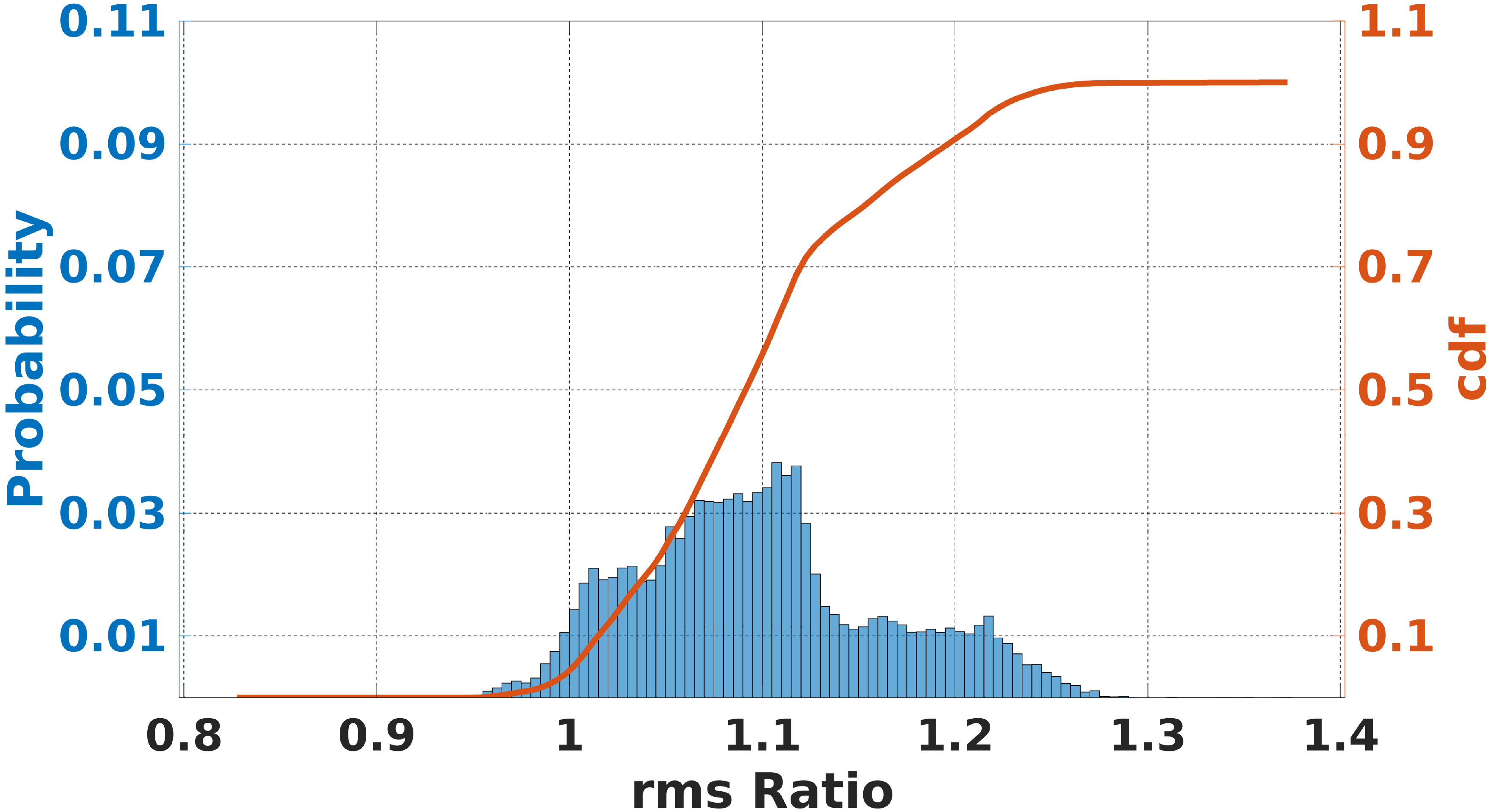}
		\caption{ rms ratio distribution}\label{9cmPwrathist8}			
	\end{subfigure}	
		\begin{subfigure}{.5\textwidth}
			\centering
			\includegraphics[width=0.95\textwidth]{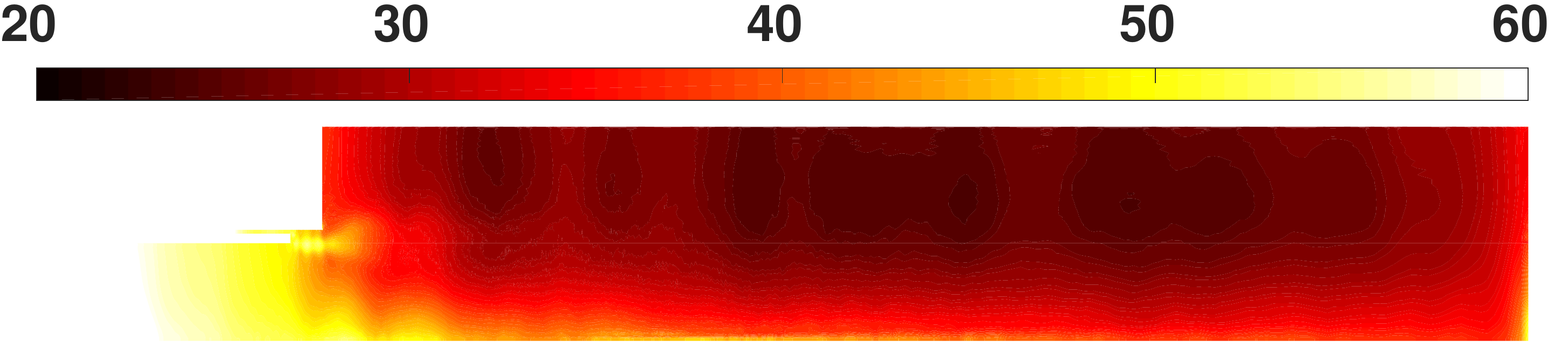}
			\caption{Table-based pressure signal rms (\si{\kilo\pascal})}\label{9cmpowTAB}
			\centering
			\includegraphics[width=0.95\textwidth]{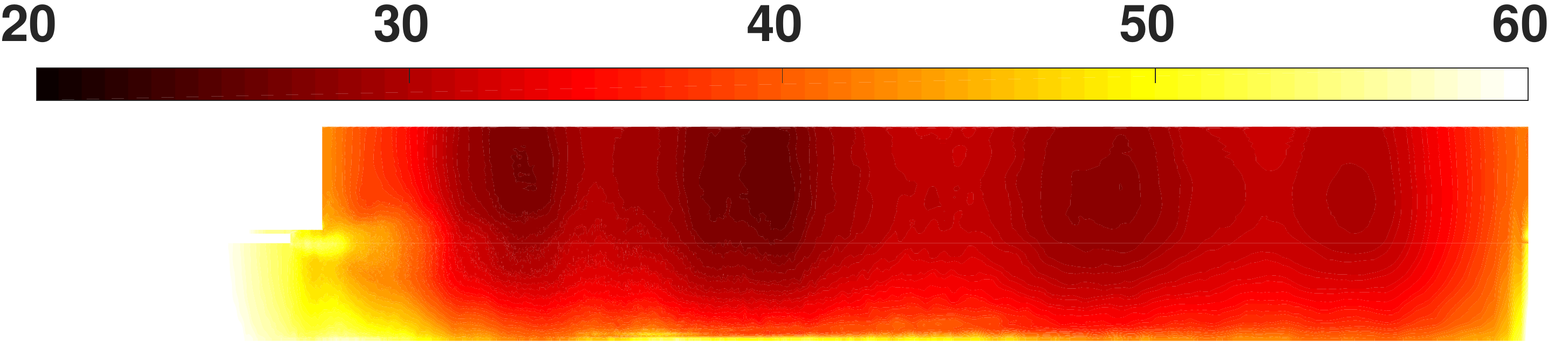}
			\caption{$NN_b$-based pressure signal rms (\si{\kilo\pascal})}\label{9cmpowNN8}
		\end{subfigure}
	\caption{Case with 9 cm oxidizer post: The distribution of $\kappa$calculated from $NN_b$-based (\ref{9cmpowNN8}) to the one calculated from table-based (\ref{9cmpowTAB}) simulations }\label{9cmcorrmeancomp1}
\end{figure}

The distribution ratio of pressure fluctuation rms in the $NN_b$-based simulation (\figurename{~\ref{9cmpowNN8}}) to that in the table-based simulation (\figurename{~\ref{9cmpowTAB}}) is shown in \figurename{~\ref{9cmPwrathist8}}. Most of the points have a similar rms value. Here again, the $NN_b$-based simulation overestimated the rms values. Both the rms values and their spatial gradients in the 9-cm case are also lower than the ones in the 14-cm unstable case. The pressure rms values are higher at the inlet, and they become smaller throughout the combustor.

\begin{figure}[hbt!]
 			\begin{subfigure}{.5\textwidth}
 				\centering
 				\includegraphics[width=0.95\textwidth]{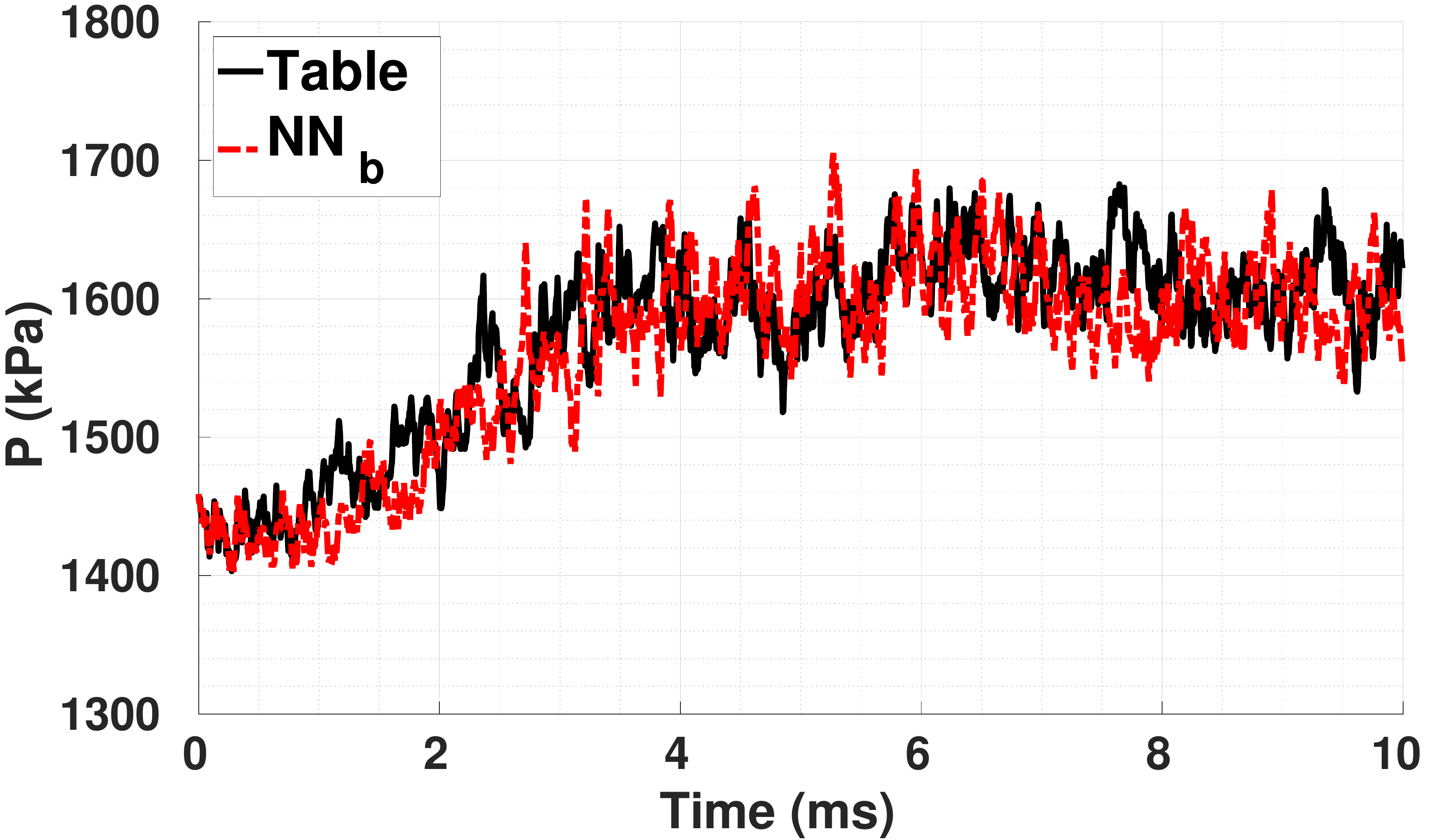}
 				\caption{ Pressure signal at $x=10$ \si{\centi\meter}}\label{9cmdyn843_404}		
 			\end{subfigure}		
 			\begin{subfigure}{.5\textwidth}
 				\centering
 				\includegraphics[width=0.95\textwidth]{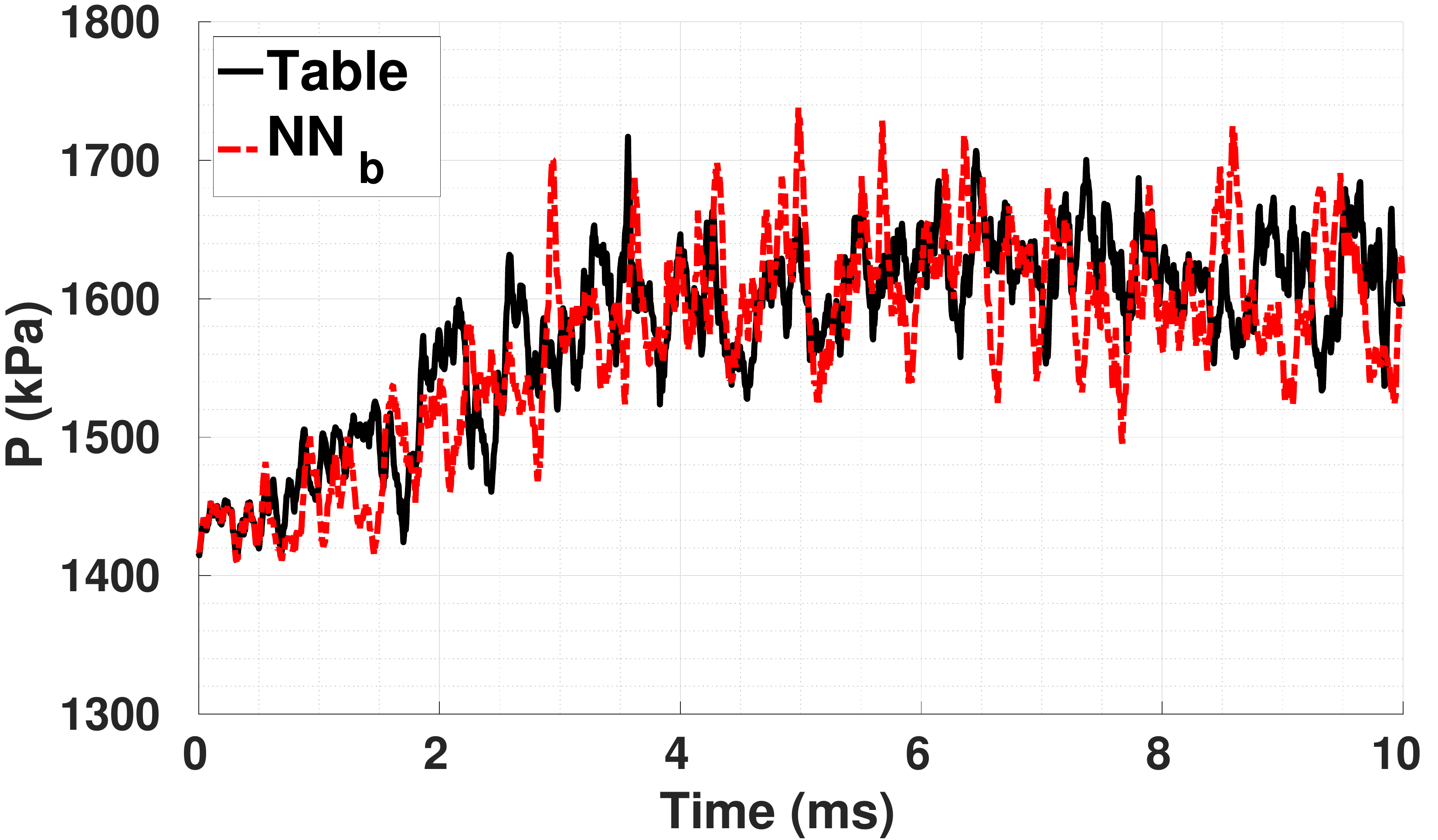}
 				\caption{ Pressure signal at $x=37$ \si{\centi\meter}}\label{9cmdyn843_645}		
 			\end{subfigure}				
 			\caption{Case with 9 cm oxidizer post: comparison of pressure signals 
 				on the wall, at pressure antinode and near the nozzle, between the $NN_b$-based and the table-based simulations} \label{9cmPsigcomp1}		
 		\end{figure}

 		Figure{ \ref{9cmPsigcomp1}} compares the pressure signal
 		at the antinode (10 \si{\centi\meter}), and near the nozzle (37 \si{\centi\meter}) on the top wall, calculated from the $NN_b$-based simulation with that from the table-based simulation. 
 		At the antinode, the correlation is -3.68\%, and near the nozzle, the correlation is 0.32\%. At these points, overall relative errors are 2.71\% and 3.25\%, respectively. The rms ratio is 108.32\%, and the mean value is estimated with 0.8\% error at $x=10$ \si{\centi\meter}. At $x=37$ \si{\centi\meter}, the rms ratio is 125.11\%, and the mean value is estimated with 0.8\% error. The low correlations between the fluctuations of signals are because of the noisy nature of the pressure waveform in this configuration.
 		Here, we are not interested in capturing noise and we did not pursue refining the NN for a better performance.
 		However, although it is harder to generate a fully correlated signal, the modal frequencies and signal trends are similar between the NN-based and table-based simulations.
 		
\begin{figure}[hbt!]
	\begin{subfigure}{.33\textwidth}
	\centering
	\includegraphics[width=1\textwidth]{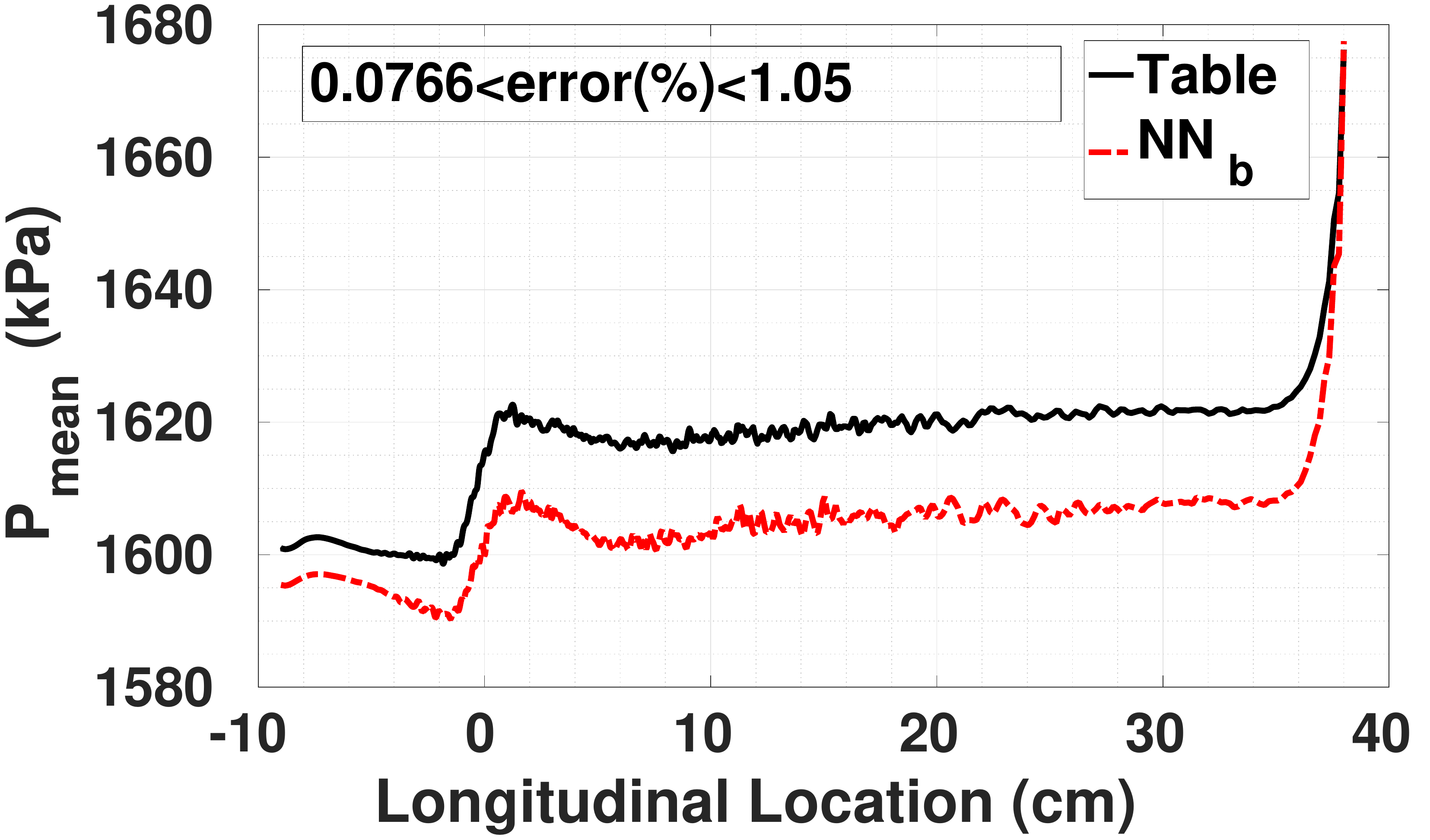}
	\caption{ Mean pressure (\si{\kilo\pascal})}\label{9cmPMS8mean}				
		\end{subfigure}
			\begin{subfigure}{.33\textwidth}
				\centering
				\includegraphics[width=1\textwidth]{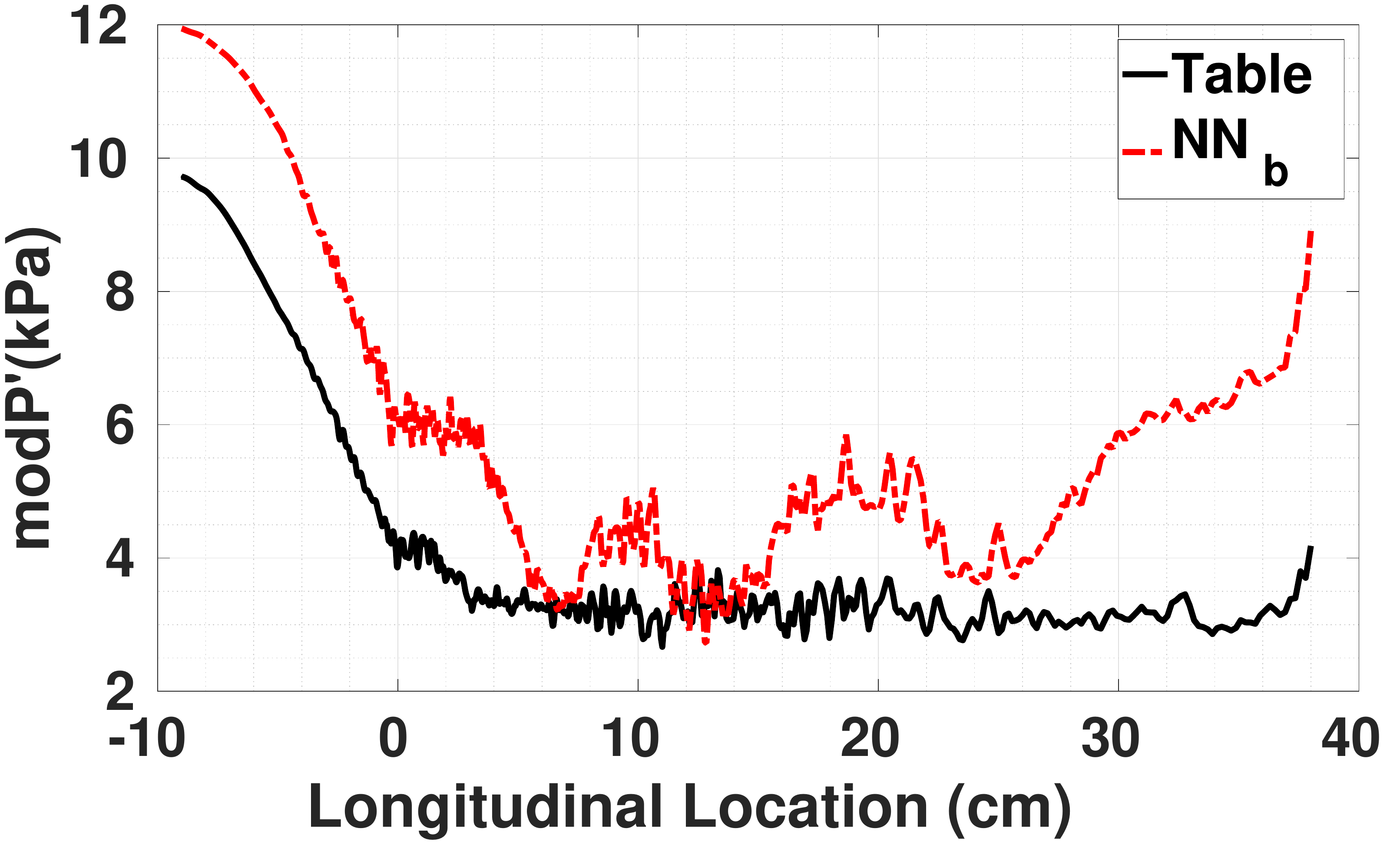}
				\caption{ First mode shape (\si{\kilo\pascal})}\label{9cmPMS8first}							
			\end{subfigure}
			\begin{subfigure}{.33\textwidth}
				\centering
				\includegraphics[width=1\textwidth]{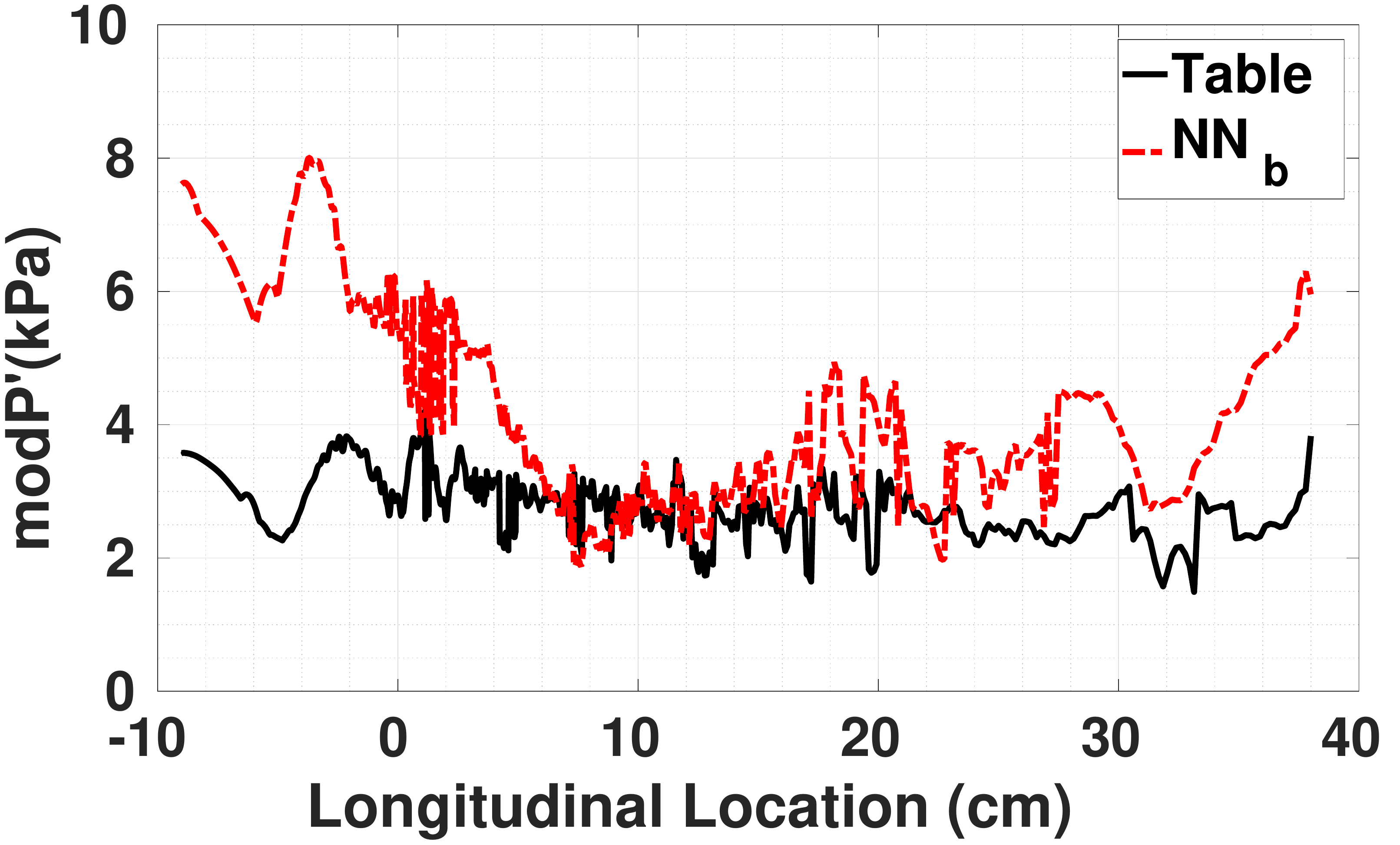}
				\caption{ Second mode shape (\si{\kilo\pascal})}\label{9cmPMS8second}							
			\end{subfigure}			
			\caption{9-cm case: comparison of pressure mean and the first and second mode shape between $NN_b$-based and table-based simulations} \label{9cmPMS8}	
\end{figure}

 The pressure mode shape shows similar behavior in \figurename{~\ref{9cmPMS8}} for the mean value and the first and the second mode shapes, where the modes are not predicted as well as the mean. 
The first and the second modes in the 9-cm case have relatively small magnitudes; also, the magnitude of the modes are approximately equally distributed. In the combustion zone, the first mode is approximately 0.31\% of the mean in the NN-based simulation, and 0.22\% in the table-based simulation. Similarly, the second mode is approximately 0.25\% of the mean in the NN-based simulation, and 0.19\% in the table-based simulation in the combustion zone.

\begin{figure}[hbt!]
				\begin{subfigure}{.5\textwidth}
					\centering
					\includegraphics[width=0.95\textwidth]{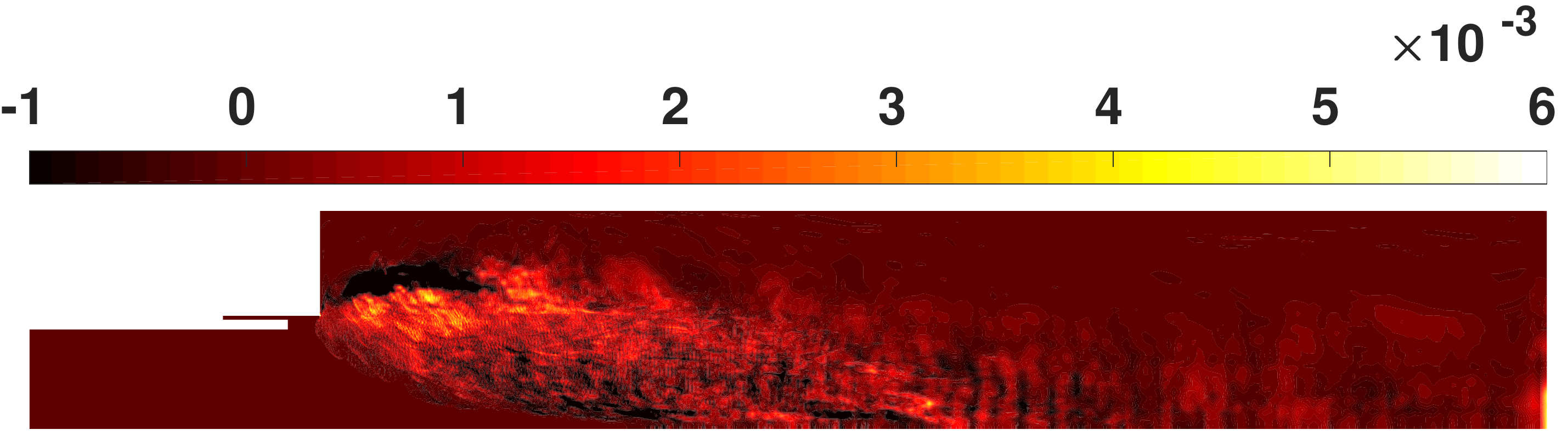}
					\caption{ Table-based CFD: \textit{mRI}}\label{9cmWCRIrr}	
					\centering
					\includegraphics[width=0.95\textwidth]{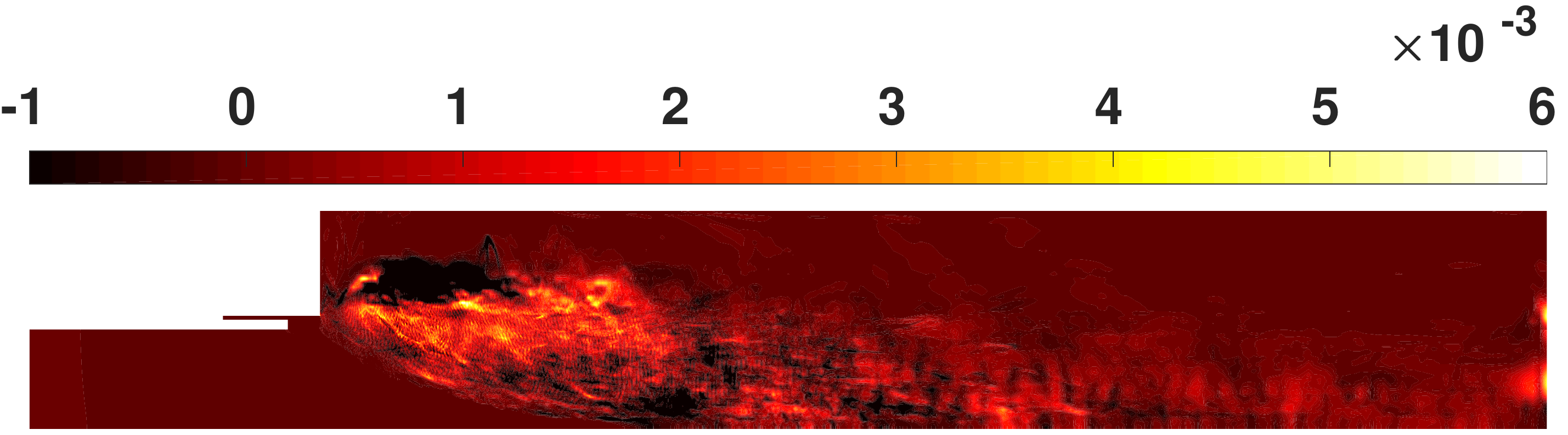}
					\caption{$NN_b$-based CFD: \textit{mRI}}\label{9cmWCRInnb}							
				\end{subfigure}		
				\begin{subfigure}{.5\textwidth}
					\centering
					\includegraphics[width=0.95\textwidth]{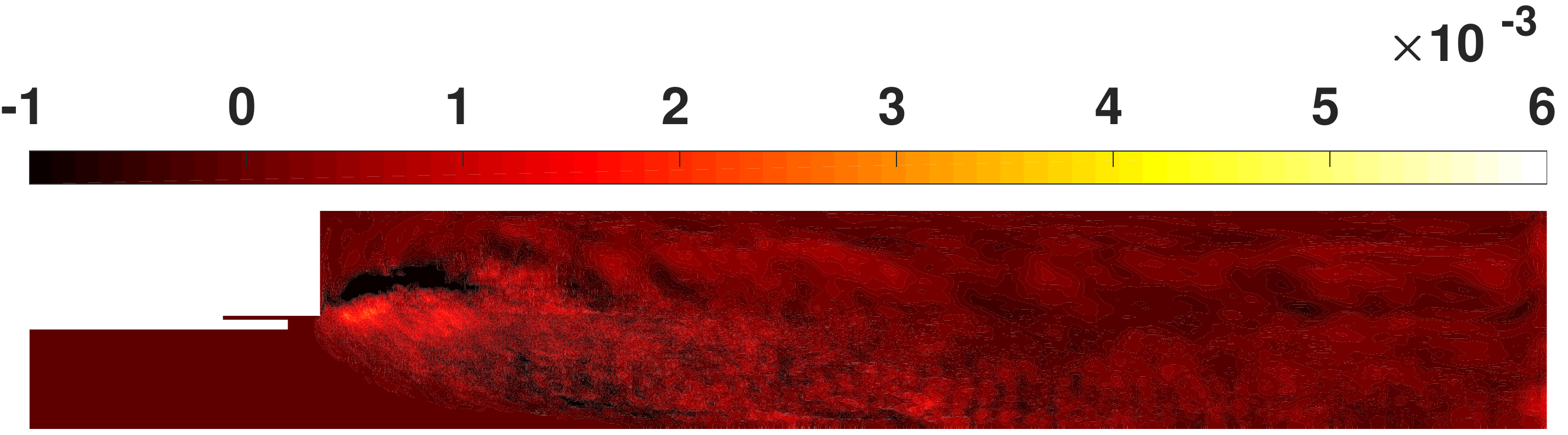}
					\caption{ Table-based CFD: \textit{RI}}\label{9cmHRRRIrr}
					\centering
					\includegraphics[width=0.95\textwidth]{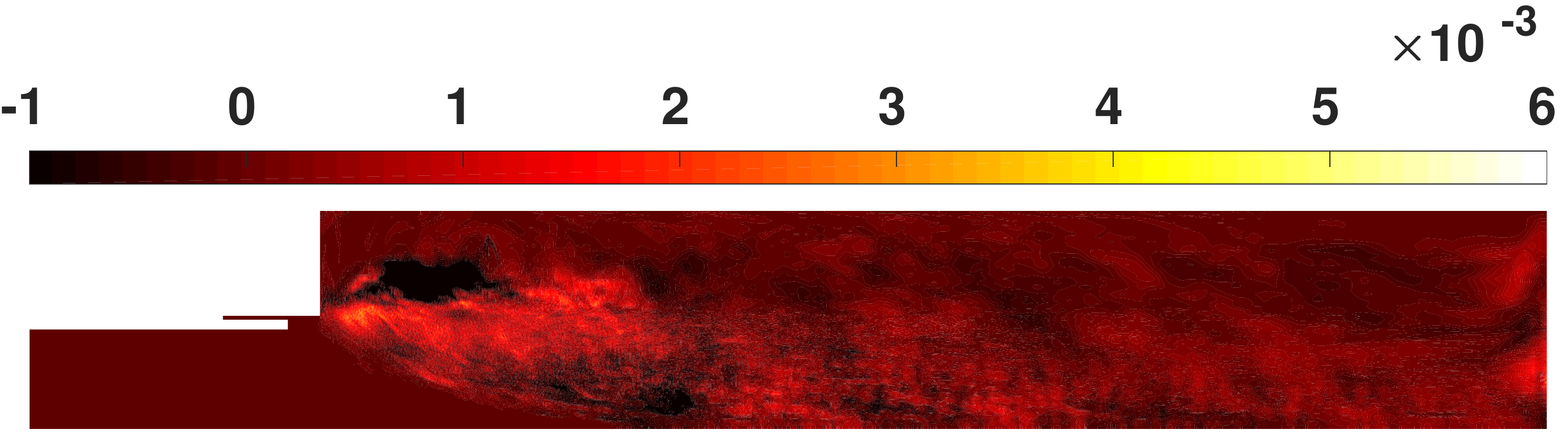}
					\caption{ $NN_b$-based CFD: \textit{RI}}\label{9cmHRRRInnb}
					\centering							
				\end{subfigure}		
				\caption{Case with 9 cm oxidizer post: comparison of \textit{RI} and \textit{mRI} from the $NN_b$-based and the table-based simulations} \label{9cmRIlabcomp1}		
			\end{figure}

		Since the 9-cm oxidizer post configuration leads to a stable configuration, the observed pressure fluctuations resembles relatively small amplitude. Essentially, the dynamic of the system is governed significantly by the turbulent combustion rather than acoustic behavior. The turbulence leads to a chaotic behavior in fluctuations for different quantities. Therefore, it is expected from a model to regenerate similar statistical behavior, such as Rayleigh Index, rather than regenerating exactly the same solution as the reference model.
The local \textit{mRI} and \textit{RI} are compared for the two simulations in \figurename{~\ref{9cmRIlabcomp1}}. Comparing \figurename{~\ref{9cmWCRIrr}} with \figurename{~\ref{9cmWCRInnb}} and comparing \figurename{~\ref{9cmHRRRIrr}} with \figurename{~\ref{9cmHRRRInnb}} show that the $NN_b$-based simulation is very similar to the table-based one in capturing the behavior near the corner and the \textit{RI} shape. Since the 9-cm case is a stable one, the magnitude of \textit{RI} and \textit{mRI} are lower than those in the 14-cm case. Low \textit{RI} shows that there is no instability occurring. 

\subsection{ Computational cost}
The computational cost for the NN-based and table-based simulations are compared in \tablename{ \ref{comptime}}. One millisecond (200 snapshots) of data is generated from the dynamic equilibrium test cases for each of the simulations. The required CPU time to generate each of the snapshots is calculated for each of the three simulations; the minimum, average, standard deviation, and the maximum of the required CPU time for generating each simulation time step are reported in \tablename{ \ref{comptime}}. The table-based simulation is considered as the reference, and the relative values of these quantities to the reference are also reported for the $NN_a$ and $NN_b$. The average CPU times per time step for the $NN_a$-based and the $NN_b$-based simulations are around 39.2 and 100.7 times the average CPU time per time step of the table-based simulation. The table-based simulation is essentially the same code that was developed and discussed in \cite{Tuan1}. \citet{Tuan1}, reported a 0.28 core-hour per millisecond for the computational cost of an axisymmetric simulation with $6.26e4$ grid points based on a flamelet model for combustion of 27 species, while a similar work, discussed in \cite{Harvazinskithesis}, reported 53 core-hour per millisecond for the computational cost of an axisymmetric simulation with $5.5e4$ grid points based on a general equation and mesh solver (GEMS) for combustion of 4 species. The current work uses $1.375e5$ grid points.
Also, one should note that the computational costs for the NN-based simulations reported here are based on a non-optimized code; there is a potential to reduce the NN-based simulation computational cost considerably by applying methods to parallelize and optimize the code. Yet, optimizing the computational cost was not in the scope of this work.

\begin{table}[hbt!]
	\centering
	\caption{CPU time per time step between two consecutive time samples statistics comparison among table-based, $NN_a$-based, and $NN_b$-based simulations for 14-cm dynamic equilibrium case} \label{comptime}
\begin{tabular}{|c|c|c|c|c|c|c|c|c|}
		\hline
		\textbf{CPU Time per }& \multicolumn{2}{c|}{\textbf{minimum}} & \multicolumn{2}{c|}{\textbf{mean}} & \multicolumn{2}{c|}{\textbf{standard deviation}} & \multicolumn{2}{c|}{\textbf{maximum}} \\ \cline{2-9} 
		\textbf{Time Step (\si{\second})} & \textbf{Absolute} & \textbf{Relative} & \textbf{Absolute} & \textbf{Relative} & \textbf{Absolute} & \textbf{Relative} & \textbf{Absolute} & \textbf{Relative} \\ \hline
		\textbf{Table} & 15.000 & 1.000 & 18.407 & 1.000 &5.574 &	1.000 & 53.000 & 1.000 \\ \hline
			\textbf{$NN_a$} & 558.000	&37.200&721.201&	39.181&	255.471	&45.832&1995.000&	37.642 \\ \hline
			\textbf{$NN_b$} & 1444.000 &	96.267	&1841.935&	100.067& 778.595&	139.681&	7072.000&	133.434 \\ \hline
	\end{tabular}
\end{table}

Although NN-based closure models increase the computational cost over a flamelet table model, they reduce the required memory significantly. 
	The flamelet model requires around 1.2 GB of the hard disk memory, while $NN_a$ and $NN_b$ require around 144.2 kB and 376 kB, respectively, to be stored on our computer machine. We estimated that loading NN-based models into CFD causes the required RAM to increase less than 1 MB, while loading the table-based model into CFD increases the required RAM around 276 MB. The difference between the volume of data when it is stored on the hard disk vs. when it is loaded in the RAM is because of the data on hard disk is stored as text file, while it is loaded as numbers in computer RAM.
	The very low required memory by NN-based models paves the way for utilization of Graphic Processing Units (GPUs) and consequently high levels of parallelization, which can significantly reduce the computational cost. Particularly, in 2010, \cite{GPUERA} provided examples of 16 to 137 times speed up in fluid mechanics related problems when they are implemented on GPU.
	
\section{Conclusion}\label{conclusion}
In this study, two deep learning neural network sets with different levels of complexity were developed to represent a flamelet model in turbulent combustion with unsteady pressure. The design of the networks was explained, including input-output and training set selection. The goal is to explore the capability of neural networks as a tool for combustion modeling. 
 The two developed models ($NN_a$ and $NN_b$) are first validated by testing on the flamelet table data, as an offline test, and then validated by being implemented in CFD simulations of different cases and compared with the table-based simulations. These simulations include dynamic equilibrium and transient simulations on an unstable rocket configuration (14-cm oxidizer CVRC), and transient simulation in a stable configuration (9-cm oxidizer CVRC). 
 $NN_b$, which contains more layers than $NN_a$, has shown a better performance both in offline and online validation. In the offline test, the difference between their error is very small (up to 2\%). $NN_b$ wins over the $NN_a$ model in the accuracy competition. However, its required computational cost is 2.5 times of that for $NN_a$ for retrieving one set of outputs. 
 
 In the dynamic equilibrium case, the two NN-based models have similar performance according to measures such as overall relative error and fluctuation correlation. The $NN_a$-based simulation results were in agreement with the table-based simulation results. However, \textit{mRI} was underestimated near the dump plane through the $NN_a$-based simulation.
In the transient case, the main shortcoming of the $NN_a$-based simulation is in predicting the pressure limit cycle or rms of pressure fluctuation. On average, only 75\% of the fluctuation energy is captured by the $NN_a$ simulation; whereas, the $NN_b$ simulation can capture the amplitude with great consistency, close to 100\%. In the transient case, $NN_b$ is the proper model as it is capable of faithfully reproducing the unstable acoustic behavior.
The $NN_b$-based model was also implemented in a case with 9-cm oxidizer post, which was characterized as a stable configuration of CVRC. Although, the mean pressure were predicted in this test, the pressure fluctuation of the NN-based simulation did not follow those from table-based simulation exactly. In the stable case, the dynamics are significantly governed by the turbulent combustion rather than acoustic behavior. The significance of turbulence combustion causes the simulation to be resemble more chaotic behavior in different quantities. Particularly, our analysis in all three simulations show that NN-based simulation provides more correlated results with the table-based simulation, when the acoustic phenomenon dominates the system over the turbulence. 

Flamelet models provide a good stepping stone, as evidenced by the encouraging results shown in this work. The authors also hold the view that the framework presented in this paper can be applied on high-quality data from sources other than the flamelet table such as high fidelity LES or even DNS, where the cost saving of using machine learning models can be highly advantageous. 
NN models require much lower memory than the flamelet table. Although the data retrieval from a NN model is more time costly than reading the data from a look-up table, the computational cost is still lower than using the chemical kinetics solver. Our data retrieval code for NN-based CFD is not optimized. In our future work, the computational cost of NN modeling will be revisited after parallelizing and optimizing the codes for a GPU implementation. 

\section*{Acknowledgments}
This research was supported by the U.S. Air Force Office of Scientific Research under Grant FA9550-18-1-0392, with Mitat Birkan as the scientific officer.

 \bibliography{thesis}
\end{document}